\documentclass[11pt,twoside]{report}

\usepackage{fancyheadings}
\usepackage{amsfonts}
\usepackage{graphicx}
\advance\voffset by -1.0cm
\advance\hoffset by -0.54cm
\renewcommand{\baselinestretch}{1.2}

\begin{document}
\thispagestyle{empty}

\begin{center}

\vspace*{3cm}

{\Huge {\bf Grand Unified Models and Cosmology}}

\vspace{2cm}

{\bf \underline{Rachel JEANNEROT }}
%\vspace{.5cm}

{\em {\normalsize {Department of Applied Mathematics and Theoretical Physics}} 

{\normalsize {Silver Street, Cambridge CB3 9EW, UK}}

and 

{\normalsize {Newnham College, }} 
{\normalsize {Sidgwick avenue, Cambridge CB3 9DF, UK}}}

\vspace{1cm}

{\bf September 1996}

\vspace{5cm}

{\em A dissertation submitted for the degree of Doctor of Philosophy 

at the University of Cambridge. }

\end{center}

\pagenumbering{roman}

\thispagestyle{empty}
\vspace*{2cm}

\begin{center}
{\Large {\bf Declaration}}
\end{center}

\vspace{.5cm}

Apart from chapter 1, in which background material is reviewed, and
chapter 2, which is the result of work in collaboration with Dr
Anne-Christine Davis, the work presented in this dissertation is my
own and includes no material which is the outcome of work done in
collaboration. The work is original, and has not been submitted for
any other degree, diploma, or other qualification. 

The results of Chapters 2, 3 and 4 have been published as three
separate papers in Physical Review D \cite{paper1,paper2,paper3}. The
results of Chapter 5 have been published in
Phys. Rev. Lett. \cite{paper4}. 
\newpage

\thispagestyle{empty}
\cleardoublepage

\thispagestyle{empty}
\vspace*{2cm}

\begin{center}
{\Large {\bf Acknowledgements}}
\end{center}
\vspace{.5cm}

I am grateful to my supervisor, Dr Nick Manton, for helpful
discussions and advice. I would like to thank Dr A-C. Davis. I would
especially like to thank Professor Qaisar Shafi for his help and his
continuous interest in my work. I would like to thank Professor
Brandon Carter for his contagious enthusiasm and for discussions. I
would like to thank Dr Robert Caldwell for discussions.

I would like to thank everybody in my department and everybody in the
d{\'e}partement d'astrophysique relativiste et cosmologie at the
Observatoire de Paris in Meudon. I would like to thank my friend
Fr{\'e}d{\'e}ric, for his patience and his care. I would like to thank
my family. I would like to thank all my friends in and outside
Cambridge, without whom I wouldn't have been able to finish my Ph.D.  

Finally, I would like to thank Newnham College for financial support
and also Trinity College.

\thispagestyle{empty}
\cleardoublepage

\thispagestyle{empty}

\vspace*{2cm}

\begin{center}
{\huge {\bf Summary}}
\end{center}

\vspace{.5cm}

The cosmological consequences of particle physics grand unified 
theories (GUTs) are studied. Cosmological models are implemented in
realistic particle physics models. Models consistent from both
particle physics and  
cosmological considerations are selected. 

After a brief introduction to the big-bang cosmology and to the
particle physics standard model, the ideas of grand unification,
supersymmetry, topological defects and inflation
are introduced. Special emphasis is given to the physics of massive neutrinos and to supersymmetric GUTs. The
elastic and inelastic scattering of fermions off cosmic strings arising
in a nonsupersymmetric SO(10) model are first studied. It is then shown
that the presence of supersymmetry does not affect the conditions for topological defect formation. By studying
the impact of the spontaneous symmetry breaking patterns from
supersymmetric SO(10) down to the standard model on the standard
cosmology, through the formation of topological defects, and by 
requiring that the model be
consistent with proton lifetime measurements, it is shown that there
are only three patterns consistent with observations. Using this
analysis, a specific model is built. It gives rise to a false vacuum
hybrid  
inflationary scenario which solves the monopole problem. At the end of
inflation, cosmic strings form. It is argued that this type of
inflationary scenario is generic in supersymmetric SO(10) models. Finally, 
a new mechanism for baryogenesis is described. This works in
unified theories with rank greater or equal to five which contain an
extra gauge ${\rm U}(1)_{B-L}$ symmetry, with right-handed neutrinos and
$B-L$ cosmic string, i.e. cosmic strings which form at the $B-L$ breaking scale, where $B$ and $L$ are baryon and lepton numbers. 

\thispagestyle{empty}
\cleardoublepage

\thispagestyle{empty}
\vspace*{7cm}

\renewcommand{\baselinestretch}{1} 

\hspace *{6cm} \begin{minipage}{10cm}
{\footnotesize
\begin{quotation}
\begin{center}
KALIAYEV,{\it {\'e}gar{\'e}.}
\end{center}
Je ne pouvais pas pr{\'e}voir... Des enfants, des enfants surtout. As-tu regard{\'e} des enfants? Ce regard grave qu'ils ont parfois... Je n'ai jamais pu soutenir ce regard... Une seconde auparavant, pourtant, dans l'ombre, au coin de la petite place, j'\'etais heureux. Quand les lanternes de la cal\`eche ont commenc\'e \`a briller au loin, mon coeur s'est mit \`a battre de joie, je te le jure. Il battait de plus en plus fort \`a mesure que le roulement de la cal\`eche grandissait. Il faisait tant de bruit en moi. J'avais envie de bondir. Je disais `` oui, oui ''...Tu comprends?
\begin{quote}
{\it Il quitte Stepan du regard et reprend son attitude affaiss\'ee. }
\end{quote}
J'ai couru vers elle. C'est \`a ce moment que je les ai vus. Ils ne riaient pas, eux. Ils se tenaient tout droit et regardaient dans le vide. Comme ils avaient l'air triste! Perdus dans leurs habits de parade, les mains sur les cuisses, le buste raide de chaque c\^ot\'e de la porti\`ere! Je n'ai pas vu la grande duchesse. Je n'ai vu qu'eux. S'ils m'avaient regard\'e, je crois que j'aurais lanc\'e la bombe. Pour \'eteindre au moins ce regard triste. Mais ils regardaient toujours devant eux.
\begin{quote}
{\em Il l\`eve les yeux vers les autres. Silence. Plus bas encore.}
\end{quote}
Alors je ne sais pas ce qui s'est pass\'e. Mon bras est devenu faible. Mes jambes tremblaient. Une seconde apr\`es, il \'etait trop tard. ({\it Silence. Il regarde \`a terre.}) Dora, ai-je r\^ev\'e, il m'a sembl\'e que les cloches sonnaient \`a ce moment l\`a?

\vspace{.5cm}

Albert CAMUS, {\em Les justes}, ACTE II.

\end{quotation}

}

\end{minipage}

\thispagestyle{empty}
\cleardoublepage

\renewcommand{\baselinestretch}{1.655} 

\tableofcontents
\newpage
\thispagestyle{empty}
\cleardoublepage

\chapter{Introduction}
\label{chap-chap1}
\pagenumbering{arabic}

\section{Particle physics and the standard cosmology}

The particle physics standard model and theories beyond, such as grand
unified theories or supersymmetry, are essential ingredients for any  
successful description of the evolution of the universe from very
early time until today. Particle physics is particularly necessary for
the understanding of the very early universe. It is also the only way
to solve  
some cosmological problems, such as the baryon asymmetry of the universe or the
dark matter problem. Particle physics also plays an important role in
inflationary cosmology. This thesis is interested in this interplay
between particle physics and cosmology.

In this introductory chapter, we  briefly review the big-bang cosmology and the
particle physics standard model and outline their main problems. We 
show the need for theories beyond these standard models, and introduce
the ideas of grand 
unification, supersymmetry, topological defects and inflation. A 
general consequence of 
unified theories is that neutrinos acquire a mass. This has
interesting cosmological consequences. We discuss the physics of
massive neutrinos in Sec. \ref{sec-maj}. We compare Dirac and Majorana
cases and review the see-saw mechanism. Supersymmetric grand unified
theories (GUTs) have more predicted power than nonsupersymmetric
ones. In Sec. \ref{sec-susyGUTs}, we discuss the main features of
supersymmetric GUTs. We introduce the notions of 
superspace, superfields and superpotential, and explain how to
construct supersymmetric Lagrangians. For a review of cosmology the
reader is referred to 
Ref. \cite{KolbTurner}, for quantum field theories to Ref. \cite{QFT},
for topological defects to Ref. \cite{ShelVil}, for grand unified
theories to Ref. \cite{Langacker81} and for supersymmetry to
Ref. \cite{SUSY}. Throughout the manuscript, we will work in units where
$\hbar = c = k_B = 1$. All quantities will be expressed in GeV =
$10^9$ eV  such that 
\begin{equation}
[Energy] = [Mass] = [Temperature] = [Length]^{-1} = [Time]^{-1} . \nonumber
\end{equation}
Some useful conversion factors are
\begin{eqnarray}
&{\rm Temperature} : & 1 {\rm GeV} = 1.16 \times 10^{13} {\rm K} \nonumber\\
&{\rm Mass} : & 1 {\rm GeV} = 1.78 \times 10^{-24} {\rm g} \nonumber\\
&{\rm Time} : & 1 {\rm GeV}^{-1} = 6.58 \times 10^{-25} {\rm sec} \nonumber\\
&{\rm Length} : & 1 {\rm GeV}^{-1} = 1.97 \times 10^{-14} {\rm cm} .\nonumber
\end{eqnarray}
In these units, the Planck mass $M_{{\rm pl}}  = 2.18 \times 10^{-5} {\rm g}
= 1.22 \times 10^{19} {\rm GeV}$ and the gravitational constant $G =
M_{{\rm pl}}^{-2}$.

The hot big-bang cosmology is based upon the Robertson-Walker metric
which describes an homogeneous and isotropic universe. It can be
written in the form 
\begin{equation}
ds^2 = dt^2 - a(t)^2 \Big \{ {d r^2 \over 1 - k r^2 } + r^2 d\theta^2
+ r^2 sin^2 \theta d \phi^2 \Big \} 
\end{equation}
where ($r, \theta , \phi $) are spatial coordinates, referred to as
comoving coordinates. $a(t)$ is the cosmic scale factor and $k$
determines the geometry of space. With an appropriate rescaling of the
parameters, $k$ can be chosen to be $+1$, $0$ or $-1$ corresponding to
open, flat and closed geometries respectively. The value of $k$ and
the time dependence of the function $a$ can be determined by solving
Einstein's equations.  

In an homogenous and isotropic cosmology the stress-energy tensor is
taken to be that of a perfect fluid, $T_{\mu \nu } = diag(\rho
,-p,-p,-p)$, where $\rho (t)$ and $p(t)$ are the energy and pressure
densities respectively. Einstein's equation then lead to the
Friedman equation 
\begin{equation}
H^2 \equiv  ({\dot{a} \over a})^2 = {8 \pi G \rho \over 3} - {k\over
a^2} + {\Lambda \over 3} , \label{eq:Fried} 
\end{equation}
where $H(t)$ is the Hubble parameter which characterises the expansion
of the universe and $\Lambda $ is the cosmological constant. The
present-day value of the Hubble parameter $H_0 = 100 h_0 \; {\rm km
s^{-1} Mpc^{-1}}$ and observational bounds give $0.4 < h_0 < 1$.  

Assuming no cosmological constant, it is convenient to define the
`critical' energy density, 
\begin{equation} 
\rho_c = {3 H^2 \over 8 \pi G } = 1.88 \times 10^{-29} h^2 \; \; \;
{\rm kg m^{-3}} . 
\end{equation}
Defining the density parameter 
\begin{equation}
\Omega = {\rho \over \rho_c} ,
\end{equation}
 the Friedman equation (\ref{eq:Fried}) can be rewritten in the form 
\begin{equation}
{k\over  H^2 a^2} \equiv \Omega - 1 ,
\end{equation}
 and hence $\Omega >1$, $<1$, $=1$ for $k =+ 1$, $-1$ ,$0$. $\Omega_0
$ is a very difficult quantity to measure. Observational limits give
$0.1 \leq \Omega_0 \leq 2$ \cite{Omega}, and hence observations do not
tell us whether the universe is open, flat, or closed. Luminous matter
(stars and associated material) contributes to $\Omega_0 $ by a very
small amount, $\Omega_{Lum}\simeq 0.005$. Hence, since observations give
$\Omega_0 \geq 0.1$, we deduce that the missing energy density must be
in the form of 
non-luminous dark-matter.

Big-bang nucleosynthesis (BBN) predicts the abundances  of the
light elements ${\rm He}^3$, D, ${\rm
He}^4$ and ${\rm Li}^7$, with
respect to hydrogen, the most abundant element in the universe, as a
function of an adjustable parameter, the 
baryon-to-photon ratio, 
$\eta = {n_B \over n_\gamma}$. BBN
only works if, within the uncertainties, $ \eta  \simeq 2 \times
10^{-10} - 7 \times 10^{-10}$ \cite{Walker}. From the  
constraints on the parameter $\eta$ follows the constraints on the baryonic
dark-matter energy density $\Omega_B h_0^2  \simeq 0.007 - 0.025$. 

While a conservative limit for $\Omega_0$ is $\Omega_0 \geq 0.1$,
other observational limits give $\Omega_0 \geq 0.3$. $\Omega_0 \simeq
1$ is also favoured by theoretical arguments. These considerations,
together with the limits on the baryonic dark matter energy density,
strongly suggest that there must be some non-baryonic dark
matter. Some non-baryonic dark-matter 
 candidates are the neutrino, if massive 
\cite{neutrinoDM}, the lightest superparticle (LSP), if stable
\cite{LSPDM}, and the axion \cite{axion}. A dark-matter candidate
is called `hot' if it was moving at relativistic speed at the time at
which galaxies started to form, and it is called `cold' if it was
non-relativistic at that time. Massive neutrinos are hot whereas the
LSP and the axion are cold dark-matter candidates.

The hot big-bang cosmology  predicts the expansion of the universe and
the present abundances of the light-elements. Its best recent success is
the predicted perfect black-body spectrum of the cosmic background
radiation (CBR) measured by COBE (Cosmic Background Explorer Satellite) \cite{COBE1}. But in spite of all its
successes, the hot big-bang model faces problems and unanswered
questions remain. Observational
bounds give $0.1 \leq \Omega_0 \leq 2$, implying that $\Omega $ was
incredibly close to unity from very early times until now. This is
known as the flatness problem. Also, which process(es) lead to the baryon
asymmetry of the 
universe? Which process(es) can predict the small ratio $\eta$? What
is the nature of the dark-matter of the universe? How can the CBR look the 
same in all directions \cite{COBE2} when it comes from causally disconnected
regions of space? This is referred to as the horizon problem. What was
the source(s) 
of small and large scale structure formations? What is the origin of
the small fluctuations in the CBR
temperature?  What happened
before the Planck epoch?  To answer some of these questions, inflation was
constructed \cite{inflconstr1,inflconstr2}. Particle physics attempts to answer
the remaining questions.

The particle physics standard model is based on the ${\rm SU}(3)_c
\times {\rm SU}(2)_L \times {\rm U}(1)_Y$ gauge theory of the strong,
weak and electromagnetic interactions. The non-abelian ${\rm SU}(3)_c$ part
describes the strong interaction and the ${\rm SU}(2)_L \times {\rm
U}(1)_Y$ part describes the electroweak interactions which combine the weak and
electromagnetic interactions. At energies $\sim 100$ 
GeV, the standard model gauge group is spontaneously broken down to
${\rm SU}(3)_c \times 
{\rm U}(1)_Q$ by the vacuum expectation value of a single Higgs
field, which is an SU(2) doublet. The electric charge $Q$ is a linear 
combination of the weak isospin $T_{3L}$ and of the weak hypercharge $Y$.
The standard model contains 12 spin-1 gauge bosons: the 8 gluons of
${\rm SU}(3)_c$, the three $W_L^{\pm,0}$ of ${\rm SU}(2)_L$ and the
${\rm U}(1)_Y$ gauge boson (the $W_L^{\pm}$ bosons, the $Z$ boson and
the photon). There are three independent gauge coupling constants 
$g_1$, $g_2$ and $g_3$ associated with the three gauge groups ${\rm
U}(1)_Y$, ${\rm SU}(2)_L$ and ${\rm SU}(3)_c$ respectively. The inverse of the `structure constants', $\alpha_1^{-1} = {4 \pi \over g_1^2}$, $\alpha_2^{-1} = {4 \pi \over g_2^2}$ and $\alpha_3^{-1} = {4 \pi \over g_3^2}$, depend
logarithmically on the 
energy scale. Interpolating their low energy values measured at
LEP/SLC to very high energies, it is found that they roughly meet at
$10^{14} - 10^{15}$ GeV with values $\alpha_1^{-1} = \alpha_2^{-1} = 43$ and $\alpha_3^{-1} \simeq 38$. In the minimal supersymmetric extension of the
standard model, the three couplings all merge in a single
point at a scale $\simeq 2 \times 10^{16}$ GeV with a value $\alpha_1^{-1} = \alpha_2^{-1} = 25 \simeq \alpha_3^{-1}$ \cite{meet}.  

Only left-handed fermions take part in weak
interactions and hence 
only left-handed fermions transform non-trivially under ${\rm
SU}(2)_L$. The fact that right-handed particles do not take part in
weak interactions means that parity is maximally violated. 
The left-handed quarks and the
left-handed leptons are grouped into, respectively, three
SU(2) doublets, $\left (\begin{array}{c}  u_L \\ d_L  \end{array}\right
)$ and $ \left (\begin{array}{c} e_L\\ \nu_L \end{array}\right )
$. The right-handed fields are SU(2) singlets, $(e_R)$, $(u_R)$ and
$(d_R)$. The standard model does not contain right-handed
neutrino. Only quarks and gluons feel the strong interaction. Quarks
form colour triplets, gluons colour octets, whereas leptons, other gauge
bosons and the Higgs particle form colour singlets.

The standard model predicts the conservation of the colour and
electric charges. It also predicts the
conservation of baryon number $B$ and lepton numbers $L_e$, $L_\mu$ and
$L_\tau$ for each 
family. The standard model has been successfully tested at high
energy colliders, with some predictions checked up to accuracy as high
as 0.1\% \cite{partphysbook}. One predicted particle is
however missing in these tables \cite{partphysbook}: the electroweak
Higgs boson.

The particle physics standard model faces problems and difficulties,
as does the big-bang cosmology. It contains 18 free parameters which
are not predicted. It does not predict the fermion masses nor the
weak mixing angle. Why is the theory left-handed? Why is the electric
charge quantised?  
 Why do we observe three families? Why is the electroweak symmetry
breaking scale ($\sim 10^2$ GeV) so small compared to the Planck scale
($\sim 10^{19}$ GeV)? Where does gravity fit in?  The particle physics
standard model also misses interesting features, such as neutrino
masses, high-energy gauge coupling constant unification, and cold
dark-matter candidates. To solve these problems, theories beyond 
the standard model have been constructed. These include grand unified
theories (GUTs) and supersymmetry.

The basic idea of grand unification is to embed the standard model
gauge group ${\rm SU}(3)_c 
\times {\rm SU}(2)_L \times {\rm U}(1)_Y$ into a larger simple group G
which has a single gauge coupling constant $g$, so that the additional
symmetries restrict 
the number of unpredicted parameters. GUTs predict the unification at
high energies of the strong, weak and electromagnetic interactions
into a 
single force.  Any grand unified gauge group
${\rm G} \supset {\rm SU}(3)_c \times {\rm SU}(2)_L \times {\rm
U}(1)_Y$ should be consistent with low energy phenomena, and should be at
least rank four.  It should contain the 8 gluons, the two $W_L^\pm$ bosons,
the $Z$ boson and the photon; G will also contain extra gauge
bosons, some of which can violate baryon and/or lepton number. As a
consequence, GUTs predict proton decay. The particle representations of G 
should contain SU(2)
doublets and SU(3) triplets. G should contain complex
representations to which fermions will be assigned, so that low energy
structures emerge. Extra scalar bosons in 
appropriate representations must be introduced in order to get an
acceptable spontaneous symmetry breaking pattern from G down to the
standard model gauge group. GUTs with rank 
greater or 
equal to five also predict extra fermions. One of them, transforming as a
singlet under the standard model gauge group, can be identified with
a right-handed neutrino. Such unified theories predict non-zero
neutrino masses. 

GUTs also provide the standard scenario for 
baryogenesis, via the out-of-equilibrium decay of heavy Higgs and
gauge bosons which violate baryon and/or lepton number
\cite{KolbTurner}. GUTs with heavy Majorana
right-handed neutrinos provide the standard scenario for
leptogenesis \cite{Y86,Luty,Y93}.

The minimal GUT which unifies all matter, without introducing
`exotic' particles, except a right-handed neutrino, is SO(10) \cite{G74}. SO(10)
has a sixteen dimensional spinorial representation to which all
fermions, both left and right-handed (precisely, the left-handed and the charge
conjugate of the right-handed ones or vice-versa), belonging to a
single family, can be assigned. The sixteenth fermion transforms as a
singlet under ${\rm SU}(3)_c 
\times {\rm SU}(2)_L \times {\rm U}(1)_Y$ and hence can be identified
with a right-handed neutrino. SO(10) predicts non zero neutrino
masses. SO(10) contains 45 gauge bosons: the 8 gluons, the three
$W_L^{\pm,0}$, the three $W_R^{\pm ,0}$ of ${\rm
SU}(2)_R$ subgroup of SO(10), 30 gauge bosons which violate baryon
number, and one gauge boson $B'$ which violates 
$B-L$. The ${\rm U}(1)_Y$ associated gauge boson is a linear
combination of $W^0_R$ and $B'$. All the gauge bosons which are not
contained in the standard model must acquire masses well above the
electroweak symmetry breaking scale. GUTs such as E(6), E(7) or higher
rank GUTs 
predict more fermions and even more gauge bosons. They contain many
more parameters and so are more complicated than SO(10), but they do not give a
better fit to the low energy data.

We should also mention that the
minimal GUT, which is SU(5), is inconsistent with low energy
data. Nonsupersymmetric SU(5) is ruled out because of the gauge
coupling constant which do not all exactly meet, and supersymmetric SU(5) because
it is inconsistent with proton lifetime measurements. It should be
pointed out that in nonsupersymmetric grand unified theories with rank
greater or equal to five, one (or more) intermediate symmetry breaking
scale can be introduced; this scale can be chosen in such a way that
the three coupling constants meet in a single point at $\sim
10^{15}$ GeV. Also, in SU(5), 
fermions must be assigned to different representations; so SU(5)
does not unify matter, and neither does it predict right-handed
neutrinos.

GUTs introduce new symmetries between different kinds of matter and
between different kind of forces. 
Supersymmetry introduces a symmetry between fermions and bosons. It introduces
new particles, new fermions and new 
bosons. Each
particle of the nonsupersymmetric theory under consideration is
associated with a particle with the same quantum numbers except from
the spin; it differs from the spin of the nonsupersymmetric
particle by half a unit. These are referred to as
superparticles. Hence fermions with spin-${1\over 2}$ are 
associated with spin-0 bosons, called sfermions, gauge bosons have
spin-1 and are associated with spin-${1\over 2}$ fermions, called
gauginos, and Higgs bosons have spin-1 and are associated with spin-$1\over
2$ fermions, called Higgsinos. The minimal supersymmetric standard model (MSSM)
is the minimal supersymmetric extension of the standard
model.  It contains the usual quarks, leptons and gauge bosons and
their supersymmetric partners. It contains two 
Higgs doublets and their supersymmetric partners, one 
of which gives a mass to
up-type quarks and one 
which gives a mass to d-type quarks. The gauge symmetry of the MSSM
is still the ${\rm SU}(3)_c 
\times {\rm SU}(2)_L \times {\rm U}(1)_Y$ gauge theory. For instance,
the left-handed ${\rm SU}(2)_L$ fermion doublet $ \left (\begin{array}{c} e_L\\
\nu_L \end{array}\right ) $ is partnered with the scalar doublet $
\left (\begin{array}{c} \tilde{e}_L\\ \tilde{\nu}_L \end{array}\right )
$. The Feynman rules for the supersymmetric partners allow the same
interactions, although one must take into account the different spins
of the particles. 

An important assumption in
the MSSM is the imposition of a discrete symmetry, known as R-parity,
which takes values $+1$ for particles and $-1$ for superparticles. The
assumption of unbroken R-parity restricts the possible interactions in
the theory. It prevents rapid proton decay and stabilises the
lightest superparticle, thus providing a cold dark-matter
candidate. Some GUTs automatically conserve R-parity; this will be discussed
in Sec. \ref{sec-susyGUTs}.

The study of
supersymmetric GUTs is well motivated. Especially since, as was mentioned above, in the MSSM,
the gauge 
coupling constants have been shown to merge in a single point at $2
\times 10^{16}$ GeV. Further motivations are discussed in
Sec. \ref{sec-susyGUTs}.

A consequence of grand unified theories, supersymmetric or not, is the
formation of topological defects, according to the Kibble 
mechanism \cite{Kibble}, at phase transitions associated with
the spontaneous symmetry 
breaking of the grand unified gauge group G down to the standard model
gauge group.  Examples of
such defects are monopoles, cosmic strings or domain walls. The Kibble mechanism relies on the fact that the Higgs field mediating the SSB of a group G down to a subgroup H of G must have taken different values in different regions of space which were not correlated. When the topology of the
vacuum manifold ${{\rm G}\over {\rm H}}$ is non trivial, topological defects then form. Monopoles form when the vacuum manifold contains
non-contractible two-spheres, cosmic strings form when it contains
non-contractible loops and domain walls form when it is
disconnected. The topology of the vacuum manifold can be determined 
using homotopy theory. Hence, topological defects forming when G breaks down to H, can be classified in terms 
of the homotopy groups of the vacuum manifold ${\rm G}\over {\rm H}$
\cite{Kibble}. 
If the fundamental homotopy group $\pi_0 ({{\rm G}\over {\rm H}}) \neq I$ is
non trivial, domain walls form when G breaks down to H. If the
first homotopy group $\pi_1({{\rm G}\over {\rm H}}) \neq I$ is non
trivial, topological 
cosmic strings form. If the second homotopy group $\pi_2({{\rm G}\over 
{\rm H}}) \neq I$ is non trivial, monopoles form. If the subgroup H of
G breaks later to a subgroup K of H, the defects formed  
when G breaks to H can either be stable or rapidly decay. The
stability conditions are as follows. If the fundamental homotopy group
$\pi_0({{\rm G}\over {\rm K}})$ is non trivial, the walls are
topologically stable,  
if $\pi_1({{\rm G}\over {\rm K}})$ is non trivial, the strings are
topologically stable 
and if $\pi_2({{\rm G}\over {\rm K}})$ is non trivial, the monopoles are 
topologically stable down to K.

As mentioned above, we are especially interested in SO(10)
GUT. Consider then the phase transition associated with the 
breaking of {\rm SO}(10) down to a subgroup G of {\rm SO}(10), and apply
the above results to this particular case; it will be useful later
on.  Note first that when we write {\rm SO}(10) we mean 
its universal covering group Spin(10) which is simply
connected. Since Spin(10) is
connected and simply connected we have $\pi_2({{\rm SO}(10) \over {\rm G}}) = \pi_1({\rm G})$ and
$\pi_1({{\rm SO}(10) 
\over {\rm G}}) = \pi_0({\rm G})$ and therefore the formation of
monopoles and strings 
during the grand unified phase transition is governed by the non
triviality of $\pi_1({\rm G})$ and $\pi_0({\rm G})$ respectively. If G
breaks down 
later to a subgroup K of G, monopoles formed during the first phase transition
will remain topologically stable after the second phase transition if
$\pi_1({\rm K}) \neq I$. Strings formed during the first phase 
transition will be topologically stable after the next phase transition if
$\pi_0({\rm K}) \neq I$.

Domain walls, because they are 
too heavy, and monopoles, because they are too abundant according 
to the Kibble mechanism, if present today, would dominate the energy 
density of the universe and hence are in conflict with
the standard cosmology \cite{ShelVil}. On the 
other hand, cosmic strings may have interesting cosmological
consequences. They may help explain the formation of large scale structure, the
baryon asymmetry of the universe, temperature fluctuations in the CBR
\cite{ShelVil} 
and could be part of the missing matter 
\cite{Brandon}. The formation of topological defects is unavoidable.

A solution to the monopole and domain wall problems is inflation
\cite{KolbTurner,inflconstr1,inflconstr2}. The basic idea of inflation
is that there was an epoch in the very early universe when the
potential, vacuum energy density dominated the energy density
of the universe, so that the cosmic scale factor grew
exponentially. Regions initially within the causal horizon were then
expanded to sizes much greater than the present Hubble
radius. 

In the following chapters, we will be mainly interested in
hybrid inflation which was first introduced by Linde \cite{Linde}. 
Hybrid inflationary scenarios use two scalar fields, ${\cal S}$ and $\phi$. 
One, ${\cal S}$, is utilised to trap the second, $\phi$, in a false vacuum state. Inflation then takes place as the field ${\cal S}$ slowly moves. 
When the field
${\cal S}$ reaches its critical value, 
the $\phi$ field quickly changes, reducing its energy density, and inflation ends. Initial conditions
in hybrid inflationary scenario are `chaotic', thermal equilibrium is
not assumed, 
and thus hybrid inflation belongs to the general class of chaotic
inflation models \cite{Lindechaos}. We will be interested in models where the ${\cal S}$ field
is a scalar singlet and the $\phi$ field is a Higgs used to lower the
rank of the group under consideration by one unit.

Inflation can solve the monopole problem, introduced by grand
unified theories, and the big-bang horizon and flatness problems. Also,
quantum fluctuations in the fields driving inflation may have been
at the origin of the density perturbations that caused structure
formation and temperature fluctuations in the CBR. Hence inflation is
very promising.

Unfortunately, inflation usually requires very fine
tuning in the parameters; the scalar field which drives inflation must
have a very small self-coupling and the potential must be almost flat. The
aim is therefore to find a theory in which such a 
scalar field with flat potential naturally emerges, so that inflation
occurs `naturally'. This goal seems to have been attained in
supersymmetric grand unified models
\cite{ShafiDvaSch,Ed1,Laz,sugrainfl,Dvali}.

If inflation ends before the last spontaneous symmetry breaking down
to the standard 
model gauge group has taken place, topological defects, such as cosmic
strings, may still be present today. The formation of topological
defects at the end of inflation has been studied in Refs. \cite{Laz,
infltop}. Models in which monopoles or domain walls form at the end of
inflation are cosmologically unacceptable. On the other hand, if cosmic
strings form at the end of inflation, there may be important
cosmological consequences.  
Recall that cosmic strings are good candidates for structure formation
and that they may be necessary to explain the spectrum of temperature
fluctuations in the CBR and the baryon asymmetry of the universe. The
formation of cosmic strings at the end of inflation may be
necessary. Hence, a unified model with both inflation, which inflates
away all unwanted defects, and cosmic strings, forming at the end of
inflation, is quite attractive. However, the construction of such a
model is not easy.

In this thesis, we investigate some interesting features of cosmic
strings. We look at the possibility that cosmic strings arising in a
specific GUT model may catalyse proton decay and also that cosmic
strings with right-handed neutrinos trapped as transverse zero modes
in their core may be candidates for baryogenesis. We are especially
interested in groups with rank greater or equal to five for the
unified models, with 
particular interest in SO(10) GUT. We construct a
grand unified model which contains both inflation and cosmic strings,
as well as hot and cold dark-matter candidates and provides a
mechanism for baryogenesis.

Several years ago, Callan and Rubakov suggested that grand
unified monopoles could catalyse proton decay with a strong
interaction cross-section \cite{Rubakov}. This is known as the
Callan-Rubakov effect. More recently,  Perkins {\em et
al.}\cite{Perkins91} showed that the same effect could occur with
cosmic 
strings; they were using a toy model. In Chap. \ref{chap-chap2}, we
investigate if this
effect occurs when a specific GUT model is used. The model we use is a
nonsupersymmetric SO(10) model with SO(10) broken down to the standard
model via an ${\rm SU}(5) \times Z_2$ symmetry. We study the
scattering of fermions off the abelian string which forms when SO(10)
is broken down to ${\rm SU}(5) \times Z_2$. The elastic scattering
cross-sections and the inelastic scattering cross-sections due to the
coupling of fermions to gauge fields which violate baryon number in
the 
core of the strings are computed.

The aim of Chapters \ref{chap-chap3} and \ref{chap-chap4} is to
construct a grand unified model consistent with observations. We
require that the model undergoes a period of inflation which arises
naturally from the theory, and that cosmic strings form at the end of
inflation. This 
work was motivated by the work of Dvali {\em et al.}\cite{ShafiDvaSch}
which ``outlined how supersymmetric models can lead to a successful
inflationary scenario without involving small dimensionless
couplings''. 

In 
Chap. \ref{chap-chap3}, we give the motivations for choosing
supersymmetric SO(10) as grand unified theory. We analyse the
conditions for topological defect formation in supersymmetric
theories. The 
formation of topological defects in all possible spontaneous symmetry breaking
patterns from supersymmetric SO(10) down to the standard
model gauge group are then studied. Assuming that the universe
underwent a period of inflation, as will be  
described in Chap. \ref{chap-chap4}, and by 
requiring that the model be consistent with proton lifetime
measurements, we select the only 
patterns consistent with 
observations. 

In Chap. \ref{chap-chap4}, we describe an inflationary scenario which
arises naturally from the theory in supersymmetric SO(10) models. We
give the 
form of the general superpotential which gives rise to an
inflationary period and implements the spontaneous symmetry breaking
pattern. We 
give a brief description of the evolution of the fields. We then
consider one of the model selected in Chap. \ref{chap-chap3}. 
 We argued above that a model with both inflation and cosmic strings
was a good framework for describing the very early universe. We
therefore choose such a model. We construct the full
superpotential. We study the evolution of the fields and the 
formation of topological defects before and at the end of
inflation. Monopoles are inflated away and cosmic strings form at the
end of inflation. The properties of the strings are presented. The
dark-matter components of the model are analysed. The properties of a mixed
scenario of cosmic string and inflationary large-scale structure
formation are briefly discussed.

Motivated by baryogenesis via leptogenesis scenarios
\cite{Y86,Luty,Y93} introduced a decade ago \cite{Y86}, we present a
new mechanism for leptogenesis in Chap. \ref{chap-chap5}. The basic
idea is that the out-of-equilibrium decay of heavy Majorana
right-handed neutrinos released by decaying cosmic string loops
produces a lepton asymmetry which is then converted into a baryon
asymmetry. Theories in which such a scenario does work are identified.

We finally conclude in Chap. \ref{chap-chap6}, summarising the results
of the works carried 
out in the previous Chapters. We also give some ideas for future
research works.

\section{Massive neutrinos}
\label{sec-maj}

\subsection{Introduction}

Grand unified theories with rank greater or equal to five predict
extra fermions in addition to the usual quarks and leptons. One of
them (one per fermion generation), a singlet under the standard model gauge
group, can be interpreted as a right-handed neutrino. As a
consequence, the observed neutrinos acquire a non-zero mass. Right-handed
neutrinos are predicted to be massive Majorana particles. For a review of
massive neutrinos, the reader is referred to
Refs. \cite{Kayserbook,Bilenky}.

The existence of non-zero neutrino mass may have important
cosmological and astrophysical implications \cite{neutrinoDM}. In
particular, if neutrinos are massive, they will account for the
invisible matter of the universe. Recent analytical work on large
scale structure formation strongly favours cosmological models with both
hot and cold dark matter \cite{HCDM}. Massive neutrinos are the
natural candidate for the hot dark matter component. 

Non-zero neutrino
masses also lead to neutrino flavour oscillations. These can explain,
via the MSW mechanism \cite{MSW}, 
 the apparent deficit in the flux of solar neutrinos as measured
by the Homestake, Kamiokande, GALLEX and SAGE experiments. They can
also explain the 
small ratio of muon to electron atmospheric neutrinos. Nevertheless, we should
point out that the solar neutrino problem may still be a terrestrial
or an astrophysical problem \cite{Dar}. Only future experiments such
as SNO, Superkamiokande, BOREXINO and HELLAZ, will be able to supply
definitive proof that the solar neutrino problem is a consequence of
physics beyond the electroweak standard model \cite{Dar}. 

We conclude that there is strong
theoretical support for neutrinos to be massive 
and probably also soon direct experimental support.  

 The experimental bounds on the neutrino masses, which may be
Dirac or Majorana type, are as follows \cite{partphysbook} 
\begin{eqnarray}
m_{\nu_e} &\leq & 15 {\rm eV} \\
m_{\nu_\mu} &\leq & 170 {\rm keV} \\
m_{\nu_\tau} &\leq & 24 {\rm MeV} .
\end{eqnarray}

In the following chapters, we will be
particularly interested in unified theories which contain an extra
${\rm U}(1)_{B-L}$ gauge symmetry, such as the 
simple U(1) extension of the standard model, ${\rm SU}(3) \times {\rm SU}(2)
\times {\rm U}(1) \times {\rm U}(1)'$, left-right models or 
SO(10) and  
E(6) GUTs. In such theories, right-handed neutrinos acquire a heavy
Majorana mass 
at the $B-L$ breaking scale. 

Majorana particles are introduced in Sec. \ref{sec-Majorana}. In
Sec. \ref{sec-masses}, we review the see-saw mechanism. In
Sec. \ref{sec-quant}, we give the expansions of both Dirac and Majorana
quantised fermion fields. The difference between Dirac and Majorana
fermion fields is then obvious.

\subsection{Majorana particles}
\label{sec-Majorana}

Massive charged spin-${1\over 2}$ particles can be described by
four-component spinor fields $\Psi$ which can be written as the sum of
two two-component Weyl (or chiral) spinors, 
\begin{equation}
\Psi = \Psi_L + \Psi_R = {1+\gamma^5 \over 2} \Psi + {1-\gamma^5 \over
2} \Psi = P_L \Psi + P_R \Psi 
\end{equation}
where, in the case that the mass of the particle can be ignored,
$\Psi_L$ ($\Psi_R$) annihilates a left (right)-handed particle or
creates a right (left)-handed particle.

Under charge conjugation, the fields $\Psi$ and $\overline{\Psi } =
\Psi^\dagger \gamma^0$ transform as follows, 
\begin{eqnarray}
\Psi &\stackrel{C}{\rightarrow} & \Psi^c \equiv C \overline{\Psi}^T =
C \gamma^0 \Psi^*  \\ 
\overline{\Psi} &\stackrel{C}{\rightarrow} & \overline{\Psi}^c \equiv
- \Psi^T C^{-1} \end{eqnarray} 
where $C$ is the charge conjugation matrix defined by $C^{-1}
\gamma_\mu C = - \gamma_\mu^T$ and satisfies $C = - C^{-1} = - C^\dagger
= - C^T$. Also we have 
\begin{eqnarray} 
\Psi_{L,R} & \stackrel{C}{\rightarrow} & \Psi^c_{L,R} \equiv P_{L,R}
\Psi^c = C \overline{\Psi}_{R,L}^T = C \gamma_0^T \Psi_{R,L}^*
\label{eq:psicLR}\\ 
\overline{\Psi}_{L,R} & \stackrel{C}{\rightarrow} &
\overline{\Psi}^c_{L,R} \equiv - \Psi^c_{R,L} C^{-1} , 
\end{eqnarray}
hence it follows that
\begin{eqnarray}
\Psi_{L,R} &=& C \overline{\Psi}^{cT}_{R,L} \\
\overline{\Psi}_{L,R} &=& - \Psi^{cT}_{R,L} C^{-1} .
\end{eqnarray}

A Majorana spinor $\Psi_M$ is by definition a spinor which is
proportional to its own charge conjugate,  
\begin{equation}
\Psi_M^c = \pm \Psi_M \label{eq:majcond}
\end{equation}
and thus has only two independent components.

 In the zero-mass limit the relation $\Psi_M^c = \Psi_M $ holds.
Obviously, a Majorana spinor can only describe
non-electrically-charged spin-${1\over 2}$ particles. We also see from
(\ref{eq:majcond}) the anti-particle of Majorana particle is the
particle itself.

\subsection{Majorana mass terms and the see-saw mechanism}
\label{sec-masses}

Recall that the left and right-handed components of a Dirac spinor
fields $\Psi_L$ and $\Psi_R$ are independent, and are usually used to
describe the left and right-handed fermions as in the standard
model. But one can as well choose $\Psi_L$ and  $\Psi_L^c$, or
$\Psi_R^c$ and $\Psi_R$ as independent fields to describe left and
right-handed particles. This is usually the case in grand unified
theories where left and right-handed particles are assigned to the
same multiplet (since gauge interactions conserve chirality, left and
right-handed fields cannot be  assigned to the same multiplet, see
App. \ref{sec-so10}).

However $\Psi_L$ and $\Psi_R^c$ or $\Psi_R$ and $\Psi_L^c$ are not
independent. They are related by Eq. (\ref{eq:psicLR}). Using $\Psi_L$
and $\Psi_R^c$ or $\Psi_R$ and $\Psi_L^c$,  one can construct two
Majorana spinors  
\begin{equation}
\Psi_{M_L} = \Psi_L + \Psi_R^c \; \; \; {\rm and} \; \; \; 
\Psi_{M_R} = \Psi_R + \Psi_L^c .
\end{equation}
Majorana mass terms can be constructed for both
$\Psi_{M_L}$ and $\Psi_{M_R}$  
\begin{eqnarray}
M_L \overline{\Psi_{M_L}}\Psi_{M_L} &=& M_L (\overline{\Psi}_L
\Psi_R^c  + h.c ) \\ 
M_R \overline{\Psi_{M_R}}\Psi_{M_R} &=& M_R (\overline{\Psi}_R
\Psi_L^c  + h.c ) . 
\end{eqnarray} 
These Majorana mass terms violate lepton number by two units, and
hence may lead to interesting lepton number violating processes in the
early universe (see Ref. \cite{Y86} and chapter 4).

Therefore the general Lagrangian for free neutrino fields can be written as
\begin{eqnarray}
L &=& \overline{\nu_L} \gamma^\mu \partial_\mu \nu_L +
\overline{\nu_R} \gamma^\mu \partial_\mu \nu_R  + {M_D}
(\overline{\nu_L} \nu_R + \overline{\nu_R} \nu_L) \nonumber \\ 
&& + {M_L\over 2} (\overline{\nu^c_R} \nu_L + h.c)  + {M_R\over 2}
(\overline{\nu^c_L} \nu_R + h.c) \label{eq:Lag1}  
\end{eqnarray}
where $M_D$ is the neutrino Dirac mass and $M_L$ and $M_R$ are the
Majorana masses of the left-handed and right-handed neutrino fields
respectively.

In order to understand the physical content of this Lagrangian, we
introduce new fields $f$ and $F$ 
\begin{equation} 
f = {\nu_L + \nu_R^c \over \sqrt{2}} \; \; \; F =  {\nu_R + \nu_L^c
\over \sqrt{2}} 
\end{equation}
and the Lagrangian given in Eq. (\ref{eq:Lag1}) is equivalent to
\begin{equation}
L = \overline{f} \gamma^\mu \partial_\mu f + \overline{F} \gamma^\mu
\partial_\mu F + (\overline{f} , \overline{F} ) \left ( 
\begin{array} {cc}
M_L & M_D \\
M_D & M_R
\end{array}
\right )  
 \left (
\begin{array} {c}
f \\
F
\end{array}
\right )
\end{equation}
and the matrix
\begin{equation} 
M = 
\left (
\begin{array} {cc}
M_L & M_D \\
M_D & M_R
\end{array}
\right ) \label{eq:Lag2}
\end{equation}
is called the neutrino mass matrix.

In the models of interest, which contain a ${\rm U}(1)_{B-L}$ symmetry
and where the see-saw mechanism can be implemented \cite{seesaw}, the left-handed
neutrino Majorana mass vanishes, $M_L = 
0$. The right-handed neutrino Majorana mass $M_R$ is generated when the rank
of the group is reduced by one unit, as a consequence of $B-L$
breaking. It comes from Yukawa coupling of the Majorana field
describing the right-handed neutrino to the Higgs field $\phi_{B-L}$
used to break $B-L$, $\lambda <\Phi_{B-L}> \overline{\nu}_L^c \nu_R +
h.c.$. Hence $M_R$ is expected to be the order of the $B-L$ 
breaking scale, but it also
depends on the value of the Yukawa coupling constant
$\lambda$. Therefore, even in supersymmetric models where the $B-L$
symmetry breaking scale can be as high as $10^{16}$ GeV, we can still get
right-handed neutrino Majorana masses in the observational bounds for
neutrinos masses by assuming very small Yukawa coupling
constant. Indeed, this is fine tuning. For nonsupersymmetric theories
however, 
$\eta_{B-L}$ usually lies in the range $10^{12} - 10^{13}$ GeV, and no
fine tuning is required. The Dirac neutrino mass is generated by the
electroweak Higgs doublet, and hence is expected to be the order of
the masses of the related charged up-type quarks (e.g. SO(10)). The
neutrino mass matrix then becomes 
\begin{equation} 
M = 
\left (
\begin{array} {cc}
0 & M_D \\
M_D & M_R
\end{array}
\right ) 
\end{equation}
with $M_R \geq M_D$.

In order to find the physical fields and masses, we must diagonalise
$M$. We find that the eigenvalues of the neutrino mass matrix are 
\begin{eqnarray}
M_N &\simeq & M_R \\
{\rm and} \; \; \; M_{\nu } &\simeq & - {M_D^2 \over M_R} ,
\end{eqnarray}
and the corresponding eigenvector fields are
\begin{eqnarray}
N &\simeq & F + {M_D \over M_R} f  \\
\nu ' &\simeq & f - {M_D \over M_R} F .
\end{eqnarray}
In order to get a positive mass for the light neutrino, we take the
physical neutrino field to be  
\begin{equation}
\nu  = \gamma^5 \nu ' .
\end{equation}
In terms of $\nu $ and $N$, the Lagrangian (\ref{eq:Lag2}) becomes
\begin{equation}
L = \overline{\nu }  \gamma^\mu \partial_\mu \nu  + \overline{N}
\gamma^\mu \partial_\mu N + M_\nu \overline{\nu} \nu + M_N
\overline{N} N 
\end{equation}
with 
\begin{equation}
M_\nu = {M_D^2 \over M_R} \; \; \; {\rm and} \; \; \; M_N = M_R .
\end{equation}

Since we expect $M_D$ to be the order of the associated charged lepton
or quark, $M_D \simeq M_{q {\rm or} l}$, we see that 
\begin{equation}
M_\nu . M_N = M_{q \, , \, l}^2 .
\end{equation}
This is the famous see-saw relation \cite{seesaw}.

The light and heavy neutrinos $\nu '$ and $N$ are Majorana particles.
\subsection{Quantised fermions fields}
\label{sec-quant}
\subsubsection{Dirac Case}

Using the normalisation
\begin{equation}
\overline{u}^\alpha (k) u(k)_\alpha = \delta_{\alpha \alpha '} \: .
\end{equation}
the Dirac fermion field operator has the form \cite{QFT}
\begin{eqnarray}
\Psi (x) &=& \int {d^3 k \over (2\pi )^3} {m \over k_0} \sum_{\alpha =
1,2} \Big [ b_\alpha (k) u^\alpha (k) e^{-i k.x} + d^\dagger_\alpha
(k) v^\alpha  (k) e^{i k.x} \Big ] \\ 
\overline{\Psi} (x) &=& \int {d^3 k \over (2\pi )^3} {m \over k_0}
\sum_{\alpha = 1,2} \Big [ b^\dagger_\alpha (k) \overline{u}^\alpha
(k) e^{i k.x} + d_\alpha (k) \overline{v}^\alpha  (k) e^{-i k.x} \Big
] 
\end{eqnarray} 
where $u(k)$ and $v(k)$ are positive and negative energy spinors
solutions of the Dirac equation. $b_\alpha^\dagger (k)$ and $b_\alpha
(k)$,  $d_\alpha^\dagger (k)$ and $d_\alpha (k)$ are, respectively,
the creation and annihilation operators of a particle and an
antiparticle with four-momentum $k$ and helicity $\alpha$.

The operators $b_\alpha (k)$ and $b_\alpha^\dagger (k)$, $d_\alpha
(k)$ and $d_\alpha^\dagger (k)$ satisfy the anti-commutation relations 
\begin{eqnarray}
\{ b_\alpha (k) , b^\dagger_{\alpha '} (k') \} &=& \{ d_\alpha (k) ,
d^\dagger_{\alpha '}(k')  \} = (2 \pi)^3 {k_0 \over m} \delta^3 (
\bf{k} - \bf{k'})\delta_{\alpha \alpha'} \\ 
\{ b_\alpha (k) , b_{ \alpha '} (k') \} &=& \{ b^\dagger_\alpha (k) ,
b^\dagger_{\alpha '}(k')  \} = 0 \\ 
\{ d_\alpha (k) , d_{\alpha '}(k')  \} &=&\{ d^\dagger_\alpha (k) ,
d^\dagger_{\alpha '} (k') \} = 0 \: . 
\end{eqnarray}

\subsubsection{Majorana case}

The Majorana field operator has the form \cite{Kayserbook,Bilenky}
\begin{eqnarray}
\Psi (x) &=& \int {d^3 k \over (2\pi )^3} {m \over k_0} \sum_{\alpha =
1,2} \Big [ b_\alpha (k) u^\alpha (k) e^{-i k.x} + \lambda
b^\dagger_\alpha (k) v^\alpha  (k) e^{i k.x} \Big ] \\ 
\overline{\Psi} (x) &=& \int {d^3 k \over (2\pi )^3} {m \over k_0}
\sum_{\alpha = 1,2} \Big [ b^\dagger_\alpha (k) \overline{u}^\alpha
(k) e^{i k.x} + \lambda b_\alpha (k) \overline{v}^\alpha  (k) e^{-i
k.x} \Big ] 
\end{eqnarray} 
where $\lambda$ is an arbitrary phase factor which is called the
creation phase factor \cite{Kayser84}. $u^\alpha (k)$ and $v^\alpha
(k)$ are positive and negative energy Dirac spinors; they satisfy the
charge conjugation relation 
\begin{equation}
v^\alpha (k) = C \Big [ \overline{u}^\alpha (k) \Big ]^T = u^\alpha (-k) .
\end{equation} 

The operators $b_\alpha (k)$ and $b^\dagger_\alpha(k) $ satisfy the
canonical anti-commutation relations  
\begin{eqnarray}
\{ b_\alpha (k) , b^\dagger_{\alpha '} (k') \} &=& (2 \pi)^3 {k_0
\over m} \delta^3 ( \bf{k} - \bf{k'})\delta_{\alpha \alpha'} \\ 
\{ b_\alpha (k) , b_{ \alpha '} (k') \} &=& \{ b^\dagger_\alpha (k) ,
b^\dagger_{\alpha '}(k')  \} = 0 
\end{eqnarray} 
and, respectively, are the creation and annihilation operators of a
Majorana particle with four-momentum $k$ and helicity $\alpha$. 

The only difference between Majorana and Dirac quantised fermion
fields is that the creation and annihilation operators are different
for the particle and the antiparticle in the Dirac case, whereas they
are the same in the case of Majorana fermions.

\section{Supersymmetry}
\label{sec-susyGUTs}

\subsection{Motivation}

In the past few years, supersymmetric GUTs have received much attention. We
give below the 
motivations for studying supersymmetric GUTs. In
Sec. \ref{sec-susylag}, we introduce the 
notion of superspace, superfields 
and superpotentials and explain how to construct a supersymmetric
Lagrangian.

Recall that the main motivation for studying
supersymmetric GUTs rather than nonsupersymmetric ones has to do with
the running of the gauge 
coupling constants. In the minimal supersymmetric standard model (MSSM), with supersymmetry broken at $\sim 10^3$
GeV, the hypercharge, weak isospin and QCD couplings meet in a single
point at $2 \times 10^{16}$ GeV, with a value $\alpha^{-1} \simeq
25$ \cite{meet}. There is no need to have an intermediate symmetry
breaking scale, 
but an intermediate symmetry breaking scale is not forbidden
either. The second main motivation for supersymmetry is to solve the
hierarchy problem, related to the fourteen orders of magnitude
separating the GUT scale Higgs vacuum expectation value (VEV) from the
electroweak scale Higgs 
VEV. 

Supersymmetric GUTs also provide a cold dark matter candidate, the
lightest superparticle (LSP) which is stable if R-parity remains
unbroken at low energy. $R = (-1)^{3(B-L)+2S}$, where $B$ and $L$ are
respectively baryonic and leptonic charges and $S$ is the particle
spin. So defined, R-parity has value $(+1)$ for all particles and value
$(-1)$ for all antiparticles. There are three candidates for
the LSP : the lightest neutralino, the sneutrino and, if gravity is
included, the gravitino. There are many arbitrary parameters in the
MSSM, and it does not predict what the LSP is. The situation is
different in supergravity models \cite{LSPDM}. Note also that
R-parity has to be kept unbroken at low energy in order to forbid
rapid proton decay. R-parity has to be artificially imposed in the MSSM,
whereas in grand-unified models which contain a ${\rm U}(1)_{B-L}$
gauge symmetry, such as SO(10), R is automatically 
conserved if all Higgs vacuum expectation values used to implement the
symmetry breaking pattern carry $3(B-L)$ \cite{Martin}. In that case, ${\rm
U}(1)_{B-L}$ is broken down to a gauged $Z_2$ subgroup which
acts as R-parity. In SO(10), safe representations, those which
conserve R, are the 10, 45, 54, {\bf 126}, 210 dimensional
representations. Unsafe representations, which break R, are the
{\bf 16} and 144 dimensional representations.

The nature of dark-matter must have influenced the density
perturbations which lead to galaxy and large scale structure
formation. Recent analysis shows that mixed cold and hot
dark-matter scenarios work well \cite{HCDM}. This strongly favours
supersymmetric particle physics models with stable LSP and massive
neutrinos.

 There is not yet experimental evidence for supersymmetry. If
supersymmetry is fundamental to nature, it has to be a broken
symmetry.  The mechanism for supersymmetry breaking remains a
fundamental problem. Searches for supersymmetry are
being (and will be) conducted at high energy colliders
\cite{colliders}, in double beta decay experiments \cite{beta}, in
proton decay experiments \cite{Suzuki} and in dark matter detection
experiments \cite{DMexpt}.

\subsection{Constructing supersymmetric Lagrangians}
\label{sec-susylag}

We give here a brief review of supersymmetry. We introduce the
supersymmetry algebra, the notion of superspace, superfields and
superpotentials. For further details on supersymmetry the reader is
referred to Ref. \cite{SUSY}. We explain how to construct
supersymmetric Lagrangians with superfields and how to derive the
effective scalar potential from a given superpotential.

In addition to the usual Lorentz-Poincar{\'e} generators, the six
generators $M^{\mu \nu }$ of the Lorentz group and the four generators
$P^\mu$  of the translation group ($\mu , \nu = 0 ... 3$),
supersymmetry introduces two-component Weyl spinor generators
$Q^A_\alpha$ ($\alpha = 1, 2$, $A = 1 ... N$) which can change the
spin of a particle by half-a-unit. In $N=1$ supersymmetry there
is one such generator, and there are $A = N$ generators in $N$-extended
supersymmetry. We will restrict ourselves to $N=1$ supersymmetry. 

The generator $Q_\alpha$ is a two-component left-handed Weyl spinor
generator and 
its adjoint hermitian conjugate denoted
$\overline{Q}_{{\dot{\alpha}}}$ is a two-component right-handed Weyl
spinor. The $N=1$ supersymmetry algebra is  
\begin{eqnarray}
&&\{ Q_\alpha, Q_\beta \} = \{ \overline{Q}_{\dot{\alpha}},
\overline{Q}_{{\dot{\beta }}} \} = 0 \\ 
&&\{ Q_\alpha, \overline{Q}_{\dot{\beta }} \} = 2 \sigma_{\alpha
{\dot{\beta }}}^\mu P_\mu \\ 
&& [ Q_\alpha, P_\mu ] = [ \overline{Q}_{\dot{\alpha}}, P_\mu ] = 0 \\
&&[ M^{\mu\nu } , Q_\alpha ] = - i (\sigma^{\mu \nu })_\alpha^\beta
Q_\beta \\ 
&&[ M^{\mu\nu } , \overline{Q}^{\dot{\alpha }} ] = - i
(\overline{\sigma }^{\mu \nu })^{\dot{\alpha }}_{\dot{\beta }}
\overline{Q}^{\dot{\beta }} \; . 
 \end{eqnarray}
The Weyl indices are lowered and raised with the help of the tensors
$\epsilon_{\alpha \beta } =  
\left ( 
\begin{array} {cc}
0 & 1 \\
-1 & 0
\end{array}
\right )$ and $\epsilon^{\alpha\beta } = (\epsilon_{\alpha \beta } )^{-1}$.

The usual representation of this algebra involves {\em vector} and
{\em chiral } supermultiplets. Each supermultiplet contains the same
number of fermions and bosons. The {\em chiral } supermultiplets
contain a chiral spin-${1\over 2}$ fermion and a spin-$0$ scalar
boson. The {\em vector} multiplets contain a spin-$1$ vector boson and
a Majorana spin-$1\over 2$ fermion.

Although it is possible to construct Lagrangians directly from the
component fields belonging to a supermultiplet, the introduction of
the notion of superspace and the use of superfields defined on the
superspace make things much easier. A function defined on the
superspace depends on the usual spacetime coordinates $x^\mu$ and on
new coordinates $\theta$ and $\overline{\theta }$ which transform as
two-component Weyl spinors elements of a Grassman algebra $\{ \theta_i
, \theta_j  \} = \{ \overline{\theta}_i ,  \overline{\theta}_j \} = \{
\theta_i , \overline{\theta}_j \} = 0$. A superfield $S (x^\mu,
\theta, \overline{\theta })$ can be expanded as a power series in
$\theta$ and $\overline{\theta }$. The coefficients of the various
powers of  $\theta$ and $\overline{\theta }$ are the `component'
fields, which are ordinary scalar, vector and Weyl spinor fields.

The action of the supersymmetry algebra on $S(x^\mu, \theta,
\overline{\theta})$ is 
\begin{eqnarray}
S(x^\mu, \theta, \overline{\theta}) &\rightarrow & \exp(i(\theta Q +
\overline{\theta } \overline{Q} - x^\mu P_\mu )) S(a^\mu, \alpha,
\overline{\alpha}) \nonumber \\ 
&&= S(x^\mu + a^\mu - i \alpha \sigma_\mu \overline{\theta} + i \theta
\sigma_\mu \overline{\alpha}, \theta + \alpha,\overline{\theta} +
\overline{\alpha})  \; ,
\end{eqnarray}
which is generated by
\begin{eqnarray}
&& P_\mu = i \partial_\mu \\
&& i Q_\alpha = {\partial \over \partial \theta^\alpha } - i
\sigma_{\alpha {\dot{\alpha}}}^\mu \overline{\theta}^{\dot{\alpha}}
\partial_\mu \\ 
&& i \overline{Q}_\alpha = -{\partial \over \partial \overline{\theta
}^{\dot{\alpha}} } + i \theta^\alpha \sigma_{\alpha
{\dot{\alpha}}}^\mu \partial_\mu \; . 
\end{eqnarray}

A superfield is not at first instance an irreducible representation of
the supersymmetry algebra. To get irreducible representations, we must
impose supersymmetric constraints.  Chiral superfields $\Phi$ are
subject to the condition that  
\begin{equation}
(-{\partial \over \partial \overline{\theta}^{\dot{\alpha}} } - i
\theta^\alpha \sigma_{\alpha {\dot{\alpha}}}^\mu \partial_\mu ) \Phi = 0
. \label{eq:superchiral} 
\end{equation} 
The component fields of a chiral superfield form a chiral multiplet.
The most general chiral superfield $\Phi$ solution of
(\ref{eq:superchiral}) is 
\begin{equation} 
\Phi(y^\mu , \theta ) = \phi (y) + \sqrt{2} \theta \psi (y) + \theta
\theta F(y) \label{eq:chiralS} 
\end{equation}
where $y^\mu = x^\mu + i \theta \sigma^\mu \overline{\theta }$, $\phi$
and $F$ are complex scalar fields and $\psi$ is a left-handed Weyl
spinor field. The field $F$ is called an auxiliary field.  

Vector superfields $V$ satisfy the condition $V^\dagger = V$. The
component fields of a vector superfield form the vector multiplet. 
In the Wess-Zumino gauge, a vector superfield can be written as
\begin{equation}
V (x, \theta, \overline{\theta }) =  \theta \sigma^\mu
\overline{\theta }V_\mu (x) +\theta \theta \overline{\theta }
\overline{\lambda }(x) + \overline{\theta }\overline{\theta } \theta
\lambda (x) + {1\over 2 } \theta \theta \overline{\theta
}\overline{\theta } D(x) \label{eq:vectorS} 
\end{equation}
where $V_\mu$ is a vector field, and $\lambda$ and $\overline{\lambda
}$ are Weyl spinor fields and $D$ is a scalar field. $D$ is called an
auxiliary field.   

The product of two or three chiral superfields is a chiral
superfield. On the other hand, the product $\Phi^\dagger \Phi$ is not
a chiral superfield. It is a vector superfield. The coefficient of
$\theta \theta$ in the product of two or three superfields is called a
$F$-term in analogy with (\ref{eq:chiralS}), and the coefficient of
$\overline{\theta} \overline{\theta}\theta\theta$ in a vector
superfield is called a $D$-term in analogous with
(\ref{eq:vectorS}). Under a supersymmetry transformation, the change
in the $F$-terms as well as in the $D$-terms is a total
divergence. The space-time integral $\int d^4 x$ of a total divergence
vanishes (provided that the fields are zero at infinity, which is
always assumed). Hence one can construct supersymmetric Lagrangians in
terms of $F$ and $D$-terms. A supersymmetric Lagrangian has the general
form 
\begin{equation}
L = \sum_i [\Phi^\dagger_i \Phi_i ] + ([W(\Phi)]_F + h.c. )
\end{equation}
where $W(\Phi )$ is a function of the chiral superfields $\Phi_i$
consistent with the considered gauge symmetry and at most cubic in the
fields (in order to be renormalisable).

Given a superpotential, the auxiliary fields may be calculated as follows,
\begin{eqnarray}
F_{\phi_i } &=& {\partial W \over \partial \phi_i } \label{eq:Fterm}\\
D^\alpha &=& g \sum_{a,b} \overline{\phi}_a T^{\alpha a}_b \phi^b
\label{eq:Dterm} 
\end{eqnarray}
where the $\phi_i$ are the scalar components of the chiral superfields
$\Phi_i$ and $T^\alpha$ are the generators of the gauge group. The
scalar potential may be calculated in terms of the $F$-terms in
Eq. (\ref{eq:Fterm}) and the $D$-terms in Eq. ({\ref{eq:Dterm}), 
\begin{equation}
V ({\phi_i }) = \sum_i |F_{\phi_i} |^2 + \sum_\alpha |D^\alpha |^2
\label{eq:scalarpot1} . 
\end{equation}
The only cases where one must pay attention to the $D$-terms is when a
U(1) symmetry is broken that is when the rank of the gauge group is
lowered by one unit.  

The energy of any non-vacuum supersymmetric state is always positive
definite, and the vacuum state must have zero energy. If the vacuum
energy is non-zero, supersymmetry is broken. It is easy to see from
Eq. (\ref{eq:scalarpot1}) that supersymmetry breaking can occur if a
$F$-term or a $D$-term gets a non-vanishing expectation value
(VEV). If we want to keep supersymmetry unbroken, we must ensure that
the $F$-terms and $D$-terms get a vanishing VEV.  

These methods will be applied in Chap. \ref{chap-chap4}, where a specific 
supersymmetric grand unified model is built.

\chapter{Scattering off an SO(10) cosmic string}
\label{chap-chap2}

\section{Introduction}

Modern particle physics and the hot big-bang model
suggest that the universe underwent a series of phase
transitions at early times at which the underlying
symmetry changed. At such phase transitions
topological defects~\cite{ShelVil} could be
formed. Such topological defects, in particular cosmic
strings, would still be around today and provide a
window into the physics of the early universe. In
particular, cosmic strings arising from a grand
unified phase transition are good candidates for the
generation of density perturbations in the early
universe which lead to the formation of large scale
structure~\cite{ShelVil}. They could also give rise to
the observed anisotropy in the microwave background
radiation~\cite{ShelVil}.
 
Cosmic strings also have interesting microphysical
properties. Like monopoles~\cite{Rubakov}, they can
catalyse baryon violating
processes~\cite{Perkins91}. This is because
the full grand unified symmetry is restored in the
core of the string, and hence grand unified, baryon
violating processes are
unsuppressed. In~\cite{Perkins91} it was shown that
the cosmic string catalysis cross-section could be a
strong interaction cross-section, independent of the
grand unified scale, depending on the flux on the
string. Unlike the case of monopoles, where there is a
Dirac quantisation condition, the string cross-section
is highly sensitive to the flux, and is a purely
quantum phenomena. Defect catalysis is potentially
important since catalysis cross-sections have been 
shown to be the order of the strong interaction 
\cite{Rubakov,Perkins91}. It has already been used to bound the
monopole flux~\cite{Warren}, and could erase a
primordial baryon asymmetry~\cite{Branden89}. It is,
thus, important to calculate the string catalysis
cross-section in a realistic grand unified
theory. In~\cite{Perkins91} a toy model based on a
${\rm U}(1)$ theory was used. In a grand unified theory the
string flux is given by the gauge group, and cannot be
tuned. 

A cosmic string is essentially a flux tube. Hence the
elastic cross-section~\cite{Alf89} is just an
Aharonov-Bohm cross-section~\cite{Aharonov}, depending
on the string flux. This gives the dominant energy
loss in a friction dominated
universe~\cite{Adrian}. Since the string flux is fixed
for any given particle species it is important to
check that the Aharonov-Bohm cross-section persists in
a realistic grand unified theory.

In this chapter, we study the scattering of fermion off the abelian
string arising in the breaking of ${\rm SO}(10)$ down to ${\rm SU}(5)
\times Z_2$ \cite{Kibble82}. ${\rm SU}(5) \times Z_2$ is then broken
down to ${\rm SU}(3)_c \times {\rm SU}(2)_L \times {\rm U}(1)_Y \times
Z_2$. These strings have 
been studied elsewhere \cite{Aryal87,Ma}. We calculate the elastic
cross-sections and the 
baryon number violating
cross-sections due to the coupling to gauge fields in
the core of the string, by both a first
quantised method and a perturbative second quantised
method.

In Sec. \ref{sec-string} we define an {\rm SO}(10) string
model. We give `top-hat' forms for the Higgs and gauge
fields forming the string, since the `top-hat' core
model does not affect the cross-sections of interest
~\cite{Perkins91}. Looking at the microscopic
structure of the string core, we introduce the baryon
number violating gauge fields of {\rm SO}(10) present in the
core of the string.  

In Sec. \ref{sec-cs} we review the method used to
calculate the scattering cross-sections. There are two
different approaches. A fundamental quantum mechanical
one and a perturbative second quantised method
~\cite{Branden88,Perkins91}. The latter consists in
calculating the geometrical cross-section, i.e. the
scattering cross-section for free fermionic
fields. The catalysis cross-section is then enhanced
by an amplification factor to the power of four.  

In Sec. \ref{sec-motion} we derive the equations of
motion. In order to simplify the calculations and to
get a fuller result, we also consider a `top-hat' core
model for the gauge fields mediating quark to lepton
transitions.  

In Sec. \ref{sec-ext} and Sec. \ref{sec-int} we
calculate the solutions to the equations of motion
outside and inside the string core respectively, and
in Sec. \ref{sec-match} we match our solutions at
the string radius. In Sec. \ref{sec-ampl} we
calculate the scattering amplitude for incoming plane
waves of linear combinations of the quark and electron
fields. 

We use these results in Sec. \ref{sec-elast} and
Sec. \ref{sec-inelast} in order to calculate the
scattering cross-sections of incoming beams of pure
single fermion fields. In Sec. \ref{sec-elast} we
calculate the elastic cross-sections. And in
Sec. \ref{sec-concl2} we calculate the baryon number
violating cross-sections. 

In Sec. \ref{sec-second} we derive the catalysis
cross-section using the second quantised method of
Refs. \cite{Branden88,Perkins91}. The second-quantised
cross-sections are found to agree with the first
quantised cross-section of Sec. \ref{sec-inelast}. 

In Appendix \ref{sec-so10} we give
a brief review on {\rm SO}(10) theory, and give an explicit
notation used everywhere in this
chapter. Appendices \ref{sec-extap} and~\ref{sec-intap}
contain the technical details of the external and
internal solutions calculations. Finally,
Appendix \ref{sec-matchap} is a discussion of the
matching conditions at the core radius.

\section{An SO(10) string}
\label{sec-string}

In this section, we describe the abelian cosmic string which arises during the 
phase transition associated with the spontaneous symmetry breaking  
of SO(10) down to ${\rm SU}(5) \times Z_2$. We define the set of Higgs
fields used to implement the symmetry breaking pattern from SO(10) down
to low energy. We then model our string using a `top-hat' core
model \cite{Perkins91} and study the
baryon number violating gauge interactions which occur inside the
core of the string. 

Kibble {\em et al.}\cite{Kibble82} first pointed out that cosmic strings
form when SO(10) breaks down to ${\rm SU}(5) \times Z_2$.
Following Ref. \cite{Kibble82}, we then assume that SO(10) is
broken down to ${\rm SU}(5) \times Z_2$ via the vacuum expectation
value (VEV) of a Higgs field in the {\bf 126} dimensional
representation of SO(10), which we call $\phi_{126}$. ${\rm SU}(5)
\times Z_2$ is then broken down to the standard model gauge group with
added $Z_2$ discrete symmetry, subgroup of the $Z_4$ centre of SO(10), with the
VEV of a Higgs field in the 45 dimensional representation, which we
call $A_{45}$. The standard model gauge group is finally broken down
to ${\rm SU}(3)_c \times 
{\rm U}(1)_Q \times Z_2$ by the VEV of a Higgs field in the 10
dimensional representation of SO(10), $H_{10}$. The symmetry breaking
pattern is therefore as follows,  
$$
{\rm SO}(10) \stackrel{<\phi_{126}>}{\rightarrow} {\rm SU}(5)
\times Z_2 \stackrel{<A_{45}>}{\rightarrow} {\rm SU}(3)_c
\times {\rm SU}(2)_L \times {\rm U}(1)_Y \times Z_2
\stackrel{<H_{10}>}{\rightarrow} {\rm SU}(3)_c \times
{\rm U}(1)_Q \times Z_2  
$$
The components of
the 10-multiplet which acquire a VEV correspond to the usual Higgs
doublet. The first transition is achieved by giving vacuum
expectation value to the component of the {\bf 126}
which transforms as a singlet under SU(5). 

The first homotopy group $\pi_1
({\rm SO}(10) / {\rm SU}(5) \times Z_2) = \pi_0({\rm SU}(5) \times Z_2
= Z_2$, and therefore 
$Z_2$ strings are formed when SO(10) breaks down to ${\rm SU}(5) \times Z_2$.
Furthermore, since $\pi_0( {\rm SU}(3)_c
\times {\rm SU}(2)_L \times {\rm U}(1)_Y \times Z_2) = \pi_0 ({\rm SU}(3)_c
\times {\rm U}(1)_Q \times Z_2 = Z_2$, the strings are topologically
stable down to low energy.

In terms of {\rm SU}(5), the 45
generators of {\rm SO}(10) can be decomposed as follows, 
\begin{equation}
45 = 24 + 1 + 10 + \bar{10} \:
\end{equation}
From the 45 generators of {\rm SO}(10), 24 belong to {\rm SU}(5),
1 generator corresponds to the ${\rm U}(1)'$ symmetry in
{\rm SO}(10) not embedded in {\rm SU}(5) and there are 20
remaining ones. Therefore the breaking of {\rm SO}(10) to
${\rm SU}(5) \times Z_2$ induces the creation of two types
of strings. An Abelian one, corresponding to the
${\rm U}(1)'$ symmetry, and a non abelian one made with
linear combinations of the 20 remaining generators. In
this paper we are interested in the abelian strings
since the non abelian version are Alice strings, and
would result in global quantum number being
ill-defined, and hence unobservable \cite{Bucher}. We
note that there is a wide range of parameters where
the non abelian strings have lower energy
\cite{Ma}. However, since the abelian string is
topologically stable, there is a finite probability
that it could be formed by the Kibble mechanism
\cite{Kibble}.

In the appendix A, we give a brief review of {\rm SO}(10) \cite{G74}. 
With that notation, the Lagrangian is 
\begin{equation}
L = {1\over 4} F_{\mu\nu} F^{\mu\nu} +
(D_\mu\Phi_{126})^{\dagger} (D^\mu\Phi_{126})  -
V(\Phi) + L_F \label{eq:lagrangian} 
\end{equation}
where $F_{\mu\nu} = -i\, F_{\mu\nu}^a \tau_a$, $\tau_a
\, \, a = 1,...,45$  are the 45 generators of
{\rm SO}(10). $\Phi_{126}$ is the Higgs {\bf 126},  the self-dual
anti-symmetric 5-index tensor of {\rm SO}(10). $L_F$ is the
fermionic part of the Lagrangian. In the covariant
derivative $D_\mu = \partial_\mu + i e A_\mu$, $A_\mu =
A_\mu^a \tau_a$ where $A_\mu^a$ a = 1,...,45 are 45
gauge fields of {\rm SO}(10).

If we call $\tau_{str}$ the generator of the abelian
string, $\tau_{str}$ will be given by the diagonal
generators of {\rm SO}(10) not lying in ${\rm SU}(5)$ that is,  
\begin{equation}
\tau_{str} = {1\over 5} \, (M_{12} + M_{34} + M_{56} + M_{78} + M_{9\, 10})
\end{equation}
where $M_{ij} : i,j = 1...10$ are the 45 {\rm SO}(10)
generators defined in appendix A in terms of the
generalised gamma matrices. Numerically, this gives, 
\begin{equation}
\tau_{str} = diag({1\over 2}\, , {1\over 10}\, ,
{1\over 10}\, , {1\over 10}\, , {1\over 10}\, ,
{1\over 10}\, , {1\over 10}\, , {-3\over 10}\, ,
{1\over 10}\, , {1\over 10}\, , {1\over 10}\, ,
{-3\over 10}\, , {1\over 10}\, , {-3\over 10}\, ,
{-3\over 10}\, ,  {-3\over 10}) \label{eq:tauall} \:
. 
\end{equation}
The results of Perkins {\em et al.}\cite{Perkins91} find
that the greatest enhancement of the cross-section is
for fermionic charges close to integer values. Thus,
from equation (\ref{eq:tauall}), we expect no great
enhancement; the most being due to the right-handed
neutrino. 

We are going to model our string as is usually done
for an abelian {\rm U}(1) string \cite{ShelVil}. That is, we take the
string along the z axis, resulting in the Higgs
$\Phi_{126}$ and the gauge fields $A_\mu$ of the
string to be independent of the z coordinate,
depending only on the polar coordinates $(r,
\theta)$. Here $A_\mu$ is the gauge field of the
string, obtained from the product $A_\mu = A_{\mu ,
str} \tau_{str}$. The solution for the abelian string
can be written as, 
\begin{eqnarray}
\Phi_{126} &=& f(r) \, e^{i \tau_{str} \theta} \,
\Phi_0 = f(r) \, e^{i \theta} \, \Phi_0 \\ 
A_\theta &=& - {g(r)\over er} \tau_{str}\nonumber\\
A_r&=& A_z=0 \label{eq:gauge}
\end{eqnarray}
where $\Phi_0$ is the vacuum expectation value of the
Higgs {\bf 126} in the $1_{10}$ direction. The functions
$f(r)$ and $g(r)$ describing the behaviour the Higgs
and gauge fields forming the string are approximately given by 
\begin{eqnarray}
\left. \begin{array}{lllllll}

f(r) &=& \left\{ \begin{array} {cr}
                  {\eta } & r \geq R\\
                   ({ \eta r\over R}) & r < R
                  \end{array}
          \right. 
& , & g(r) &=& \left\{ \begin{array} {cr}
                  1 & r\geq R\\
                  ({r\over R})^2 & r < R
                  \end{array}
          \right.
\end{array}
 \right.
\end{eqnarray} 
where R is the radius of the string. $R \sim
\eta^{-1}$, where $\eta$ is the grand unified scale,
assumed to be $\eta \sim 10^{15}$ GeV. In order to
simplify the calculations and to get a fuller result
we use the top-hat core model, since it has been shown
not to affect the cross-sections of interest. The
top-hat core model assumes that the Higgs and gauge
fields forming the string are zero inside the string
core. Hence, $f(r)$ and $g(r)$ are now given by, 
\begin{eqnarray}
\left. \begin{array}{lllllll}
f(r) &=& \left\{ \begin{array} {cr}
                  {\eta } & r \geq R\\
                   0 & r < R
                  \end{array}
          \right. 
& , & g(r) &=& \left\{ \begin{array} {cr}
                  1 & r\geq R\\
                  0 & r < R
                  \end{array}
          \right.  \: .
 \end{array}
 \right. \label{eq:tophat}
\end{eqnarray}

The full ${\rm SO}(10)$ symmetry is restored in the core of
the string. ${\rm SO}(10)$ contains 30 gauge bosons
leading to baryon decay. These are the bosons $X$ and
$Y$, and their conjugates, of {\rm SU}(5) plus 18 other gauge 
bosons usual called $X'$, $Y'$ and $X_s$, and their 
conjugates. Therefore inside the
string core, there are quark to lepton transitions
mediated by the gauge bosons $X$, $X'$, $Y$, $Y'$ and
$X_s$ and we expect the string to catalyse baryon
number violating processes in the early universe.  

The $X$, $X'$, $Y$, $Y'$ and $X_s$ gauge bosons are
associated with non diagonal generators of {\rm SO}(10). For
the electron family, the relevant part of the
Lagrangian is given by, 
\begin{equation}
L_x = \bar{\Psi}_{16} \, ( i e \gamma^\mu (X_\mu \tau^X + X'_\mu \tau^{X'}
 + Y_\mu \tau^Y  + Y'_\mu \tau^{Y'} + {X_s}_\mu
\tau^{X_s})) \, \Psi_{16} \label{eq:baryon} 
\end{equation} 
where $\tau^X$, $\tau^{X'}$, $\tau^Y$, and $\tau^{Y'}$
and $\tau^{X_s}$ are the non diagonal generators of
{\rm SO}(10) associated with the  $X$, $X'$, $Y$, $Y'$ and
$X_s$ gauge bosons respectively. 

Expanding equation (\ref{eq:baryon}) gives ~\cite{Marie},
\begin{eqnarray}
L_x &=& {g\over\sqrt{2}} X^{\alpha }_\mu [-\epsilon_{\alpha \beta \gamma } \bar{u}^{c\gamma }_L \gamma^\mu u^\beta_L + \bar{d}_{L \alpha}
\gamma^\mu e^+_L +  \bar{d}_{R \alpha} \gamma^\mu e^+_R
]\nonumber \\ 
&& + {g\over\sqrt{2} } Y^{\alpha }_\mu [-\epsilon_{\alpha \beta \gamma } \bar{u}^{c\gamma }_L \gamma^\mu d^\beta_L - \bar{d}_{R \alpha}
\gamma^\mu \nu_e^c - \bar{u}_{L\alpha } \gamma^\mu e^+_L ]\nonumber
\\ 
&& + {g\over\sqrt{2} } X^{\alpha '}_\mu  [-\epsilon_{\alpha
\beta \gamma } \bar{d}^{c\gamma }_L \gamma^\mu d^\beta_L -
\bar{u}_{R\alpha} \gamma^\mu \nu_R^c - \bar{u}_{L\alpha} \gamma^\mu
\nu_L^c ] \nonumber\\ 
&&  + {g\over\sqrt{2} } Y^{\alpha '}_\mu [\epsilon_{\alpha \beta \gamma } \bar{d}^{c\gamma }_L \gamma^\mu u^\beta_L - \bar{u}_{R\alpha }
\gamma^\mu e_R^+ - \bar{d}_{L\alpha } \gamma^\mu \nu_L^c ]
\nonumber\\ 
&& + {g\over\sqrt{2} } X^{\alpha }_{s \mu} [\bar{d}_{L \alpha }
\gamma^\mu e_L^- + \bar{d}_{R\alpha } \gamma^\mu e_R^- +
\bar{u}_{L\alpha } \gamma^\mu \nu_L + \bar{u}_{R \alpha } \gamma^\mu
\nu_R ]  \label{eq:lagi} 
\end{eqnarray} 
where $\alpha$, $\beta$ and $\gamma$ are colour indices. The
$X_s$ does not contribute to nucleon decay except by
mixing with the $X'$ because there is no vertex
$qqX_s$. We consider baryon violating processes
mediated by the gauge fields $X$, $X'$, $Y$ and $Y'$
of {\rm SO}(10). In previous papers ~\cite{Perkins91,Alf89},
baryon number violating processes resulting from the
coupling to scalar condensates in the string core have
been considered. In our {\rm SO}(10) model we do not have
such a coupling.

\section{Scattering of fermions from the abelian string}
\subsection{The scattering cross-section}
\label{sec-cs}
Here, we will briefly review the two methods used to
calculate the scattering cross-section. The first is a
quantum mechanical treatment. From the fermionic
Lagrangian $L_F$, we derive the equations of motion
inside and outside the string core. We then find
solutions to the equations of motion inside and
outside the string core and we match our solutions at
the string core. Considering incoming plane waves of
pure quarks, we then calculate the scattering
amplitude. The matching conditions together with the
scattering amplitude enable us to calculate the
elastic and inelastic scattering cross-sections. The
second method is a quantised one, where one calculates
the geometrical cross-section $({d\sigma \over d
\Omega})_{geom}$, i.e. using free fermions spinors
$\psi_{free}$. The catalysis cross-section is enhanced
by a factor ${\cal A}^4$ over the geometrical
cross-section, 
\begin{equation}
\sigma_{inel} = {\cal A}^4 \; ({d\sigma \over d \Omega})_{geom}
\end{equation}
where the amplification factor ${\cal A}$ is defined by,
\begin{equation}
{\cal A} = {\psi (R) \over \psi_{free} (R) } \: .
\end{equation}
where R is the radius of the string, $R \sim
\eta^{-1}$. This method has been applied in
Refs. \cite{Perkins91} and~\cite{Michael}.

\subsection{The equations of motion}
\label{sec-motion}
The fermionic part of the Lagrangian $L_F$ is given in
terms of 16 dimensional spinors as defined in Appendix
A. We shall consider here only one family, the electron family. The generalisation to three families is straight-forward. The fermionic
Lagrangian for only one family, 
\begin{equation}
L_F = L^{(e)}_F = \bar{\Psi}_{16} \gamma^\mu D_\mu \Psi_{16} + L_M + L_x 
\end{equation}
where $L_M$ is the mass term and $L_x$ is the
Lagrangian describing quark to lepton transitions
through the $X$, $X'$, $Y$, $Y'$ and $X_s$ gauge
bosons in {\rm SO}(10) and given by
(equation~\ref{eq:lagi}). The covariant derivative is
given by $D_\mu = \partial_\mu - i e A_{\mu , str}
\tau_{str}$ where $A_{\mu , str}$ is the gauge field
forming the string and $\tau_{str}$ is the string
generator given by Eq. (\ref{eq:tauall}). Therefore, since $\tau_{str}$ is
diagonal, there will be no mixing of fermions around
the string. The Lagrangian $L_F$ will split in a sum
of eight terms, one for each fermion of the family. In
terms of 4-spinors, this is 
\begin{equation}
L_F = \sum_{i = 1}^{8} L_f^i + L_x \label{eq:fer}
\end{equation}
where $L_f^i = i \bar{\psi}_L^i \gamma^\mu D_\mu^L
\psi_L^i + i \bar{\psi}_L^{c,i} \gamma^\mu D_\mu^{Lc}
\psi_L^{c,i} + L_m^i \label{eq:lagf}$, and i runs over
all fermions of the given family. One can show that
$i \bar{\psi}_L^{c,i} \gamma^\mu D_\mu^L \psi_L^{c,i} = i
\bar{\psi}_R^{i} \gamma^\mu D_\mu^R \psi_R^{i}$ and
$\tau_{str}^{Lc \: i} = \tau_{str}^{R \: i}$. Finally,
$L_x$ is given by Eq. (\ref{eq:lagi}). It is easy
to generalise to more families.

From Eqs. (\ref{eq:fer}) and (\ref{eq:lagi}) we
derive the equations of motions for the fermionic
fields. We take the fermions to be massless inside and
outside the string core. This a relevant assumption
since our methods apply for energies above the
confinement scale. We consider the case of free quarks
scattering from the string and coupling with electrons
inside the string core. Outside the string core, the
fermions feel the presence of the string only by the
presence of the gauge field.We are interested in the
elastic cross sections for all fermions and in the
cross-section for these quark decaying into
electron. The fermionic Lagrangian given by equations
(\ref{eq:fer}) and (\ref{eq:lagi}) becomes, 
\begin{eqnarray}
L_F(e, {q}) &=& i \bar{e}_L \gamma^\mu D_\mu^{e,L} e_L + i
\bar{e}_R \gamma^\mu D_\mu^{e,R} e_R \nonumber \\ 
&&+  i \bar{q}_L \gamma^\mu D_\mu^{q,L} q_L + i \bar{q}_R
\gamma^\mu D_\mu^{q,R} q_R \nonumber \\ 
&&- {g G^\mu \over 2\sqrt{2}} \bar{q}_L \gamma_\mu e^+_L
- {g G^{' \mu} \over 2\sqrt{2}} \bar{q}_R \gamma_\mu
e^+_R + H.C. 
\end{eqnarray}
giving the following equations of motion,
\begin{eqnarray}
i \gamma^\mu D^{e,L}_\mu e_L + {g G'_\mu \over 2\sqrt{2}}
\gamma^\mu q^c_L &=&0 \nonumber \\ 
i \gamma^\mu D^{e,R}_\mu e_R + {g G_\mu \over 2\sqrt{2}}
\gamma^\mu q^c_R&=&0 \nonumber \\ 
i \gamma^\mu D^{{q^c},L}_\mu q^c_L + {g G'_\mu \over 2
\sqrt{2}} \gamma^\mu e^-_L &=&0 \nonumber \\ 
i \gamma^\mu D^{{q^c},R}_\mu q^c_R+ {g G_\mu \over 2
\sqrt{2}} \gamma^\mu e^-_R &=&0 \label{eq:motion} 
\end{eqnarray}
which are valid everywhere. The covariant derivatives
$D^{e,(L,R)}_\mu = \partial_\mu + ie A_{\mu , str}
\tau_{str}^{e,(L,R)}$ and $D^{{q^c},(L,R)}_\mu =
\partial_\mu + ie A_{\mu , str} \tau_{str}^{q,(L,R)}$.  We
have $\tau_{str}^{R, u} = \tau_{str}^{L, u} =
\tau_{str}^{L, e} = \tau_{str}^{L, d} = {1 \over 10}$
and $\tau_{str}^{R, e} = \tau_{str}^{R, d} = {-3 \over
10}$ together with $\tau_{str}^{Lc, \: i} =
\tau_{str}^{R, \: i}$ and $\tau_{str}^{L, \: i} =
\tau_{str}^{Rc, \: i}$. $G_\mu$ and $G'_\mu$ stand for
$X_\mu$, $X'_\mu$, $Y'_\mu$ or $Y'_\mu$, depending on
the chosen quark.  

Since these equations involve quarks and lepton
mixing, we do not find independent solution for the
quark and lepton fields. However, we can solve these
equations taking linear combinations of the the quark
and lepton fields, ${q_L^c \pm e_L}$ and ${q_R^c \pm
e_R}$. In this case, the effective gauge fields are  
\begin{equation} 
e \, (A_{\mu , str} \tau_{str}^{f_L} \pm G_\mu)
\end{equation}
 and 
\begin{equation}
e \, (A_{\mu , str} \tau_{str}^{f_R} \pm G'_\mu)
\end{equation}
 respectively. 

In order to make the calculations easier, we use a
top-hat theta component for $G$ and $G'$ within the
string core, since Perkins {\em et al.}\cite{Perkins91}
have shown that the physical results are insensitive
to the core model used for the gauge fields mediating
baryon violating processes.

\subsection{The external solution}
\label{sec-ext}

Outside the string core, the gauge field of the string
$A_{\mu , str}$  has only, from
equations~\ref{eq:gauge} and~\ref{eq:tophat}, a non
vanishing component $A_\theta = {1 \over er}
\tau_{str}$, and the effective gauge fields $G$ and
$G'$ are set to zero. Therefore the equations of
motion (\ref{eq:motion}) for $r > R$ become, 
\begin{eqnarray}
i \gamma^\mu D^{e,L}_\mu e_L  &=&0 \nonumber \\
i \gamma^\mu D^{e,R}_\mu e_R &=&0 \nonumber \\
i \gamma^\mu D^{u,L}_\mu q^c_L &=&0 \nonumber \\
i \gamma^\mu D^{u,R}_\mu q^c_R &=&0  \label{eq:ext}
\end{eqnarray}
where the covariant derivatives $D^{e,(L,R)}_\mu =
\partial_\mu + ie A_{\mu , str} \tau_{str}^{e,(L,R)}$ and
$D^{{q^c},(L,R)}_\mu = \partial_\mu + ie A_{\mu , str}
\tau_{str}^{q,(L,R)}$.  

We take the usual Dirac representation $e_L =
(0,\xi_e)$ , $e_R = (\chi_e,0)$ , $q^c_L =
(0,\xi_{q^c})$ and $q^c_R = (\chi_{q^c},0)$ and the
mode decomposition for the spinors $\xi_{q^c}$,
$\xi_e$, $\chi_{q^c}$ and $\chi_e$, 
\begin{eqnarray}
\chi_{(e, {q^c})}(r,\theta) &=& \sum_{n=-\infty}^{n=+\infty}
\left ( \begin{array}{rl}
	&\chi_{1\, (e, {q^c})}^n (r)\\
         i &\chi_{2\, (e, {q^c})}^n (r) \, e^{i\theta}
\end{array}
\right )
e^{in\theta} \nonumber\\
\xi_{(e, {q^c})}(r,\theta) &=& \sum_{n=-\infty}^{n=+\infty}
\left ( \begin{array}{rl}
	&\xi_{1\, (e, {q^c})}^n (r)\\
         i &\xi_{2 \, (e, {q^c})}^n (r) \, e^{i\theta}
\end{array}
\right )
e^{in\theta} \label{eq:mode} \; .
\end{eqnarray}
From appendix~\ref{sec-extap} we see that the fields
$\xi_{1,(e, {q^c})}^n$, $\xi_{2,(e, {q^c})}^n$,
$\chi_{1,(e, {q^c})}^n$ and $\chi_{2,(e, {q^c})}^n$
satisfy Bessel equations of order $n -
\tau_{str}^{R\, (e, {q^c})}$,  $n + 1 -
\tau_{str}^{R\, (e, {q^c})}$, $n -  {\tau_{str}^{L \,
(e, {q^c})}}$ and $n -  \tau_{str}^{R\, (e, {q^c})}$
respectively. The external solution becomes, 
\begin{eqnarray}
&&	\left ( \begin{array}{l}
\xi_{(e, {q^c})} (r,\theta)\\
\chi_{(e, {q^c})} (r, \theta)
	\end{array}
	\right )
 = \label{eq:external}
\\
&&\sum_{n=-\infty}^{n=+\infty}
	\left ( \begin{array}{rlcrl}
& \! (v_n^{(e, {q^c})} J_{n - \tau_{str}^{R\, (e, {q^c})}}
(\omega r) &+& v_n^{(e, {q^c})'} J_{-(n - \tau_{str}^{R\,
(e, {q^c})})} (\omega r)) & \! e^{i n \theta} \nonumber\\ 
i& \! (v_n^{(e, {q^c})} J_{n + 1 - \tau_{str}^{R\, (e,
{q^c})}} (\omega r) &-& v_n^{(e, {q^c})'} J_{-(n + 1-
\tau_{str}^{R\, (e, {q^c})})} (\omega r)) & \! e^{i (n + 1)
\theta} \nonumber\\ 
& \! (w_n^{(e, {q^c})} J_{n - \tau_{str}^{L \, (e,
{q^c})}} (\omega r) &+& w_n^{(e, {q^c})'} J_{-(n -
\tau_{str}^{L \, (e, {q^c})})} (\omega r)) & \! e^{i n
\theta} \nonumber\\ 
i& \! (w_n^{(e, {q^c})} J_{n + 1 - \tau_{str}^{L \, (e,
{q^c})}} (\omega r) &-& w_n^{(e, {q^c})'} J_{-(n + 1 -
\tau_{str}^{L \, (e, {q^c})})} (\omega r)) & \! e^{i (n +
1)\theta} 
       \end{array}
	\right ) \; . \nonumber
\end{eqnarray}
Therefore, outside the string core, we have got
independent solutions for the quark and electron
fields.

\subsection{The internal solution}
\label{sec-int}

Inside the string core, the gauge field of the string,
$A_\mu$, is set to zero whereas $G_\theta$ and
$G'_\theta$ take the value $2\sqrt{2} A$ and $2
\sqrt{2} A'$ respectively. Therefore, the equations of
motion (\ref{eq:motion}) become, 
\begin{eqnarray}
i \gamma^\mu \partial_\mu e_L + {g G'_\mu \over 2\sqrt{2}}
\gamma^\mu q^c_L &=&0 \nonumber \\ 
i \gamma^\mu \partial_\mu e_R + {g G_\mu \over 2\sqrt{2}}
\gamma^\mu q^c_R&=&0 \nonumber \\ 
i \gamma^\mu \partial_\mu q^c_L + {g G'_\mu \over 2\sqrt{2}}
\gamma^\mu e^-_L &=&0 \nonumber \\ 
i \gamma^\mu \partial_\mu q^c_R+ {g G_\mu \over 2\sqrt{2}}
\gamma^\mu e^-_R &=&0 \; . \label{eq:fint} 
\end{eqnarray}
Since these equations of motions involve quark-lepton
mixing, there are no independent solutions for the
quarks and electron fields. However, we get solutions
for the fields $\rho^\pm$ and $\sigma^\pm$ which are linear
combinations of the quarks and electron fields, 
\begin{equation}
\rho^\pm = \chi_{q^c} \pm \chi_e \label{eq:ro}
\end{equation} 
and
\begin{equation}
\sigma^\pm = \xi_{q^c} \pm \xi_e \; . \label{eq:sigma}
\end{equation}

Using the mode decomposition (\ref{eq:mode}) for the
fields $\rho^\pm$ and $\sigma^\pm$, the internal solution
becomes, 
\begin{equation}
\left ( \begin{array} {rl}
	&\rho_{n1}^\pm \, e^{in\theta}\\
	i & \rho_{n2}^\pm \, e^{i(n+1)\theta}\\
	&\sigma_{n1}^\pm \, e^{in\theta}\\
	i & \sigma_{n2}^\pm \, e^{i(n+1)\theta}
	\end{array}
	\right )
\end{equation}
where $\rho_{n1}^\pm$ and $\rho_{n2}^\pm$ and
$\sigma_{n1}^\pm$ and $\sigma_{n2}^\pm$ are the upper and
lower components of the fields $\rho^\pm$ and  $\sigma^\pm$
respectively. They are given in terms of
hyper-geometric functions. From
appendix~\ref{sec-intap} we get, 
\begin{equation}
\rho_{n1}^\pm = (kr)^{|n|} e^{-ikr} \sum_{j =
0}^{n=+\infty} \alpha^\pm_j {(2ikr)^j \over j!} 
\end{equation}
where $k^2 = w^2 - (e A)^2$, $e = {g \over 2
\sqrt{2}}$. $\alpha^\pm_{j+1} = {(a^\pm + j) \over (b
+ p)} \alpha^\pm_j$ with $a^\pm = {1\over 2} + |n| \pm
{e A (2n+1) \over 2ik}$ and $b = 1 +
2|n|$. $\rho_{n2}^\pm$ can be obtained using the coupled
equation (\ref{eq:int}.2) of
appendix~\ref{sec-intap}. We find, 
\begin{equation}
\rho_{n2}^\pm = - {1\over \omega} (kr)^{|n|} e^{-ikr} \sum_{j
= 0}^{n=+\infty} \alpha^\pm_j {(2ikr)^j \over j!} \,
\,({|n| - n \over r} -ik + {j\over r} \pm e A) \; . 
\end{equation}
We get similar hyper-geometric functions for the
fields $\sigma_{n1}^\pm$ and $\sigma_{n2}^\pm$.

\subsection{Matching at the string core}
\label{sec-match}

From now on, we will do calculations for the
right-handed fields, the calculations for the
left-handed ones being straight-forward. Once we have
our internal and external solutions, we match them at
the string core. We must take the same linear
combinations of the quark and lepton fields outside
and inside the core, and must have continuity of the
solutions at $r = R$. The continuity of the solutions
at $r=R$ implies, 
\begin{eqnarray}
(\chi_{1,q}^n \pm \chi_{1,e}^n)^{out} &=& \rho_{n1}^{\pm \: in} \\
(\chi_{2,q}^n \pm \chi_{2,e}^n)^{out} &=& \rho_{n2}^{\pm \: in} \; .
\end{eqnarray}
Nevertheless, we will have discontinuity of the first derivatives, 
\begin{eqnarray}
({d\over dr} \mp eA ) \, \rho_{n2}^{\pm \: in} &=&
({d\over dr} - {\tau_{str}^{R\, (e, {q^c})} \over R})
\, (\chi_{2,q}^n \pm \chi_{2,e}^n)^{out} \\ 
({d\over dr} \pm eA ) \, \rho_{n1}^{\pm \: in} &=&
({d\over dr} + {\tau_{str}^{R\, (e, {q^c})} \over R})
\, (\chi_{1,q}^n \pm \chi_{1,e}^n)^{out} \; . 
\end{eqnarray}
These equations lead to a relation between the
coefficients of the Bessel functions for the external
solution, as derived in Appendix~\ref{sec-matchap},   
\begin{equation}
{v_n^{q'} \pm v_n^{e'} \over v_n^{q} \pm v_n^{e}} = {
w \, \lambda_n^\pm J_{n + 1 - \tau_R}(\omega R) + J_{n -
\tau_R}(\omega R) \over w \, \lambda_n^\pm J_{-(n + 1 -
\tau_R)}(\omega R) + J_{-(n - \tau_R)}(\omega R)}
\label{eq:relation} 
\end{equation}
where 
\begin{equation}
\lambda_n^\pm = {\sum_{j = 0}^{n=+\infty} \alpha^\pm_j
{(2ikr)^j \over j!} \over \sum_{j = 0}^{n=+\infty}
\alpha^\pm_j {(2ikr)^j \over j!} ({|n| - n \over r}
-ik + {j\over r} \pm e A)} \; .\label{eq:lambda} 
\end{equation}
The relations (\ref{eq:relation}) and (\ref{eq:lambda}) are the matching conditions at $r = R$.

\subsection{The scattering amplitude}
\label{sec-ampl}

In order to calculate the scattering amplitude, we
match our solutions to an incoming plane wave plus an
outgoing scattered wave at infinity. However, since
the internal solution, and therefore the matching
conditions at $r = R$, are given in terms of linear
combinations of quarks and leptons, we consider
incoming waves of such linear combinations. Let
$f_n^\pm$ denote the scattering amplitude for the mode
n, $f_n^+$ if we consider the scattering of (quarks +
electrons) and $f_n^-$ if we consider the scattering
of (quarks - electrons). Then the matching conditions
at infinity are, 
\begin{eqnarray}
\lefteqn{ (-i)^n
\left ( \begin{array}{l}
	 J_n \\
        i  J_{n+1} \, e^{i \theta}\\
\end{array}
\right ) +
{f_n^\pm e^{ikr} \over\sqrt{r}}
\left ( \begin{array}{rl}
& 1\\
i & e^{i \theta}\\
\end{array}
\right ) = } \nonumber \\
&& \left ( \begin{array}{rlcrl}
&(v_n^q \pm v_n^{e}) J_{n - \tau_R } &+& (v_n^{q'} \pm
v_n^{e'}) J_{-(n - \tau_R)}& \\ 
i & ((v_n^q \pm v_n^{e }) J_{n + 1 - \tau_R} & + &
(v_n^{q'} \pm v_n^{e'}) J_{-(n + 1 - \tau_R)}) & e^{i
\theta}\\ 
\end{array}
\right ) \; .
\end{eqnarray}
Using then the large r forms for the Bessel functions,
\begin{equation}
J_\mu (\omega r) =\sqrt{2 \over \pi \omega r} \cos(\omega r -
{\mu \pi \over 2} - {\pi \over 4}) \: , \nonumber 
\end{equation}
and matching the coefficients of $e^{i\omega r}$ we find,
\begin{eqnarray}
\lefteqn{\sqrt{2 \pi \omega} f_n^\pm e^{i {\pi \over 4}} =} \nonumber \\
&&\left \{ \begin{array}{l}
e^{-i{n\pi}} (e^{i\tau_R \pi} -1) + (v_n^{q'} \pm v_n^{e'}) e^{i(n-\tau_R ) {\pi \over 2}} (1- e^{-2i (n-\tau_R ) \pi}) \\
 e^{i{n\pi}} (e^{i(n -\tau_R) \pi} - e^{-in\pi }) + ( v_n^q \pm v_n^{e\pm }) e^{-i(n-\tau_R ) {\pi \over 2}} (1- e^{2i (n-\tau_R ) \pi}) 
\end{array}
\right. \; . \label{eq:ampl}
\end{eqnarray}
Matching the coefficients $e^{-i \omega r}$, we get
relations between the Bessel functions coefficients, 
\begin{equation}
(v_n^q \pm v_n^e) = (1 - (v_n^{q'} \pm v_n^{e'}) e^{-
i (n - \tau_R) {\pi \over 2}}) \: e^{- i (n - \tau_R)
{\pi \over 2}} \; . \label{eq:vnp} 
\end{equation}
The relations (\ref{eq:ampl}), (\ref{eq:vnp}),
(\ref{eq:relation}) and (\ref{eq:lambda}) determine
the scattered wave.

\vspace{.75cm}
\section{The elastic cross-section}
\label{sec-elast}
When there is no baryon number violating processes
inside the string core, when the gauge fields
mediating quark to lepton transitions are set to zero,
we have elastic scattering. In this case, the
scattering amplitude reduces to, 
\begin{equation}
f_n^{elast} = {1 \over\sqrt{2 \pi \omega }} \, e^{-i {\pi \over 4}} \left \{ \begin{array}{ll}
 e^{- i n \pi } \, (e^{i \tau_R \pi } - 1) & n\geq 0 \\
e^{i n \pi} \, (e^{- i \tau_R \pi} - 1) & n \leq -1
\end{array}
\right. \; .
\end{equation}
The elastic cross-section per unit length is given by
\begin{equation}
\sigma_{elast} = | \sum_{n = -\infty}^{+ \infty}
f_n^{elast} e^{i n \theta} |^2 \; .  
\end{equation}
Using the relations $\sum_{n = a}^{+ \infty } e^{i n
x} = {e^{iax} \over 1 - e^{ix}}$ and  $\sum_{n = -
\infty}^{b} e^{i n x} = {e^{i b x} \over 1 - e^{-
ix}}$, we find the elastic cross-section to be 
\begin{equation}
\sigma_{elast} = {1 \over2 \pi \omega} {\sin^2{\tau_R \pi} \over \cos^2{\theta \over 2}} \; .
\end{equation}
This is an Aharonov-Bohm cross-section, and $\tau_R$
is the flux in the core of the string.  

Now, remember that $\tau_{str}^{Lc, u} =
\tau_{str}^{L, u} = \tau_{str}^{Lc, e} =
\tau_{str}^{L, d} = {1 \over 10}$ and $\tau_{str}^{L,
e} = \tau_{str}^{Lc, d} = {-3 \over 10}$ and
$\tau_{str}^{Lc, \: i} = \tau_{str}^{R, \: i}$ and
$\tau_{str}^{L, \: i} = \tau_{str}^{Rc, \: i}$. Hence,

\begin{equation}
\sigma_{elast}^{e_L} = \sigma_{elast}^{d_{R}}  >
\sigma_{elast}^{e_R} = \sigma_{elast}^{u_{R}} =
\sigma_{elast}^{d_L} = \sigma_{elast}^{u_{L}} \; .  
\end{equation}
We therefore have a marked asymmetry between
fermions. We have got a marked asymmetry between left
and right handed electrons, left and right handed down
quarks or, since $\sigma_{elast}^{i_{Lc}} =
\sigma_{elast}^{i_{R}}$ and $\sigma_{elast}^{i_{Rc}} =
\sigma_{elast}^{i_{L}}$, between left handed particle and
antiparticle, respectively right handed, for the
electron and the down quark. But we have equal cross
sections for right handed particles and left handed
antiparticles for the electrons and the down quark,
and equal cross sections for both left handed and
right handed up quarks an anti-quarks. This is a
marked feature of grand unified theories. If cosmic
strings are found it may be possible to use this
asymmetry to identify the underlying gauge symmetry.

\section{The inelastic cross-section}
\label{sec-inelast}
The gauge fields $X$, $X'$, $Y$ and $Y'$ are now
`switched on'. In this case we are calculating the
baryon number violating cross-section. If we consider
identical beams of incoming pure  $\rho^+$ and $\rho^-$,
recalling that $\rho^\pm = \chi_{q^c} \pm \chi_e$, this
will ensure that we will have an incoming beam of pure
quark. Therefore, the scattering amplitude for the
quark field is given by half the difference of $f_n^+$
and $f_n^-$, and the scattering amplitude for the
electron field is given by half the sum of $f_n^+$ and
$f_n^-$. From equation (\ref{eq:ampl}) we get, 
\begin{equation}
\frac{1} {2} \,\sqrt{2 \pi \omega } \, (f_n^+ - f_n^- ) \, e^{i {\pi \over 4}} = v_n^e \, e^{- i (n -\tau_R){\pi \over 2}} \, (1 - e^{- \tau_R 2 \pi}) \; . \label{eq:fn}
\end{equation}
The inelastic cross-section for the quark field is given by,
\begin{equation}
\sigma_{inel} =  | \sum_{n = -\infty}^{+ \infty} (f_n^+ - f_n^-) \,  e^{i n \theta} |^2 .
\end{equation}
Hence, from equation (\ref{eq:fn}),
\begin{equation}
\sigma_{inel} \sim  {1 \over \omega} | \sum_{n = -\infty}^{+ \infty} v_n^e \,  e^{- i n ({\pi \over 2} - \theta)} |^2 \; . \label{eq:inel}
\end{equation}
Using equations (\ref{eq:relation}), (\ref{eq:lambda}) and (\ref{eq:vnp}), we find,
\begin{equation}
v_n^e = {e^{i(n - \tau_R) {\pi \over 2}} \over 2} ({1 \over \delta_n^+ + e^i (n - \tau_R \pi)} - {1 \over \delta_n^- + e^i (n - \tau_R \pi)}) \label{eq:vninel}
\end{equation}
where
\begin{equation}
\delta_n^\pm= { w \, \lambda_n^\pm J_{n + 1 - \tau_R}(\omega R) + J_{n - \tau_R}(\omega R) \over w \, \lambda_n^\pm J_{-(n + 1 - \tau_R)}(\omega R) + J_{-(n - \tau_R)}(\omega R)} \label{eq:delta}
\end{equation}
and $\lambda^\pm$ are given by equations
(\ref{eq:lambda}). Equations (\ref{eq:inel}),
(\ref{eq:vninel}) and (\ref{eq:delta}) determine the
inelastic cross-section.This is given in terms of a
power series. However, using small argument expansions
for Bessel functions, we conclude that this power
series involves always one dominant term, the other
terms being suppressed by a factor $(\omega R)^n$ where n
is an integer such that $n \ge 1$. Therefore the
inelastic cross-section involves one dominant mode,
the other modes being exponentially suppressed. If $d$
denotes the dominant mode we get $\sigma_{inel} \sim
{1\over \omega} \, |v_{d}^e|^2$. The value of the dominant
mode depends on the sign of the the fractional flux
$\tau_{str}$. Our results can be summarised as
follow. 

For $0 < \tau_R < 1$, the mode $n = 0$ is enhanced, and the other modes are exponentially suppressed. Hence,
\begin{equation}
\sigma_{inel} \sim {1\over \omega} \, |v_0^e|^2 \; . 
\end{equation}
Using small argument expansions for Bessel functions, this yields
\begin{equation}
\sigma_{inel} \sim {1\over \omega} \, (eA R)^2 \, (\omega R)^{4(1 - \tau_R)}
\end{equation}
where $A$ is the value of the gauge field inside the
string core, $e$ is the gauge coupling constant, and
$R \sim \eta$, $\eta$ being the the grand unified
scale$\sim 10^{15} GeV$. The greater amplification
occurs for $eA R \sim 1$, giving $\sigma_{inel} \sim
{1\over \omega} \, (\omega R)^{4(1 - \tau_R)}$.  

For $-1 < \tau_R < 0$, the mode $n = -1$ is enhanced, and the other modes are 
exponentially suppressed. Hence,
\begin{equation}
\sigma_{inel} \sim {1\over \omega} \, |v_{-1}^e|^2 \; . 
\end{equation}
Using small argument expansions for Bessel functions, this yields
\begin{equation}
\sigma_{inel} \sim {1\over \omega} \, (eA R)^2 \, (\omega R)^{4 (1 + \tau_R)} \; .
\end{equation}
The greater amplification occurs for $eA R \sim 1$,
giving $\sigma_{inel} \sim {1\over \omega} \, (\omega R)^{4(1 +
\tau_R)}$. Thus, the baryon number violating
cross-section is not a strong interaction
cross-section but is suppressed by a factor depending
on the grand unified scale $\eta \sim R^{-1} \sim
10^{15} GeV$. The baryon number violation
cross-sections are very small. For $u_L$ and $d_L$ we
obtain, 
\begin{equation}
\sigma_{inel} \sim {1\over \omega} \, (\omega R)^{3.6} \; .
\end{equation}
Whereas for $d_R$ we get,
\begin{equation}
\sigma_{inel} \sim {1\over \omega} \, (\omega R)^{2.8} \; .
\end{equation}
Here again we have a marked asymmetry between left and
right handed fields. We find an indeterminate solution
for the left-conjugate up quark because its phase
around the string ($1 \over 10$) differs from the
phase of the left-handed electron $({-3\over 10})$ by
a fractional value different from a half.

\section{The second quantised cross-section}
\label{sec-second}

We now derive the baryon number violating
cross-sections using the perturbative method
introduced in Sec. \ref{sec-cs}.  

Firstly, we calculate the geometrical
cross-section. This is the cross-section for free
fields $\psi_{free}$, where $\psi_{free}$ is a
2-spinor. In the case of gauge fields mediating
catalysis it is given by, 
\begin{equation}
({d \sigma \over d \Omega })_{geom} = {1 \over \omega } \, (\omega R)^4 \, (eAR)^2
\end{equation}
where $\omega$ is the energy of the massless field
$\psi_{free}$, A is the value of the gauge field
mediating quark to lepton transitions, $e$ is the
gauge coupling constant and R is the radius of the
string with $R\sim \eta^{-1}$ with $\eta \sim 10^{15}$
GeV. 

The second step is to calculate the amplification
factor $ {\cal A} = {\psi \over \psi_{free}}$, $\psi$
and $\psi_{free}$ being two 2-spinors. The catalysis
cross-section is enhanced by a factor ${\cal A}^4$
over the geometrical cross-section, 
\begin{equation}
\sigma_{inel} \sim {\cal A}^4 \, ({d \sigma \over d \Omega})_{geom} \; .
\end{equation}
We now use the results of sections~\ref{sec-ext}, ~\ref{sec-int} and~\ref{sec-match} where we have solved the equations of motion for the fields $\psi$ and calculated the matching conditions. Using equation (\ref{eq:external}), we get the wave function $\psi$ at the string core, and for the mode n,
\begin{equation}
\psi^n  =
\left ( \begin{array}{lllll}
&((v_n^q \pm v_n^e) \, J_{n - \tau_{str}}(\omega R) &+& (v_n^{q'} \pm v_n^{e'}) \, J_{- (n  - \tau_{str})}(\omega R)) & e^{i n \theta}\\
i&((v_n^q \pm v_n^e) \, J_{n + 1- \tau_{str}}(\omega R) &+& (v_n^{q'} \pm v_n^{e'}) \, J_{-(n + 1 - \tau_{str})}(\omega R)) & e^{i (n + 1) \theta}
\end{array}
\right ) 
\end{equation}
Using equations (\ref{eq:relation}) and
(\ref{eq:lambda}) and using small argument expansions
for Bessel functions, we conclude that for $n \ge 0$,
$(v_n^q \pm v_n^e) \gg (v_n^{q'} \pm v_n^{e'})$, and
for $n < 0$, $(v_n^q \pm v_n^e) << (v_n^{q'} \pm
v_n^{e'})$. Now, from equation (\ref{eq:vnp}), we see
that one coefficient dominates which will be the
$O(1)$. Hence, for $n \ge 0$, $(v_n^q \pm v_n^e) \sim
1$, and for $n < 0$, $(v_n^{q'} \pm v_n^{e'}) \sim
1$. Therefore, using small argument expansions for
Bessel functions we get for $n \ge 0$, 
\begin{equation}
\psi^n \sim
\left ( \begin{array}{l}
(\omega R)^{n - \tau_{str}} \\
(\omega R)^{n + 1 - \tau_{str}}
\end{array}
\right ) 
\end{equation}
which is to be compared with $\psi_2^{free} \sim 1$ for free spinors. The upper component of the spinor is amplified while the other one is suppressed by a factor $\sim (\omega R)$. For $n < 0$ we have,
\begin{equation}
\psi^n \sim
\left ( \begin{array}{l}
(\omega R)^{- (n - \tau_{str, R})} \\
(\omega R)^{- (n + 1 - \tau_{str, R})}
\end{array}
\right ) 
\end{equation}
Hence we conclude that for $n < 0$ the lower component is amplified while the upper one is suppressed by a factor $\sim \omega R$.

Therefore, for $\tau_{str} = {- 3 \over 10}$, the amplification occurs for the lower component and for the mode $n = -1$. The amplification factor is
\begin{equation}
{\cal A} \sim (\omega R)^{\tau_{str}}
\end{equation}
leading to the baryon number violating cross-section,
\begin{equation}
\sigma_{inel} \sim {1\over \omega} \, (eA R)^2 \, (\omega R)^{4 \, (1 + \tau_{str})} \; . \label{eq:s1}
\end{equation}
In the case $\tau_{str} = {1 \over 10}$, the
amplification occurs for the upper component and for
the mode $n = 0$. The amplification factor is, 
\begin{equation}
{\cal A} \sim (\omega R)^{- \tau_{str}}
\end{equation}
leading to the baryon number violating cross-section,
\begin{equation}
\sigma_{inel} \sim {1\over \omega} \, (eA R)^2 \, (\omega R)^{4 \, (1 - \tau_{str})} \; . \label{eq:s2}
\end{equation}
This method shows explicitly which component of the
spinor and which mode are enhanced. The results agree
with scattering cross-sections derived using the first
quantised method.

\section{Conclusions}
\label{sec-concl2}

We have investigated elastic and inelastic scattering
off abelian cosmic strings arising during the phase
transition ${\rm SO}(10)
\stackrel{<\phi_{126}>}{\rightarrow} {\rm SU}(5) \times Z_2$
induced by the vacuum expectation value of a Higgs field in the {\bf 126} representation. The cross-sections were calculated
using both first quantised and second quantised
methods. The results of the two methods are in good
agreement. 

During the phase transition ${\rm SO}(10) \rightarrow {\rm SU}(5)
\times Z_2$, only the right-handed neutrino gets a
mass. This together with the fact that we are
interested in energies above the electroweak scale
allows us to consider massless particles. 

The elastic cross-sections are found to be
Aharonov-Bohm type cross-sections. This is as
expected, since we are dealing with fractional
fluxes. We found a marked asymmetry between
left-handed and right-handed fields for the electron
and the down quark fields. But there is no asymmetry
for the up quark field. This is a general feature of
grand unified theories. If cosmic strings were
observed it might be possible to use Aharonov-Bohm
scattering to determine the underlying gauge group. 

The inelastic cross-sections result from quark to
lepton transitions via gauge interactions in the core
of the string. The catalysis cross-sections are found
to be quite small, and here again we have a marked
asymmetry between left and right handed fields. From
Eqs. (\ref{eq:tauall}), (\ref{eq:s1}) and (\ref{eq:s2}), we
see that they
are suppressed by a factor $\sim \eta^{-3.6}$ for
the left-handed up and down quarks and by a factor
$\sim \eta^{-2.8}$ for the right-handed down quarks.   

Previous calculations have used a toy model to
calculate the catalysis cross-section. Here the string
flux could be `tuned' to give a strong interaction
cross-section. In our case the flux is given by the
gauge group, and is fixed for each particle
species. Hence, we find a strong sensitivity to the
grand unified scale. Our small cross-sections make it
less likely that grand unified cosmic strings could
erase a primordial baryon asymmetry, though they could
help generate it~\cite{Hindmarsh}.  If cosmic strings
are observed our scattering results, with the
distinctive features for the different particle
species, could help tie down the underlying gauge
group.

\chapter{Constraining supersymmetric SO(10) models through cosmology}
\label{chap-chap3}

\section{Introduction}

Supersymmetric {\rm SO}(10) models have received much interest in the
past ten years. {\rm SO}(10) is the minimal grand unified gauge group which
unifies all kinds of matter, thanks to its 16-dimensional spinorial
representation to which all fermions belonging to a single family can
be assigned. As mentioned in Chap. \ref{chap-chap1}, the running of
the gauge coupling constants measured at 
LEP in the minimal supersymmetric standard model with supersymmetry
broken at $10^3$ GeV merge in a single point at $2 \times 10^{16}$ GeV
\cite{meet}, hence strongly favouring supersymmetric versions of grand
unified theories ({\rm GUT}s). Supersymmetric {\rm SO}(10)  predicts
the ratio of the two electroweak Higgs vacuum expectation values, $\tan \beta$, an
unknown factor 
within the minimal supersymmetric standard model, giving $\tan \beta =
m_t/m_b$ \cite{tanB}. Natural doublet-triplet splitting can be
achieved in supersymmetric {\rm SO}(10) via the Dimopoulos-Wilczek mechanism
\cite{DW}. Note that the doublet-triplet splitting is related to the 
gauge hierarchy problem: coloured Higgs triplets, which can mediate baryon and lepton
number violation, must acquire a vacuum expectation value (VEV)
comparable with the GUT scale in order to prevent rapid proton decay,
whereas the usual Higgs doublets must acquire a VEV of the order of
$M_Z$. In supersymmetric SO(10), the derivation of fermion
masses and mixings can be achieved \cite{masses}. The gauge hierarchy
problem can be solved \cite{shafiun}. A $Z_2$ symmetry subgroup of the
$Z_4$ centre of 
{\rm SO}(10) can be left unbroken down to low energies, provided  only `safe'
representations \cite{Martin} are used to implement the symmetry breaking 
from {\rm SO}(10) down to the standard model gauge group. The $Z_2$
symmetry can suppress rapid proton decay and provide a cold 
dark matter 
candidate, stabilising the lightest superparticle (LSP). Finally,
introducing a pair of Higgs fields in the ${\bf 126 + \overline{126}}$
representations 
can give a superheavy Majorana mass to the right-handed neutrino, thus
providing a hot dark-matter candidate and solving the solar neutrino
problem through the Mikheyev-Smirnov-Wolfenstein (MSW) mechanism
\cite{MSW}. Supersymmetric SO(10) is also a good candidate for baryogenesis 
\cite{Y86,baryon}.

Thus, supersymmetric {\rm SO}(10) is very attractive from a particle physics
point of view and can also help to solve some cosmological problems.
One would therefore like to be able to select one of the
symmetry breaking patterns. Unfortunately, there is considerable freedom in
doing so, and the only way out from a particle physics point of view
would be from string compactification.      
      
On the other hand, by considering the implications of the symmetry
breaking patterns on the standard cosmology and by requiring that the
model be consistent with proton lifetime measurements, we can select
few of them. As mentioned in Chap. \ref{chap-chap1}, when symmetries
spontaneously 
break down, according to the 
Kibble mechanism \cite{Kibble}, topological defects, such as
monopoles, strings 
or domain walls, may form. Recall that monopoles, because they would be too
abundant, and domain 
walls, because they are too heavy, if present today would dominate the
energy density of the universe and lead to a cosmological catastrophe.
On the other hand, cosmic strings can explain structure formation and
part of the baryon asymmetry of the universe.        

We derive below the cosmological constraints on the symmetry breaking
schemes of 
supersymmetric {\rm SO}(10) down to the standard model due to the formation
of topological defects. In Sec. \ref{sec-breakings} we list the
possible symmetry breaking pattern involving at most one 
intermediate symmetry breaking scale. In Sec. \ref{sec-topoform}, we review the
conditions for the formation of topological defects, 
giving systematic conditions in
supersymmetric {\rm SO}(10). In Sec. \ref{sec-inflation} we briefly discuss the
hybrid inflationary scenario which can naturally arise in supersymmetric
{\rm SO}(10) models, see Sec. \ref{sec-inflation2}. In sections
\ref{sec-su5}, \ref{sec-leftr}, \ref{sec-three} and 
\ref{sec-direct} we give a systematic analysis of the cosmological
implications for the different symmetry breaking scenarios listed
in Sec. \ref{sec-breakings}. We conclude
in Sec. \ref{sec-concl3}, pointing out the only models not in conflict 
with the standard cosmology.

\section{Breaking down to the standard model}
\label{sec-breakings}
In this section, we give a list of all the symmetry breaking patterns
from supersymmetric {\rm SO}(10) down to the standard model, using no
more than one intermediate breaking scale. The main differences
between supersymmetric 
 and nonsupersymmetric {\rm SO}(10) models is in the symmetry breaking
scales as 
 we shall see and in the choice for the intermediate symmetry groups. 
In nonsupersymmetric models, at least one intermediate symmetry 
breaking is needed in order to obtain consistency with the measured value 
of $\sin^2 \theta_w$ and with the gauge coupling constants interpolated 
to high energy to meet around $10^{15}$ GeV. On the other hand, in 
supersymmetric {\rm SO}(10) models, we can break directly down to the 
standard model, breaking supersymmetry at $\sim 10^3 \: {\rm GeV}$, 
predicting the measured value of $\sin^2 \theta w$ and having 
the gauge coupling constant joining in a single point at $2 
\times 10^{16} \: {\rm GeV}$.

We shall consider the following symmetry breaking patterns from 
supersymmetric {\rm SO}(10) down to the standard model,

{\raggedleft
\begin{eqnarray}
&{\rm SO}(10) &\stackrel{M_{{\rm GUT}}}{\rightarrow} {\rm SU}(5)
\times {\rm U}(1)_X 
\stackrel{M_{{\rm G}}}{\rightarrow}  SM \label{eq:SSB1} \\ 
 &{\rm SO}(10) &\stackrel{M_{{\rm GUT}}}{\rightarrow} {\rm SU}(5)
\stackrel{M_{{\rm G}}}{\rightarrow}  SM  \\ 
  & {\rm SO}(10) & \stackrel{M_{{\rm GUT}}}{\rightarrow} {\rm SU}(5) \times
\widetilde{{\rm U}(1)} \stackrel{M_{{\rm G}}}{\rightarrow}  SM  \\  
&{\rm SO}(10)& \stackrel{M_{{\rm GUT}}}{\rightarrow} {\rm SU}(4)_c \times 
{\rm SU}(2)_L \times
{\rm SU}(2)_R  \stackrel{M_{{\rm G}}}{\rightarrow} SM \\ 
 &{\rm SO}(10) &\stackrel{M_{{\rm GUT}}}{\rightarrow}{\rm SU}(3)_c \times 
{\rm SU}(2)_L \times
{\rm SU}(2)_R \times {\rm U}(1)_{B\! -\! L}  \stackrel{M_{{\rm G}}}{\rightarrow} 
SM \label{SSB6}\\ 
 &{\rm SO}(10) &\stackrel{M_{{\rm GUT}}}{\rightarrow} {\rm SU}(3)_c \times
{\rm SU}(2)_L  
\times
{\rm U}(1)_R \times {\rm U}(1)_{B\! -\! L}   \stackrel{M_{{\rm G}}}{\rightarrow} SM 
\\ 
&{\rm SO}(10)& \stackrel{M_{{\rm GUT}}}{\rightarrow} SM \label{eq:SSB7}\\
&{\rm SO}(10) &\stackrel{M_{{\rm GUT}}}{\rightarrow} {\rm SU}(5) \times {\rm U}(1)_X
\stackrel{M_{{\rm G}}}{\rightarrow}  SM \times Z_2 \label{eq:SSB8}\\ 
&{\rm SO}(10) &\stackrel{M_{{\rm GUT}}}{\rightarrow} {\rm SU}(5) \times
 Z_2 \stackrel{M_{{\rm GUT}}}{\rightarrow} SM \times 
 Z_2 \\
  & {\rm SO}(10) & \stackrel{M_{{\rm GUT}}}{\rightarrow} {\rm SU}(5) \times
\widetilde{{\rm U}(1)} 
\stackrel{M_{{\rm G}}}{\rightarrow} SM \times Z_2 \\ 
&{\rm SO}(10) &\stackrel{M_{{\rm GUT}}}{\rightarrow} {\rm SU}(4)_c \times
{\rm SU}(2)_L  
\times
{\rm SU}(2)_R  \stackrel{M_{{\rm G}}}{\rightarrow} SM 
\times Z_2 \\ 
&{\rm SO}(10) &\stackrel{M_{{\rm GUT}}}{\rightarrow}{\rm SU}(3)_c \times
{\rm SU}(2)_L \times 
{\rm SU}(2)_R \times {\rm U}(1)_{B\! -\! L} \stackrel{M_{{\rm G}}}{\rightarrow} SM 
\times Z_2 \\
 &{\rm SO}(10) &\stackrel{M_{{\rm GUT}}}{\rightarrow} {\rm SU}(3)_c \times 
{\rm SU}(2)_L \times
{\rm U}(1)_R \times {\rm U}(1)_{B\! -\! L}
\stackrel{M_{{\rm G}}}{\rightarrow} SM \times Z_2 \label{eq:SSB13}\\ 
 &{\rm SO}(10) &\stackrel{M_{{\rm GUT}}}{\rightarrow}  SM \times Z_2
\label{eq:SSB14} 
\end{eqnarray}
}
where $SM$ stands for the standard model gauge group ${\rm SU}(3)_c \times
 {\rm SU}(2)_L \times {\rm U}(1)_Y$. In models (\ref{eq:SSB1})
to (\ref{eq:SSB7}), supersymmetry must be broken at $\sim 10^3$ GeV, and the symmetry 
group
${\rm SU}(3)_c \times {\rm SU}(2)_L \times {\rm U}(1)_Y$ is broken by the usual Higgs mechanism down to ${\rm
SU}(3)_c \times {\rm U}(1)_Q$ 
at $\sim M_{{\rm Z}}$. The process by which supersymmetry is broken is not considered. In models (\ref{eq:SSB8}) to (\ref{eq:SSB14}), supersymmetry must also
be broken at $M_{{\rm s}}\sim 
10^3$ GeV, and the group symmetry ${\rm SU}(3)_c \times {\rm SU}(2)_L
\times {\rm U}(1)_Y \times Z_2$ is broken by the usual Higgs mechanism down to ${\rm SU}(3)_c \times {\rm
U}(1)_Q \times Z_2$ 
at $\sim M_{{\rm Z}}$.  In the latter cases, the $Z_2$ symmetry remains
unbroken down to low energy, and acts as matter parity. It preserves
large values for the proton lifetime and stabilises the lightest 
supersymmetric particle (LSP), thus
providing a good hot dark matter candidate.

In order to satisfy LEP data, we must have $M_{{\rm GUT}} \sim M_{{\rm G}}$ (see Langacker 
and Luo in Ref. \cite{meet}). For nonsupersymmetric models, the value of 
the $B \! - \! L$ symmetry breaking scale is anywhere between $10^{10}$ and
$10^{13.5}$ GeV \cite{moha}. For the supersymmetric case it is around
$10^{15} - 10^{16}$ GeV. Indeed, the scale $M_{{\rm G}}$ is fixed by the
unification of the gauge couplings, and in the absence of particle
threshold corrections is $M_{{\rm G}} \sim 10^{16}$ GeV \cite{meet}. But, as
in the nonsupersymmetric case, threshold corrections can induce
uncertainties of a factor $10^{\pm 1}$ GeV. These corrections
vary with the     
intermediate subgroup considered, but in any cases, we can assume that
$M_{{\rm G}} \sim 10^{15} - 10^{16}$ GeV. The scale $M_{{\rm GUT}}$ must be greater than
the unified scale $M_{{\rm G}}$ and below the Planck scale, therefore we must
have $10^{19} \, {\rm GeV} \geq M_{{\rm GUT}} \geq 10^{15}-10^{16}$ GeV.

In order to simplify the notation, we shall use  
\begin{eqnarray} 
&& 4_c 2_L 2_R \equiv {\rm SU}(4)_c \times {\rm SU}(2)_L \times {\rm
SU}(2)_R \label{eq:notation1}\\ 
&& 3_c 2_L 2_R 1_{B-L} \equiv {\rm SU}(3)_c \times {\rm SU}(2)_L
\times {\rm SU}(2)_R \times {\rm U}(1)_{B-L}  \\ 
&&  3_c 2_L 1_R 1_{B-L} \equiv {\rm SU}(3)_c \times {\rm SU}(2)_L
\times {\rm U}(1)_R \times {\rm U}(1)_{B-L} \\ 
&&   3_c 2_L 1_Y (Z_2) \equiv {\rm SU}(3)_c \times {\rm SU}(2)_L
\times {\rm U}(1)_Y (\times Z_2) \\ 
&&   3_c 1_Q (Z_2) \equiv {\rm SU}(3)_c \times {\rm U}(1)_Q (\times
Z_2) \label{eq:notation2} 
\end{eqnarray}

\section{Topological defect formation in supersymmetric models}
\label{sec-topoform}

In this section, we show that the conditions for topological defect 
formation well known in non supersymmetric theories (see Ref. \cite{Kibble} and also Chap. \ref{chap-chap1}) are not affected by the presence of supersymmetry. We then study the
formation of hybrid defects, such as monopoles connected by strings
or domain walls bounded by strings, which can arise in SO(10) models, particularly looking at their
cosmological impact \cite{Kib,Vil}.

\subsection{Defect formation in supersymmetric models}

We study here the conditions for defect formation in supersymmetric 
models. We show that the conditions for topological defect formation
in nonsupersymmetric theories \cite{Kibble}, are not affected by the 
presence of supersymmetry.

In nonsupersymmetric theories, the conditions for topological 
defect formation during the spontaneous
symmetry breaking of a nonsupersymmetric Lie group G to a
nonsupersymmetric Lie group H are well known; they are associated
with the connection of the vacuum manifold ${{\rm G}\over {\rm H}}$
\cite{Kibble}. 
 Now one may worry
about the non Lie nature of the superalgebra. Fortunately, it has been
shown ~\cite{Srivastava} that the superalgebra is Lie admissible and
that the infinitesimal transformations of the superalgebra can be
exponentiated to obtain a Lie superalgebra. The Lie admissible algebra
is an algebraic covering of the Lie algebra, and it was first
identified by Albert ~\cite{Albert}. It is such a covering that
allows  a Lie admissible infinitesimal behaviour while preserving the
global structure of the Lie group. The graded Lie algebra is Lie 
admissible and therefore much of the Lie
algebra theory may be extended to it with the appropriate
modification. In particular, a connected (super)Lie group structure
persists ~\cite{Santilli}. Hence, the formation of topological defects
in supersymmetric models 
will be the same as in nonsupersymmetric ones. Whether or not
supersymmetry is broken at the phase transition will not affect
the conditions under which topological defects form. These conditions
are reviewed in Chap. \ref{chap-chap1}. In this chapter and in the
following one, when we denote a group G, we really mean the 
supersymmetric version of this group, and when we write {\rm SO}(10) we mean
its universal covering group Spin(10) (supersymmetric) which is simply
connected.

\subsection{Hybrid defects}
\label{sec-topodw}

When we have an intermediate breaking scale, we
can also get mixed defects. There are two kinds of mixed defects that 
we can get in supersymmetric {\rm SO}(10) models; they are monopoles connected
 by strings and domain walls bounded by strings. Their cosmological 
evolutions have been studied in a nonsupersymmetric general case 
\cite{Kib,Vil}.

\subsubsection{Monopoles connected by strings}
\label{sec-mncs}

In supersymmetric {\rm SO}(10) models, we can have monopoles 
connected by strings \cite{Vil}. If the first phase 
transition leaves an unbroken {\rm U}(1) symmetry which
later breaks to unity, that is if the breaking pattern proceeds as 
\begin{equation}
{\rm G} \rightarrow{\rm H}\times {\rm U}(1)_x \rightarrow{\rm H}\label{eq:monstr}
\end{equation} 
where G and H are both simply connected, then monopoles form 
at the first phase transition, and then get connected by strings at the
following one. Indeed, the second homotopy group
$\pi_2({{\rm G}\over { {\rm H}\times {\rm U}(1)}}) = \pi_1({\rm H}\times {\rm U}(1)) = Z$
indicates the formation of monopoles during the first phase transition
in (\ref{eq:monstr}). These monopoles carry a ${\rm U}(1)_x$ magnetic
charge, and are 
topologically unstable. Now the first homotopy group
$\pi_1({{{\rm H}\times {\rm U}(1)} \over {\rm H}})$ is also non trivial, 
hence cosmic strings form at 
the second stage of symmetry breaking in (\ref{eq:monstr}). The strings
connect monopole/antimonopole pairs of the first phase
transition \cite{Vil}. Because the whole system of strings rapidly 
decays \cite{Vil}, 
monopoles connected by strings do not seem to affect the standard 
cosmology in any essential way. On the other hand, if the universe 
undergoes a period of inflation between the two phase transitions, or 
if the phase transition leading to the formation of monopoles is 
itself inflationary, then the picture is very different. The decay of the 
system of strings  is negligible. If the monopoles are inflated 
beyond the horizon, 
the strings form according to the Kibble mechanism and their evolution 
is that of topologically stable cosmic strings \cite{Vil}. In this class 
of scenarios, with inflation and cosmic strings, temperature fluctuations 
in the CBR measured by COBE give
constraints on the scale of the  
phase transition leading to the string formation and on the scalar 
coupling constant (see next chapter).

\subsubsection{Walls bounded by strings}

The other kind of topological mixed defect that we can get in {\rm
SO}(10) models  
is domain walls connected by strings. A first phase transition leaves 
an unbroken discrete symmetry, and cosmic strings form. At a subsequent 
phase transition, this discrete symmetry breaks, leading to the formations 
of domain walls. They are bounded by the strings previously
formed. Specifically, consider a symmetry 
breaking pattern of the form 
\begin{equation}
{\rm G} \rightarrow{\rm H}\times Z_2 \rightarrow{\rm H}\label{eq:csdw}
\end{equation}
where G and H are both simply connected. The first homotopy group 
$\pi_1({{\rm G}\over { {\rm H}\times Z_2}}) = \pi_0({\rm H}\times Z_2) = Z_2$; thus, 
$Z_2$-strings form during the first phase transition in (\ref{eq:csdw}), and 
they are topologically unstable. The discrete
$Z_2$ symmetry breaking leads to the formation of domain walls at the
second stage of symmetry breaking bounded by strings
of the first phase transition. Such extended objects have been first 
studied by Kibble et al. ~\cite{Kib}. 
They have shown that, in the nonsupersymmetric case, the cosmological 
relevance of these mixed objects depends on whether inflation 
occurs between the time when strings form and the time when the 
symmetry breaking leading to the formation of these walls occurs. The 
presence of supersymmetry does not affect the above conclusions. 
Following Ref. \cite{Kib}, we get the following results. If the 
transition leading to the formation of the walls takes place without
supercooling, the walls lose their energy by friction and disappear  
in a time $t_d \sim (t_W t_*)^{1 \over 2}$ where $t_W$ is the cosmic 
time corresponding to the the scale $T_W$ at which the walls 
form and $t_* = {3 \alpha_G \eta_0 \over 32 \pi \eta_3} {M_p^2 \over
 M_{{\rm G}}^3}$, where $\eta_3$ is the effective massless degrees of freedom 
reflected by the walls and $\eta_0$ is the effective number of degrees 
of freedom in the supersymmetric $3_c 2_L 1_Y (Z_2 )$ phase. With
$\eta_3 = 33.75$ and $\eta_0 = 228.75$  we find $t_d \sim 10^{-33} -
10^{-36}$ sec  
for $T_W \sim 10^{15} - 10^{16}$ GeV and the  corresponding 
scale $T_* \sim 10^9 - 10^{12}$ GeV. Therefore these extended objects 
do not seem to affect the standard cosmology in any essential way. But 
if there is a period of inflation between the two phase transitions, 
the strings can be pushed to arbitrarily large scales; the walls form according
to the Kibble mechanism and their evolution is that of topologically 
stable walls. The only difference from topologically stable $Z_2$-walls
 is that the walls can now decay by the quantum nucleation of holes 
bounded by strings. Hole nucleation however is a tunnelling process and 
is typically suppressed by a large exponential factor. The corresponding 
decay time is much larger than the time at which the walls come to 
dominate the universe, thereby upsetting standard cosmology.

\section{Inflation in supersymmetric {\rm SO}(10) models}
\label{sec-inflation}

Since {\rm SO}(10) is simply connected and the standard model gauge group 
involves an unbroken {\rm U}(1) symmetry which remains unbroken down to low 
energy, all symmetry breaking patterns from supersymmetric {\rm SO}(10) 
down to the standard model automatically involve the formation of 
topologically stable monopoles. Even if some monopoles are connected 
by strings, a large fraction of them will remain stable down to low
 energy. Hence some mechanism has to be invoked in order to obtain 
consistency with standard cosmology, such as an inflationary 
scenario. In this section, we briefly discuss a false vacuum
 hybrid inflationary scenario which is the most natural mechanism 
for inflation in global supersymmetric {\rm SO}(10) models, as will be
shown in the next chapter. 

The superpotential in the inflaton sector
is similar to that studied in Refs. \cite{ShafiDvaSch,Ed1}. 
It 
involves a scalar field ${\cal S}$ singlet under SO(10) and a pair of Higgs 
fields $\Phi + \overline{\Phi}$ in the ${\bf 16} + {\bf
\overline{16}}$ or in the ${\bf 126} + {\bf \overline{126}}$
dimensional representations of SO(10). These Higgs fields are used to lower the
rank of SO(10) by one unit, they must get a VEV the order of the GUT scale. If  
the $Z_2$ parity is to be kept unbroken, as in models (\ref{eq:SSB8})
to (\ref{eq:SSB14}), a pair of ${\bf 126} + {\bf \overline{126}}$
must be used. The superpotential can be written as
\begin{equation}
W = \alpha {\cal S} \overline{\Phi} \Phi - \mu^2 {\cal S} 
\end{equation}
where $\mu$ and $\alpha$ are two positive constants such that 
${\mu \over {\sqrt {\alpha}}} = M_{{\rm GUT}}$. If the rank of the group is 
lowered at $M_{\rm G}$, we have ${\mu \over {\sqrt {\alpha}}} = M_{{\rm G}}$.

The evolution of the fields is as follows  (a complete discussion
of the potential in a general supersymmetric case is studied in 
Ref. \cite{ShafiDvaSch} and in a specific supersymmetric SO(10) model 
is studied in the next chapter). The fields take random 
initial values, just
subject to the constraint that the energy density be at the Planck
scale. The inflaton field is distinguished from the other fields
from the fact that the gradient of the GUT potential with respect to the
inflaton field is very small. Therefore the non inflaton fields, except
the $\Phi$ and $\overline{\Phi}$ fields, will roll very
quickly down to their minimum at an approximately fixed value for the
inflaton. Inflation occurs as the inflaton rolls slowly down
the potential. The symmetry breaking implemented with the $\Phi +
\overline{\Phi}$ 
fields occurs at the end of inflation and associated topological defects
are not inflated away, see \cite{ShafiDvaSch} and Chap. \ref{chap-chap4}.

\section{{\rm SU}(5) as intermediate scale} 
\label{sec-su5}

We shall describe in this section the symmetry breaking patterns 
from supersymmetric {\rm SO}(10) involving an {\rm SU}(5) intermediate
symmetry. 
When the intermediate scale involves {\rm SU}(5) as a subgroup, 
 the scale $M_{{\rm G}}$ has to be $\sim 
10^{16}$ GeV, and consequently the scale $M_{{\rm GUT}}$ is pushed close to 
the string compactification scale. ${\rm SO}(10)$ can break via ${\rm
SU}(5)$ in four different ways. It can break via ${\rm SU}(5) \times
{\rm U}(1)_X$, ${\rm SU}(5)$ , via ${\rm SU}(5)  
\times \widetilde{{\rm U}(1)}$ and via ${\rm SU}(5) \times Z_2$.

\subsection{Breaking via ${\rm SU}(5) \times {\rm U}(1)_X$}

We consider here two symmetry breaking patterns, 
\begin{eqnarray}
&{\rm SO}(10)& \stackrel{M_{{\rm GUT}}}{\rightarrow} {\rm SU}(5) \times {\rm
U}(1)_X\\ 
&&\stackrel{M_{{\rm G}}}{\rightarrow} {\rm SU}(3)_c \times {\rm SU}(2)_L
\times {\rm U}(1)_Y (\times Z_2) \\ 
&& \stackrel{M_{{\rm Z}}}{\rightarrow} {\rm SU}(3)_c \times {\rm U}(1)_Q (\times Z_2)
\end{eqnarray}
with and without the $Z_2$ symmetry unbroken down to low energy. The latter 
is necessary to preserve large values for the proton lifetime and to 
stabilise the LSP. It can arise only if a pair of $({\bf 126} + {\bf
\overline{126}})$  
dimensional Higgs representations is used to lower the rank of the group. 

The ${\rm U}(1)_X$ symmetry commutes with {\rm SU}(5). 
The X and Y directions are orthogonal 
to each other, and thus the ${\rm U}(1)_X$ symmetry breaks down to unity at 
$M_{{\rm G}}$ (or to $Z_2$ if a pair of ${\bf 126} + {\bf \overline{126}}$
Higgs fields are used to  
break ${\rm SU}(5) \times {\rm U}(1)_X$). This feature is going to affect the 
formation of topological defects.

The first homotopy group $\pi_1 ({\rm SU}(5) \times {\rm U}(1)_X) = Z$
is non trivial  
and thus topological monopoles form when {\rm SO}(10) breaks. They have a mass 
$M_m  
\geq 5 \times 10^{17}$ GeV. At the following phase transition the
${\rm U}(1)_X$  
symmetry breaks to unity (to $Z_2$) and hence cosmic strings 
($Z_2$-strings) form. They  connect monopole-antimonopole pairs 
previously formed (see section \ref{sec-mncs}). They have a mass 
per unit length $\sim 10^{32} \: {\rm GeV}^2$.

When ${\rm SU}(5) \times {\rm U}(1)_X$ breaks down to $3_c 2_L 1_Y
(Z_2)$ new lighter monopoles form. Indeed, since ${\rm U}(1)_X$ breaks
down to unity (to $Z_2$) we  
consider the second homotopy group $\pi_2 ({ {\rm SU}(5) \over 3_c 2_L 1_Y})$ 
to look for monopoles formations at $M_{{\rm G}}$. Hence topologically stable 
monopoles form. They have a mass $M_m \sim 10^{17}$ GeV. They are
topologically stable. Their topological charge may change from Y to
Q. 

Since monopoles form at both phase transitions and since the lighter ones 
are topologically stable, the inflationary scenario, as in section
\ref{sec-inflation}, is unable to solve the monopole problem. Hence
these two models are inconsistent with observations.

\subsection{Breaking via {\rm SU}(5)}
\label{sec-alone}

Here, {\rm SO}(10) breaks down to the standard model with intermediate
{\rm SU}(5) symmetry  
alone. In this case, there is no interest in going to a larger grand
unified group. The breaking scheme is 
\begin{eqnarray}
&{\rm SO}(10)& \stackrel{M_{{\rm GUT}}}{\rightarrow} {\rm SU}(5) \\
&&\stackrel{M_{{\rm G}}}{\rightarrow} {\rm SU}(3)_c \times {\rm SU}(2)_L
\times {\rm U}(1)_Y \\ 
&& \stackrel{M_{{\rm Z}}}{\rightarrow} {\rm SU}(3)_c \times {\rm U}(1)_Q 
\end{eqnarray}
which is that of model (\ref{eq:SSB1}). Since ${\rm SO}(10)$ and {\rm
SU}(5) are both simply connected, no topological defects form during
the first stage of symmetry breaking.  

The second homotopy group $\pi_2({{\rm SU}(5) \over 3_c 2_L 1_Y}) = Z$ 
hence topological monopoles form when ${\rm SU}(5)$ breaks down to the 
standard model. 
The monopoles carry Y topological charge. The second homotopy 
group $\pi_2({{\rm SU}(5) \over 3_c 1_Q}) = Z$ which shows that the monopoles 
are topologically stable. They have a mass $M_m \sim \: 10^{17}$ GeV. 
Their topological charge may change from Y to Q. 

Since the rank of {\rm SO}(10) is 5 and the rank of ${\rm SU}(5)$ is
4, if we use an inflationary scenario as described in
Sec. \ref{sec-inflation} to solve  
the monopole problem, the inflaton field will
couple to a pair of ${\bf 16} + {\bf \overline{16}}$ Higgs fields
representations which will be used used to break {\rm SO}(10).  
The monopoles described above will form at the end of inflation, 
and their density will be high enough to dominate the universe. Hence 
this model is in conflict with the standard cosmology. It is also inconsistent
with the actual data on the proton lifetime.

\subsection{Breaking via ${\rm SU}(5) \times \widetilde{{\rm U}(1)}$}

More interesting is the breaking via flipped ${\rm SU}(5)$
\begin{eqnarray}
&{\rm SO}(10) &\stackrel{M_{{\rm GUT}}}{\rightarrow} {\rm SU}(5) \times
\widetilde{{\rm U}(1)} \\ &&\stackrel{M_{{\rm G}}}{\rightarrow} {\rm SU}(3)_c
\times {\rm SU}(2)_L \times {\rm U}(1)_Y  \\ 
&& \stackrel{M_{{\rm Z}}}{\rightarrow} {\rm SU}(3)_c \times {\rm U}(1)_Q 
\end{eqnarray}
Note that with flipped {\rm SU}(5), rather than using {\rm SO}(10) for
the grand  
unified gauge group, the monopole problem is avoided \cite{nano}. 
The $\widetilde{{\rm U}(1)}$ contains part of the electromagnetic gauge 
group ${\rm U}(1)_Q$. The above symmetry breaking can only be implemented 
in supergravity {\rm SO}(10) models \cite{nano}.

The first homotopy group $\pi_1({\rm SU}(5) \times \widetilde{{\rm
U}(1))} = Z$ and  
therefore the 
first phase transition leads to the formation of topological monopoles 
when {\rm SO}(10) breaks. Furthermore, since $\pi_1(3_c 2_L 1_Y) = 
\pi_1(3_c 1_Q) = Z$ and $\widetilde{{\rm U}(1)}$ contains part of the
${\rm U}(1)_Y$ and ${\rm U}(1)_Q$ symmetries, these monopoles are
topologically stable. They have a  
mass $M_m \geq 5 \times 10^{17}$ GeV. They carry $B\! -\! L$, and 
their topological charge may change to Y and then to Q. Embedded cosmic 
strings form after the second stage of symmetry breaking \cite{Nathan}.

We should be able to cure the monopole problem with an hybrid inflationary
scenario for supergravity models. Indeed, since 
the rank of ${\rm SU}(5) \times \widetilde{{\rm U}(1)}$ is 5, the
inflaton field can couple  
to the Higgs needed to break ${\rm SU}(5) \times \widetilde{{\rm
U}(1)}$, and embedded strings  
will form at the end of inflation. 
Hence from a defect point of view the model is interesting, but appears 
to be inconsistent  with the actual data for proton lifetime 
\cite{Martin} and does not provide 
any Majorana mass for the right-handed neutrino. The latter problems are 
solved if we break ${\rm SU}(5) \times
\widetilde{{\rm U}(1)}$ down to $3_c 2_L 1_Y Z_2$. In that case, a
$({\bf 126} + {\bf \overline{126}})$ dimensional Higgs representation
is used to break  
${\rm SU}(5) \times \widetilde{{\rm U}(1)}$. Since the 
first homotopy groups $\pi_1( {{\rm SU}(5) \times \widetilde{{\rm U}(1)} 
\over 3_c 2_L 1_Y Z_2}) 
= Z_2$ and $\pi_1( {{\rm SU}(5) \times {\rm U}(1) \over 3_c 1_Q Z_2}) 
= Z_2$, topologically stable $Z_2$-strings also form. They have a mass
per unit length $\sim 10^{32} \: {\rm GeV}^2$.

\subsection{Breaking via ${\rm SU}(5) \times Z_2$}

We consider here the breaking of {\rm SO}(10) via {\rm SU}(5) with
added parity. The symmetry breaking is 
\begin{eqnarray}
&{\rm SO}(10) &\stackrel{M_{{\rm GUT}}}{\rightarrow} {\rm SU}(5) \times Z_2
\\ &&\stackrel{M_{{\rm G}}}{\rightarrow} {\rm SU}(3)_c \times {\rm SU}(2)_L
\times {\rm U}(1)_Y \times Z_2 \\ 
&& \stackrel{M_{{\rm Z}}}{\rightarrow} {\rm SU}(3)_c \times {\rm U}(1)_Q \times Z_2
\end{eqnarray}
 where the unbroken $Z_2$ symmetry is a subgroup of the $Z_4$ centre of 
{\rm SO}(10). It plays the role of matter parity. It preserves large
values for  
the proton lifetime and stabilises the LSP; thus, the model is consistent 
with the actual data on proton decay and provides a good hot dark 
matter candidate.

Now the fundamental homotopy group $\pi_0({\rm SU}(5) \times Z_2) = Z_2$ and 
therefore $Z_2$ cosmic strings form
during the first phase transition. They have a mass per
unit length $10^{38} {\rm GeV}^2 \geq \mu \geq 10^{32} \: {\rm
GeV}^2$. Since the $Z_2$  
symmetry is kept unbroken down to low energy, these strings
remain topologically stable. They have been widely
studied in the nonsupersymmetric case, see Ref. \cite{Aryal87,Ma} and
also Chap. \ref{chap-chap2}.  

As in section \ref{sec-alone}, it is clear that topologically stable 
monopoles form during the second phase transition with mass $M_m 
\sim 10^{17}$ GeV. Hence as in section \ref{sec-alone}, the model 
is in contradiction with observations.

We conclude that the only symmetry breaking pattern from {\rm SO}(10)
down to the standard model with intermediate {\rm SU}(5) symmetry
consistent with observations,  
is 
\begin{equation}
 {\rm SO}(10) \rightarrow {\rm SU}(5) \times \widetilde{{\rm U}(1)}
\rightarrow 3_c 2_L 1_Y Z_2  
\rightarrow 3_c 1_Q Z_2
\end{equation}
where the $Z_2$ symmetry must be kept unbroken in order to 
preserve large values for the proton lifetime. The above symmetry 
breaking can only be implemented in supergravity models.

\section{Patterns with a left-right intermediate scale} 
\label{sec-leftr}

In this section we study the symmetry breaking patterns from 
supersymmetric {\rm SO}(10) down to the standard model involving an 
${\rm SU}(2)_L \times {\rm SU}(2)_R$ intermediate symmetry. These are the 
symmetry breaking patterns with intermediate $4_c 2_L 2_R (Z_2)$ 
or $3_c 2_L 2_R 1_{B-L} (Z_2)$ symmetry groups. We show that these models, due 
the unbroken ${\rm SU}(2)_L \times {\rm SU}(2)_R$ symmetry share a property, 
which can make them cosmologically irrelevant, depending on the Higgs field
chosen to implement the symmetry breaking.  We then give a full study
of the formation of the topological defects in each model. 

\subsection{Domain walls in left-right models}
\label{sec-DW}

We study here a property shared by the symmetry breaking schemes from 
{\rm SO}(10) down to the standard model, with or without unbroken parity $Z_2$,
\begin{equation}
{\rm SO}(10)  \stackrel{M_{{\rm GUT}}}{\rightarrow}{\rm G} \stackrel{M_{G}}{\rightarrow}
3_c 2_L 1_Y (Z_2) \label{eq:LR1}
\end{equation}
where G is either $4_c 2_L 2_R$ or $3_c 2_L 2_R 1_{B-L}$. In these models, 
the intermediate scale involves an unbroken ${\rm SU}(2)_L \times {\rm
SU}(2)_R$  
symmetry, and consequently the intermediate symmetry group can be 
invariant under the charge conjugation operator, depending on the 
Higgs multiplet chosen to break {\rm SO}(10). The latter leaves an unbroken 
discrete $Z_2^c$ symmetry which breaks at the following phase transition. 
In this case, the general symmetry breaking scheme given in Eq. (\ref{eq:LR1}) should really be written as   
\begin{equation}
{\rm SO}(10) \stackrel{M_{{\rm GUT}}}{\rightarrow} {\rm G}
\times Z_2^c \stackrel{M_{{\rm G}}}{\rightarrow} SM (\times Z_2)  \label{eq:LR3} \: .
\end{equation}
If ${\rm G} = 4_c 2_L 2_R$, the discrete $Z_2^c$ symmetry appears if the Higgs 
used to break {\rm SO}(10) is a single 54-dimensional
representation \cite{54}.  
If ${\rm G}= 3_c 2_L 2_R 1_{B-L}$ the $Z_2^c$ symmetry appears if a single 210 
dimensional Higgs representation is used, with appropriate parameter 
range in the Higgs potential \cite{210}. The appearance of the discrete 
$Z_2^c$ symmetry leads to a cosmological problem \cite{Kib}. Indeed, 
since Spin(10) is simply connected, $\pi_1({{{\rm SO}(10)} \over {{\rm G}\times 
Z_2^c}}) = \pi_0({\rm G}\times Z_2^c ) = Z_2$ and therefore $Z_2$ strings 
form during the first phase transition associated with the breaking of 
{\rm SO}(10). They have a mass per unit length $\sim 10^{32} - 10^{34}
\: {\rm GeV}^2$.  
When  
the discrete $Z_2^c$ symmetry breaks, domain walls form bounded by the
strings of the previous phase transition. Some closed walls can also
form. As shown in Sec. \ref{sec-topodw}, these domain walls do not
affect the standard cosmology in any essential way. On the other
hand, if a period of inflation occurs between the two phase transition, or 
if the phase transition leading to the walls formation is itself 
inflationary, then the evolution of the walls is that of topologically 
stable $Z_2$ walls. They dominate the universe, destroying the
standard cosmology.

\subsection{Breaking via $4_c 2_L 2_R$}

We now consider the symmetry breaking of {\rm SO}(10) via the
Pati-Salam gauge group  
$4_c 2_L 2_R$ subgroup of {\rm SO}(10) which later breaks down to the 
standard model gauge group with or without matter parity
\begin{eqnarray}
&{\rm SO}(10)& \stackrel{M_{{\rm GUT}}}{\rightarrow} {\rm SU}(4)_c \times
{\rm SU}(2)_L \times  
 {\rm SU}(2)_R \\
&&\stackrel{M_{{\rm G}}}{\rightarrow} {\rm SU}(3)_c \times {\rm SU}(2)_L
\times {\rm U}(1)_Y  
(\times Z_2) \\
&&\stackrel{M_{{\rm Z}}}{\rightarrow} {\rm SU}(3)_c \times {\rm U}(1)_Q
(\times Z_2) \label{eq:PS} 
\end{eqnarray}
with supersymmetry broken at $\simeq 10^3$ GeV and the scales $M_{{\rm GUT}}$ 
and $M_{{\rm G}}$, respectively, satisfy $M_{pl} \geq M_{{\rm GUT}} \geq 10^{16}$ GeV 
and $M_{{\rm G}} \sim 10^{15}-10^{16}$ GeV.
The discrete $Z_2$ symmetry is kept unbroken if we use a pair of 
$({\bf 126} + {\bf \overline{126}})$ dimensional Higgs representation
to break $4_c 2_l 2_R$,  
and is broken if we use a pair of $({\bf 16} + {\bf \overline{16}})$
dimensional 
Higgs representation. The unbroken $Z_2$ symmetry plays the role of
matter parity,  
preserving large values for the proton lifetime and stabilising the LSP. 
Hence only the model with unbroken $Z_2$ at low energy is consistent
with the actual value for proton lifetime.

If a single 54 dimensional Higgs representation is used to break
{\rm SO}(10),  
equation (\ref{eq:PS}) should really be written as \cite{Kib}
\begin{eqnarray}
&Spin(10) &\stackrel{M_{{\rm GUT}}}{\rightarrow} ({({\rm Spin}(6) \times
{\rm Spin}(4)) 
\over Z_2 }) \times Z_2^c \label{eq:PS2}\\
&&\stackrel{M_{{\rm G}}}{\rightarrow} {\rm SU}(3)_c \times {\rm SU}(2)_L
\times {\rm U}(1)_Y  
(\times Z_2) \\
&&\stackrel{M_{{\rm Z}}}{\rightarrow} {\rm SU}(3)_c \times {\rm U}(1)_Q (\times Z_2) 
\end{eqnarray}
where we have explicitly shown the hidden symmetry. A $Z_2$ symmetry has 
to be factored out in Eq. (\ref{eq:PS2}) since Spin(6) and Spin(4) 
have a non trivial intersection. The overall $Z_2^c$  is generated by the 
charge conjugation operator; it is unrelated to the previous $Z_2$ one. 
Subsequently, the $Z_2^c$ discrete symmetry is broken. If a pair of Higgs 
fields 
in the ${{\bf 126} + {\bf \overline{126}}}$ representation are used to
break $4_c 2_l 2_R$,  
then a new $Z_2$ symmetry emerges, as described above; it is unrelated 
to the previous ones. The standard model gauge group is broken with 
Higgs fields in the {\bf 10} dimensional representation of {\rm SO}(10). 

If a single {210}-Higgs multiplet is used to break $4_c 2_l 2_R$, with 
appropriate range in the parameters of the Higgs potential, the $Z_2^c$ 
does not appear \cite{210}. 

\subsubsection{Monopoles}
\label{sec-1}

The non trivial intersection of Spin(6) and Spin(4) leads to the 
production of superheavy monopoles \cite{Vil} when {\rm SO}(10) breaks to
$4_c 2_L 2_R$. These monopoles are superheavy with a mass $M_m \geq 
10^{17}$ GeV. They are topologically unstable.

Since the second homotopy group $\pi_2({ 4_c 2_L 2_R \over  3_c 2_L 1_Y 
(Z_2)}) = Z$ is non trivial, new monopoles form when $4_c 2_L 2_R $ breaks 
down to the standard model gauge group. They are unrelated to the
previous monopoles. Furthermore, since the second homotopy group
$\pi_2({ 4_c 2_L 2_R  
\over  3_c 1_Q (Z_2)}) = Z$ is also non trivial, these lighter monopoles
 are topologically stable. They have a mass $M_m \sim 10^{16}-10^{17}$ GeV.
These monopoles form according to the Kibble mechanism, and
their density is such that, if present today, they would dominate the
energy density of the universe.

\subsubsection{Domain walls}
\label{sec-ps}

If a 54 dimensional Higgs representation is used to break {\rm SO}(10) 
down to $4_C 2_L 2_R$, the symmetry breaking is given by 
Eq. (\ref{eq:PS2}) which is of the form of Eq. (\ref{eq:LR3}) 
with ${\rm G}= 4_c 2_L 2_R$, so that a discrete $Z_2^c$ symmetry emerges at the 
intermediate scale. Thus, as shown in Sec. \ref{sec-DW}, $Z_2$-strings 
form during the first phase transition. (They are unrelated 
to any of the monopole just discussed above.) During the second stage of 
symmetry breaking, this $Z_2^c$ breaks, leading to the formation 
of domain walls which connect the strings previously formed. 
These walls bounded by strings do not affect the standard cosmology
in any essential way. But if there is a period of inflation before
the phase transition leading to the walls formation takes place (see 
section \ref{sec-topodw}), the walls would dominate the energy density 
of the universe, leading to a cosmological catastrophe.

\subsubsection{Cosmic strings}

Now we consider the models where $4_c 2_L 2_R$ breaks down to the standard 
model gauge group with added $Z_2$ parity, as in model ${\bf 8}$. Then a 
new $Z_2$ symmetry emerges at $M_{{\rm G}}$, which is unrelated to the previous ones. 
Since $\pi_1({4_c 2_L 2_R  \over 3_c 2_L 1_Y Z_2}) = Z_2$, $Z_2$-strings form 
when $4_c 2_L 2_R$ breaks. They have a mass per unit length $\mu \sim 
10^{30} - 10^{32} \: {\rm GeV}^2$. Since the $Z_2$ symmetry is then
kept unbroken  
down to low energy, we break the standard model gauge group with a
Higgs 10-plets. The strings are topologically stable down to low
energy. 

Density perturbations in the early universe and temperature fluctuations 
in the CBR generated by these strings could be computed.

\subsubsection{Solving the monopole problem}

In order to solve the monopole problem, we use
a hybrid inflationary scenario, as discussed in section
\ref{sec-inflation}. The rank of both $4_c 2_L 2_R$ and $4_c 2_L 2_R Z_2$ 
is four. Therefore the inflaton field will couple to a pair of 
Higgs fields which will break $4_c 2_L 2_R$. The primordial monopoles formed 
when {\rm SO}(10) breaks are diluted by the inflation. But then
lighter monopoles  
form at the end of inflation when $4_c 2_L 2_R$ breaks, which are 
topologically stable. In the case 
of unbroken $Z_2$ parity, cosmic strings also form. Monopole creation at this
later stage makes the model inconsistent with observations.

If {\rm SO}(10) is broken with a 54-dimensional Higgs
representation, domain  
walls will form through the Kibble mechanism at the end of inflation, 
which will dominate the universe, as shown in Sec. \ref{sec-DW}, hence 
leading to a cosmological catastrophe. 

We conclude that the model is cosmologically inconsistent with observations. 
It is inconsistent whether or not the discrete 
$Z_2^c$ symmetry is unbroken at the intermediate scale.

\subsection{Breaking via $3_c 2_L 2_R 1_{B-L}$}

We can break via $3_c 2_L 2_R 1_{B-L}$ and then down to the standard model 
with or without the discrete $Z_2$ symmetry preserved at low energy
\begin{eqnarray}
& {\rm SO}(10) & \stackrel{M_{{\rm GUT}}}{\rightarrow}{\rm SU}(3)_c \times
{\rm SU}(2)_L  
\times  {\rm SU}(2)_R \times {\rm U}(1)_{B\! -\! L} \label{eq:above}
\\
&&\stackrel{M_{{\rm G}}}{\rightarrow} {\rm SU}(3)_c \times {\rm SU}(2)_L
\times {\rm U}(1)_Y  
(\times Z_2) \\
&& \stackrel{M_{{\rm Z}}}{\rightarrow} {\rm SU}(3)_c \times {\rm U}(1)_Q (\times Z_2) 
\end{eqnarray}
The $Z_2$ symmetry, which can be kept unbroken down to low energy if only 
safe representations are used to implement the symmetry breaking, plays 
the role of matter parity. It preserves large values for the proton
lifetime. Hence only models with unbroken $Z_2$ parity at low energy
are consistent  
with the actual values of proton decay. If {\rm SO}(10) is broken with
a single  
210-Higgs multiplet, with the appropriate range of the parameters in the 
Higgs potential \cite{210}, then there appears a discrete $Z_2^c$ symmetry at 
the intermediate scale which is generated by the charge conjugation operator, 
and the symmetry breaking really is
\begin{eqnarray}
& {\rm SO}(10) & \stackrel{M_{{\rm GUT}}}{\rightarrow}{\rm SU}(3)_c \times
{\rm SU}(2)_L \times  
 {\rm SU}(2)_R \times {\rm U}(1)_{B\! -\! L} \times Z_2^c \label{eq:LR2}\\
&&\stackrel{M_{{\rm G}}}{\rightarrow} {\rm SU}(3)_c \times {\rm SU}(2)_L
\times {\rm U}(1)_Y  
(\times Z_2)\label{eq:21}\\
&& \stackrel{M_{{\rm Z}}}{\rightarrow} {\rm SU}(3)_c \times {\rm U}(1)_Q
(\times Z_2)\label{eq:22} \: . 
\end{eqnarray}
The $Z_2^c$ is unrelated to the $Z_2$ symmetry which can be added to
the standard model gauge group in Eqs. (\ref{eq:21}) and (\ref{eq:22}).

If one uses a combination of a 45 dimensional Higgs representation
with a 54 dimensional one to break {\rm SO}(10),  
then the symmetry breaking is that of equation (\ref{eq:above}), and no 
discrete symmetry appears as in (\ref{eq:LR2}) \cite{45}. The rest of 
the symmetry 
breaking is implemented with a pair of ${\bf 16} + {\bf
\overline{16}}$ Higgs multiplets  
or with a pair of ${\bf 126} + {\bf \overline{126}}$ Higgs multiplets
if matter parity  
is preserved at low energy. $3_c 2_L 1_Y$ is broken with a 10-Higgs 
multiplet.

\subsubsection{Monopoles}

The first homotopy groups $\pi_1(3_c 2_L 2_R 1_{B-L}) = Z$, $\pi_1(3_c
2_L 1_{Y}) = Z$ and $\pi_1(3_c 1_{Q}) = Z$, showing that topologically
stable monopoles are produced during 
the first phase transition from {\rm SO}(10) down to $3_c 2_L 
2_R 1_{B-L}$. They have a mass $M_m \geq 10^{17}$ 
GeV. These monopoles are in conflict with cosmological observations.

\subsubsection{Domain walls}

If {\rm SO}(10) is broken with a single 210-dimensional Higgs
representation, 
then the symmetry breaking is that of Eq. (\ref{eq:LR}). Hence, as 
in the breaking pattern (\ref{eq:PS2}), the appearance of the 
discrete $Z_2^c$ symmetry leads to the 
formation of non stable cosmic strings during the first symmetry breaking
 and to the formation of domain walls in the breaking of  $3_c 2_L 
2_R 1_{B-L}$ down to the standard model gauge group. The cosmological
relevance of these walls bounded by strings depends upon the presence of an 
inflationary epoch before the phase transition leading to the walls formation
has taken place; see Sec. \ref{sec-DW}.

\subsubsection{Embedded Defects}

In these models with intermediate $3_c 2_L 2_R 1_{B-L}$ symmetry, the 
breaking schemes are equivalent to
\begin{equation}
{\rm SO}(10) \stackrel{M_{{\rm GUT}}}{\rightarrow}{\rm G}\times  {\rm SU}(2)_R \times
{\rm U}(1)_{B\! -\! L} \stackrel{M_{{\rm G}}}{\rightarrow}{\rm G}\times {\rm
U}(1)_Y (\times Z_2)  
\stackrel{M_{{\rm Z}}}{\rightarrow} 3_c 1_Q (Z_2)  
\end{equation}
where ${\rm G} = {\rm SU}(3)_c \times {\rm SU}(2)_L$. In direct analogy with
electroweak 
strings \cite{Tanmay}, it is easy to see that embedded defects form 
during the second stage
of symmetry breaking. They have a mass per unit length $\mu \sim
10^{30}-10^{32} \: {\rm GeV}^2$. The stability conditions for these 
strings can be computed. If these strings are dynamically stable, 
they may generate density perturbations 
in the early universe and temperature anisotropy in the CBR.

\subsubsection{Cosmic Strings}

Consider the model where $3_c 2_L 2_R 1_{B-L}$ breaks down to 
$3_c 2_L 1_Y Z_2$. The first homotopy group $\pi_1({3_c 
2_L 2_R 1_{B-L} \over 3_c 
2_L 1_Y Z_2}) = Z_2$ is non trivial which shows the 
formation of topological $Z_2$ strings. Since the $Z_2$ parity symmetry is 
kept unbroken down to low energy, the strings are 
topologically stable. They have a mass per unit length $\mu 
\sim 10^{30}-10^{32} \: {\rm GeV}^2$. These strings will generate
density perturbations in the early universe and temperature anisotropy
in the  
CBR.

\subsubsection{Solving the monopole problem}

One can use an inflationary scenario as described in Sec. \ref{sec-inflation}
to dilute the monopoles formed at $M_{{\rm GUT}}$. Since the rank of $3_c 
2_L 2_R 1_{B-L} (Z_2^c)$ is four, the inflaton field will couple to a 
pair of ${\bf 16} + {\bf \overline{16}}$ or
${\bf 126} + {\bf \overline{126}}$ which will break $3_c 2_L 2_R
1_{B-L}$, (see Sec. \ref{sec-inflation}). Cosmic strings (if unbroken
$Z_2$ symmetry at low energy) and/or domain walls (if unbroken $Z_2^c$
symmetry at the intermediate  
scale) will form at the end of inflation. As shown in Sec. \ref{sec-DW}
the presence of this inflationary epoch between the two phase
transitions at $M_{{\rm GUT}}$ and $M_{{\rm G}}$, respectively, would make the walls
dominate the energy density of the universe, (see
Sec. \ref{sec-DW}). Now the unbroken $Z_2$ symmetry is necessary to
preserve large values for the proton lifetime; hence, the only
symmetry breaking pattern consistent with cosmology with intermediate
$3_c 2_L 2_R 1_{B-L}$ symmetry is
\begin{eqnarray}
& {\rm SO}(10) & \stackrel{<45> + < 54>}{\rightarrow}{\rm
SU}(3)_c \times {\rm SU}(2)_L \times  
 {\rm SU}(2)_R \times {\rm U}(1)_{B\! -\! L} \label{eq:LR}\\
&&\stackrel{<126> + < \overline{126}>}{\rightarrow} {\rm
SU}(3)_c \times {\rm SU}(2)_L \times {\rm U}(1)_Y  
\times Z_2\\
&& \stackrel{<10>}{\rightarrow} {\rm SU}(3)_c \times {\rm U}(1)_Q \times Z_2
\end{eqnarray}
where {\rm SO}(10) is broken with a combination of a 45 dimensional
Higgs representation and 54 dimensional one,  $3_c  
2_L 2_R 1_{B-L}$ is broken with pair of ${\bf 126} + {\bf
\overline{126}}$ dimensional Higgs representation and  $3_c 2_L 2_Y
Z_2$ is broken with a 10 Higgs multiplet.

\section{Breaking via $3_c 2_L 1_R 1_{B-L}$} 
\label{sec-three}

We shall consider first the symmetry breaking with intermediate  $3_c
2_L 1_R 1_{B-L}$ symmetry without conserved matter parity at low
energy : 
\begin{eqnarray}
& {\rm SO}(10)& \stackrel{M_{{\rm GUT}}}{\rightarrow} {\rm SU}(3)_c \times
{\rm SU}(2)_L \times {\rm U}(1)_R  
\times {\rm U}(1)_{B\! -\! L} \\
&&\stackrel{M_{{\rm G}}}{\rightarrow} {\rm SU}(3)_c \times {\rm SU}(2)_L
\times {\rm U}(1)_Y\\ 
&& \stackrel{M_{{\rm Z}}}{\rightarrow} {\rm SU}(3)_c \times {\rm U}(1)_Q 
\end{eqnarray}
The first homotopy group $\pi_1(3_c 2_L 1_R 1_{B\! -\! L}) = Z \oplus Z$ and 
therefore
topological monopoles form during the first phase transition 
from supersymmetric {\rm SO}(10) down to $3_c 2_L 1_R 1_{B-L}$. These monopoles
carry $R$ and $B\! -\! L$, and have a mass $M_m \geq
(10^{16}-10^{17})$ GeV. Now $\pi_1(3_c 2_L 1_Y)$  
and  $\pi_1(3_c \times 1_Q)$ are both non trivial
and hence, from an homotopy point of view, the monopoles are
topologically stable. But as we are going to show below, some of these
monopoles are indeed topologically stable, but some others will decay.
During the second phase transition, the formation of strings is
governed by the first homotopy group $\pi_1({3_c 2_L
1_R 1_{B\! -\! L} \over 3_c 2_L 1_Y}) = Z $, showing the formation of
cosmic strings during the 
second phase transition. These are associated with the breaking of
${\rm U}(1)_R \times {\rm U}(1)_{B\! -\! L}$ down to ${\rm U}(1)_Y$
where the unbroken 
${\rm U}(1)_R \times {\rm U}(1)_{B\! -\! L}$ symmetry in the first stage of
symmetry breaking is responsible for the formation of monopoles. Now
the weak hypercharge ${Y \over 2}$ is a linear combination of $B\! -\! L$ 
and
$R$, ${Y \over 2} = ({{B\! -\! L} \over 2}  + R)$. Therefore primordial
monopoles with topological charge ${{B\! -\! L} \over 2} -  R  \neq 0$
get connected by  
the strings at the second stage of symmetry breaking. Some infinite
and closed strings can also form. These cosmic strings are
topologically unstable. They can break producing
monopole-antimonopole pairs at the free ends. The
monopole-antimonopole pairs connected by strings annihilate in less
than a Hubble time and could produce the observed baryon asymmetry of
the universe. Other monopoles formed during the first phase
transition do not
get connected by strings and remain stable down to low energy.

The monopole problem can be solved with an inflationary scenario 
as described in Sec. \ref{sec-inflation}. Since the rank of
$3_c 2_L 1_R 1_{B-L}$ is five, the inflaton field will couple to the
Higgs representation mediating the second phase transition associated
with the breaking of 
$3_c 2_L 1_R 1_{B-L}$. The monopoles
can be pushed beyond the present horizon, and the monopole problem 
solved. Furthermore, since all the monopoles are inflated away, the
string decay probability is negligible and the evolution of strings is
identical to that of topologically stable strings. We therefore have a
very interesting breaking scheme, where monopoles are created during a
first transition, inflated away, and cosmic strings form at the end of
inflation.  

This model where $3_c 2_L 1_R 1_{B\! -\! L}$ breaks down to the
standard model without matter parity is in conflict with 
the actual data for proton lifetime. The solution to this problem is
therefore that the intermediate subgroup break down to $3_c
2_L 1_Y Z_2$ as in model (\ref{eq:SSB13}). In this case, topologically
stable $Z_2$-strings will form during the second phase
transition. They have a mass per unit length $\mu \sim (10^{30} - 
10^{32}) \; {\rm GeV}^2$.      
This interesting model with inflation and cosmic strings is studied in
detail in the next chapter.

\section{Breaking directly down to the standard model}
\label{sec-direct}

Supersymmetric {\rm SO}(10) can break directly down to the standard model as 
in model (\ref{eq:SSB7})
\begin{equation}
{\rm SO}(10) \stackrel{M_{{\rm GUT}}}{\rightarrow}  {\rm SU}(3)_c \times
{\rm SU}(2)_L \times {\rm U}(1)_Y \stackrel{M_{{\rm Z}}}{\rightarrow} {\rm
SU}(3)_c \times {\rm U}(1)_Q \label{eq:d2} 
\end{equation}
or as in model (\ref{eq:SSB14})
\begin{equation}
{\rm SO}(10) \stackrel{M_{{\rm GUT}}}{\rightarrow} {\rm SU}(3)_c \times {\rm
SU}(2)_L \times {\rm U}(1)_Y \times Z_2  
 \stackrel{M_{{\rm Z}}}{\rightarrow} {\rm SU}(3)_c \times {\rm U}(1)_Q \times
Z_2 \label{eq:d1} 
\end{equation}
with (\ref{eq:d1}) or without (\ref{eq:d2}) the $Z_2$ symmetry, subgroup of 
the $Z_4$ centre of {\rm SO}(10), unbroken down to low energy. The
latter plays  
the role of matter parity, giving large values for the proton lifetime 
and stabilising the LSP. The symmetry breaking occurs at $M_{{\rm GUT}} \simeq 
2 \times 10^{16}$ GeV. The scenario without the unbroken $Z_2$ symmetry 
(\ref{eq:d2}) is not, with the present data for proton decay, relevant 
phenomenologically.

In model (\ref{eq:d1}), the $Z_2$ symmetry remains unbroken
down to low energy, preserving large values for the proton
lifetime. Furthermore, the first homotopy group $\pi_1( {{{\rm SO}(10)} \over 
{3_c 2_L 1_Y Z_2}}) = \pi_0 ( 3_c 2_L 1_Y Z_2) \\ = Z_2$ and therefore 
cosmic strings form when {\rm SO}(10) breaks. They are
associated with the unbroken $Z_2$ symmetry; since the latter
remains unbroken down to low energy, the strings are topologically
stable down to low energy. They have a mass per unit length $\mu \sim
10^{32} \; {\rm GeV}^2$. The latter could account for the density perturbations
produced in the early universe which lead to galaxy formation and to
temperature fluctuations in the CBR.

Again, due to the unbroken ${\rm U}(1)_Y$   
symmetry, monopoles form at the grand unified phase transition. They 
carry $Y$ topological charge and are topologically stable down to low
energy. Their topological charge may change from $Y$ to $Q$.

Since monopoles form in both models, the potential conflict with the
standard big-bang cosmology is again not avoided. Nevertheless, in
model (11), if the Higgs field leading to monopole production takes
its vacuum expectation value (VEV) before inflation ends and the latter ends
before the Higgs field leading to cosmic string formation acquires its VEV
then we are left with a very attractive scenario.  

Unfortunately, it does not seem possible to achieve this. If one attempts
to inflate away the monopoles with a superpotential of the form given 
in Sec. \ref{sec-inflation}, an intermediate  scale is introduced. 
Thus, one is either left with the monopole problem in cosmology or 
loses the simplicity of this breaking scheme.

\section{Conclusions}
\label{sec-concl3}

The aim of this chapter is to constrain supersymmetric {\rm SO}(10) models 
which lead to the formation of topological defects through
cosmological considerations. The main reason for  
considering supersymmetric versions of the grand unified gauge group 
{\rm SO}(10), rather than nonsupersymmetric ones, is to predict 
the measured values of $\sin^2 \theta_w$ and the gauge coupling 
constants merging in a single point at $\simeq 2 \times 10^{16}$
GeV. Spontaneous symmetry breaking (SSB) patterns from supersymmetric
{\rm SO}(10) down to the  
standard model differ from
nonsupersymmetric ones first in the scale of $B-L$
symmetry breaking and second in the ways of breaking from {\rm SO}(10) down to
the standard model. For nonsupersymmetric models the scale of $B-L$
breaking has to be anywhere between $10^{10}$ and $10^{13.5}$ GeV,
whereas it is $\sim 10^{15}-10^{16}$ GeV in supersymmetric models.
Furthermore, in the supersymmetric case, we can break directly down to
the standard model without any intermediate breaking scale, and not
more than one intermediate scale is expected. We have given a
systematic analysis of topological defect formation  
and their cosmological implications in each model. We found that the
rules for topological defect formation are not affected by the
presence of supersymmetry, and since {\rm SO}(10) is simply connected and the 
standard model gauge group involves an unbroken {\rm U}(1) symmetry, all 
SSB patterns from supersymmetric {\rm SO}(10) down to the standard model 
involve automatically the formation of topologically stable monopoles. 
In tables 3.1, 3.2, 3.3 and 3.4 we give a summary of all the defects formed in 
each model. In the models where $Z_2$-walls arise at the second phase
transition, we have in fact hybrid defects. The walls are bounded by
the $Z_2$-strings previously formed and are unstable. In order to
solve  
the monopole problem, we propose an hybrid inflationary scenario
\cite{ShafiDvaSch,Ed1}  which arises in supersymmetric {\rm SO}(10)
models  
without imposing any external symmetry and without imposing any external 
field (see Chap. \ref{chap-chap5}). The inflationary scenario can cure
the monopole problem, but  
then stabilises the $Z_2$ walls previously discussed. Hence these 
cases lead to another cosmological problem. Imposing also that the models
satisfy the actual data on the proton lifetime,
we found that there are only two SSB 
patterns consistent with cosmological considerations. Breaking directly 
to the standard model at first sight seems attractive. Unfortunately, 
one is unable to inflate away the monopoles without the introduction of 
an intermediate scale. The only breaking schemes consistent with 
cosmology  correspond to the intermediate symmetry groups 
$3_C 2_L 2_R 1_{B-L}$, where {\rm SO}(10) is broken 
with a combination of a 45 dimensional Higgs representation and a 
 54 dimensional one, and $3_C 2_L 1_R 1_{B-L}$. These intermediate 
symmetry groups 
must later break down to the standard model gauge group with unbroken
matter parity;  
the symmetry breaking must be implemented with only Higgs fields in `safe'
representations \cite{Martin}, hence the rank of the group must be lowered 
with a pair of Higgs fields in the 
$({\bf 126} + {\bf \overline{126}})$ dimensional representation, and
the standard model  
gauge group broken with two {\bf 10}-dimensional ones. The model with 
intermediate $3_C 2_L 1_R 1_{B-L}$, inflation, and cosmic strings, is 
studied in detail in the next chapter. In supergravity {\rm SO}(10) models, 
the breaking of {\rm SO}(10) via flipped {\rm SU}(5) is also possible.

%{\raggedleft

\begin{tabular}{||p{3cm}@{}||p{3.5cm}|p{3.5cm}|p{3.5cm}||}
\hline
\hline
G & ${\rm SO}(10) \rightarrow G$ & $G \rightarrow 3_c 2_L 1_Y$ &  Cosmological problems \\ \hline \hline
${\rm SU}(5) \times {\rm U}(1)_X$ & monopoles-1 &    monopoles-2  
+ strings &
 monopoles-2  
+ proton lifetime   ($Z_2$ broken)
\\ \hline
{\rm SU}(5) & no defects & monopoles & 
 monopoles 
+ proton lifetime ($Z_2$ broken)
\\ \hline
${\rm SU}(5) \times \widetilde{{\rm U}(1)}$ & monopoles & embedded strings & 
proton lifetime   ($Z_2$ broken)
 \\ \hline
$4_c 2_L 2_R$ &  monopoles-1 &  monopoles-2 & 
monopoles-2 + proton lifetime   ($Z_2$ broken)
\\ \hline
$4_c 2_L 2_R Z^c_2$ & 
  monopoles-1 + $Z_2$-strings 
&  monopoles-2 +
$Z_2$-walls & $Z_2$-walls and monopoles-2 + proton lifetime ($Z_2$ broken)
\\ \hline
$3_c 2_L 2_R 1_{B-L}$ & monopoles & embedded strings & 
proton lifetime  ($Z_2$ broken)
\\ \hline
$3_c 2_L 2_R 1_{B-L}  Z^c_2 $ & monopoles + $Z_2$-strings 
& embedded strings + $Z_2$-walls 
 & $Z_2$-walls + proton lifetime  ($Z_2$ broken)
\\ \hline
$3_c 2_L 1_R 1_{B-L}$ & monopoles & strings & proton lifetime   ($Z_2$ broken)
\\ \hline \hline
\end{tabular}
\vspace{.5cm}

%}

\underline{Table 3.1} : Formation of topological defects 
in the possible symmetry breaking patterns from supersymmetric {\rm SO}(10) 
down to the standard model with broken matter parity. These models are 
inconsistent with proton lifetime measurements. The table also 
shows the relevant cosmological problems associated with each symmetry 
breaking pattern, when occuring within a hybrid inflationary scenario. 
From a topological defect point of view, models with intermediate 
${\rm SU}(5) \times \widetilde{{\rm U}(1)}$, $3_c 2_L 2_R 1_{B-L}$ and 
$3_c 2_L 1_R 1_{B-L}$ symmetry groups are compatible with 
observations. The model with an intermediate ${\rm SU}(5) \times 
\widetilde{{\rm U}(1)}$ symmetry is only possible in supergravity 
{\rm SO}(10) models. 

%{\raggedleft

\begin{tabular}{||p{3cm}@{}||p{3.5cm}|p{3.5cm}|p{3.5cm}||}
\hline \hline
G & ${\rm SO}(10) \rightarrow G$ & $G \rightarrow 3_c 2_L 1_Y Z_2$ & Cosmological problems \\ \hline \hline
${\rm SU}(5) \times {\rm U}(1)_X$ & monopoles-1 &  monopoles + $Z_2$-strings &  monopoles-2 
\\ \hline
${\rm SU}(5) \times Z_2$ & $Z_2$-strings & monopoles-2 &  monopoles-2
\\ \hline
${\rm SU}(5) \times \widetilde{{\rm U}(1)}$ & monopoles & $Z_2$-strings & 
no problem, monopoles inflated away
\\  \hline
$4_c 2_L 2_R$ &  monopoles-1 & monopoles-2 + $Z_2$-strings &   monopoles-2\\ \hline
$4_c 2_L 2_R Z^c_2$ & 
  monopoles-1 + $Z_2$-strings 
 &  
 monopoles-2 + $Z_2$-strings
 + $Z_2$-walls 
 &  monopoles-2 +  $Z_2$-walls  \\ \hline
$3_c 2_L 2_R 1_{B-L}$ & monopoles & embedded strings 
+ $Z_2$-strings 
& 
no problem,  monopoles inflated away
\\ \hline
$3_c 2_L 2_R 1_{B-L} Z^c_2$ & 
 monopoles + $Z_2$-strings  &  embedded strings  + $Z_2$-strings  
+ $Z_2$-walls 
&    $Z_2$-walls \\ \hline
$3_c 2_L 1_R 1_{B-L} $ & monopoles & $Z_2$-strings & 
no problem, 
 monopoles inflated away
 \\ \hline \hline
\end{tabular}
%}
\vspace{0.5cm}

\underline{Table 3.2} : Formation of topological defects
in the possible symmetry breaking patterns from supersymmetric {\rm SO}(10) 
down to the
standard model with unbroken matter parity. These models are
consistent with proton life time measurements and can
provide a superheavy Majorana mass to the right-handed neutrinos. 
The table also shows the relevant cosmological problems associated 
with each symmetry breaking pattern, when occurring within a
hybrid inflationary scenario. The
models with intermediate ${\rm SU}(5) \times \widetilde{{\rm U}(1)}$, 
$3_c 2_L 2_R 1_{B-L}$ and $3_c 2_L 1_R
1_{B-L}$ symmetry groups are consistent with observations. The model with
intermediate ${\rm SU}(5) \times \widetilde{{\rm U}(1)}$ symmetry is only possible
in supergravity {\rm SO}(10) models.

\vspace{1cm}

\begin{tabular}{||p{5cm}|p{5cm}||}
\hline \hline
${\rm SO}(10) \rightarrow 3_c 2_L 1_Y$ & Cosmological problems\\ \hline
 monopoles-2 & 
monopoles-2 
+ proton lifetime ($Z_2$ broken) 
 \\ \hline \hline
\end{tabular}

\vspace{.5cm}

\underline{Table 3.3} : Formation of topological defects
in models where supersymmetric {\rm SO}(10) breaks directly down to the 
MSSM with broken matter parity. The table also shows the relevant 
cosmological problems associated with the symmetry breaking pattern, 
when occurring within a hybrid inflationary scenario. These models 
are inconsistent with observations.  

\vspace{1cm}

\begin{tabular}{||p{5cm}|p{5cm}||}
\hline \hline
${\rm SO}(10) \rightarrow 3_c 2_L 1_Y Z_2$& Cosmological problems \\ \hline
 monopoles-2 + $Z_2$-strings & monopoles-2 \\ \hline \hline
\end{tabular}

\vspace{.5cm}

\underline{Table 3.4} : Formation of topological defects
in models where supersymmetric {\rm SO}(10) breaks directly down to the MSSM 
with unbroken matter parity. The table also shows the relevant 
cosmological problems associated with the symmetry breaking pattern, 
when occurring within a hybrid inflationary scenario. These models 
are inconsistent with observations.

\chapter{ Supersymmetric SO(10) model with inflation and cosmic strings}
\label{chap-chap4}

\section{Introduction}

In the previous chapter, we have constrained supersymmetric {\rm SO}(10)
models using both cosmological and particle physics arguments. We have
in particular studied the 
formation of topological defects in all possible symmetry breaking
patterns from supersymmetric {\rm SO}(10) down to the standard model,
considering no more than one intermediate symmetry breaking
scale.   
Recall that domain walls and monopoles are in conflict with the standard cosmology whereas
cosmic strings may have interesting cosmological consequences. Since {\rm SO}(10) is simply connected and the
standard model gauge group involves an unbroken {\rm U}(1) symmetry,
which remains 
unbroken down to low energy, all symmetry breaking patterns from
supersymmetric {\rm SO}(10) down to the standard model automatically lead to
the formation of topologically stable monopoles. All supersymmetric {\rm SO}(10) models are therefore
cosmologically irrelevant without invoking some mechanism for the 
removal of the monopoles, such as an inflationary scenario. 
The conclusion in Chap. \ref{chap-chap3} is that there are only two
possibilities  
for breaking {\rm SO}(10) down to the standard model which are consistent 
with observations. {\rm SO}(10) can break  
via ${\rm SU}(3)_c \times {\rm SU}(2)_L \times {\rm SU}(2)_R \times
{\rm U}(1)_{B-L}$, here {\rm SO}(10)  
must be broken with a combination of a 45-dimensional Higgs 
representation and a 54-dimensional one, and via ${\rm SU}(3)_c 
\times {\rm SU}(2)_L \times {\rm U}(1)_R \times {\rm U}(1)_{B-L}$. In
these models,   
the intermediate symmetry group must be broken down to the standard model 
gauge group with unbroken matter parity, ${\rm SU}(3)_c 
\times {\rm SU}(2)_L \times {\rm U}(1)_Y \times Z_2$. In supergravity
{\rm SO}(10) models,  
the breaking of {\rm SO}(10) via flipped {\rm SU}(5) is also possible. 

In this chapter, 
we study a supersymmetric {\rm SO}(10) model
involving an intermediate ${\rm SU}(3)_c \times {\rm SU}(2)_L \times
{\rm U}(1)_R \times 
{\rm U}(1)_{B-L}$ symmetry. The resultant cosmological model is
compatible with observations.

In Sec. \ref{sec-inflation2}, we describe a hybrid inflationary scenario,
already introduced in Sec. \ref{sec-inflation}, 
 and we argue that this type of 
inflationary scenario occurs naturally in global supersymmetric {\rm
SO}(10) models. 
Neither any external field nor any external symmetry has to be imposed. 
We give the general form of the potential which, in supersymmetric SO(10)
models, leads to the required spontaneous symmetry breaking pattern and 
gives rise to such a period of inflation.

In the next sections, we construct a specific supersymmetric {\rm
SO}(10) model, 
as mentioned above. The latter aims to be consistent with
observations. In Sec. \ref{sec-modelcosmo}  
we study the symmetry breaking pattern. We conclude on the proton lifetime 
and on a hot dark-matter candidate 
provided by the model. Using homotopy theory, we find topological defects 
which form according to the Kibble mechanism \cite{Kibble}.

In Sec. \ref{sec-building}, we explain how to implement the symmetry 
breaking pattern which solves the doublet-triplet splitting and includes 
the inflationary scenario described in Sec. \ref{sec-inflation2}. We 
write down the superpotential and find its global minimum with 
corresponding Higgs VEVs.

In Sec. \ref{sec-infl}, we evaluate the dynamics of the symmetry 
breaking and inflationary scenario, studying the scalar potential. It
is shown that the monopole  
problem may be solved and that cosmic strings form at the end of inflation. 

In Sec. \ref{sec-strings}, we give general properties of the strings
formed at the end of inflation. In particular, we study the possibility
that the strings may be superconducting.

In Sec. \ref{sec-observations}, we estimate the observational 
consequences. The temperature fluctuations 
in the CBR due to the mixed inflation-cosmic 
strings scenario are evaluated. 
Using the temperature fluctuations measured by COBE we find values 
for the scalar coupling constant, the scale at which the strings 
formed and the strings mass per unit length. We specify the dark-matter 
present in the model and give a 
qualitative discussion of the large-scale structure formation scenario
in this model.   

We finally conclude in Sec. \ref{sec-conclusion}. 

We shall use the notations given in
Eqs. (\ref{eq:notation1})-(\ref{eq:notation2}).

\section{Inflation in supersymmetric {\rm SO}(10) models}
\label{sec-inflation2}

In this section, we argue that false vacuum hybrid inflation, with a
superpotential in the inflaton  
sector similar to that studied in Refs. \cite{ShafiDvaSch,Ed1}, is a natural
mechanism for inflation in global supersymmetric {\rm SO}(10)
models. Neither any external  
field nor any external symmetry has to be imposed, it can just be a 
consequence of the theory.

The first thing to note in {\rm SO}(10) models, is that the rank 
of {\rm SO}(10) is greater than one unit from the rank of the standard model 
gauge group. The rank of {\rm SO}(10) is five, whereas the rank of the
standard  
model gauge group ${\rm SU}(3)_c \times {\rm SU}(2)_L 
\times {\rm U}(1)_Y (\times Z_2)$ is four. In other words, {\rm SO}(10)
has an additional {\rm U}(1) symmetry, named ${\rm U}(1)_{B-L}$,
compared to the standard model gauge group. Therefore the rank of the
group must be lowered by one unit at some stage of the symmetry
breaking pattern, i.e., ${\rm U}(1)_{B-L}$ must be broken.  This can
be  
done using a pair of ${\bf 16 + \overline{16}}$ Higgs 
representations or by a pair of ${\bf 126 + \overline{126}}$ 
representations. If a ${\bf 16 + \overline{16}}$ pair 
of Higgs fields are used, then the $Z_2$ symmetry, subgroup of both the $Z_4$ 
centre of {\rm SO}(10) and of ${\rm U}(1)_{B-L}$ is broken.  
On the other hand, if a ${\bf 126 + \overline{126}}$ pair of Higgs
fields are used, 
 then the $Z_2$ symmetry can be kept unbroken down to low energy 
if only safe representations \cite{Martin} are used to implement 
the full symmetry breaking pattern, such as the 10, the 45, the 54 or the 
210-dimensional representations. If a ${\bf 126 + \overline{126}}$ pair of 
Higgs fields are used, the right-handed neutrino can get
a superheavy Majorana mass, and the solar neutrino problem can be solved 
via the MSW mechanism \cite{MSW}.

In order to force the VEVs of the  ${\bf 16 + \overline{16}}$ or ${\bf
126 + \overline{126}}$ 
pair of Higgs fields, needed to lower the rank of the group, to be the
order of the {\rm GUT} scale, we can use a scalar field ${\cal S}$
singlet under {\rm SO}(10). The superpotential can be written as
follows, 
\begin{equation}
W = \alpha {\cal S} \overline{\Phi} \Phi - \mu^2 {\cal S} 
\label{eq:inflation} 
\end{equation}
where $\Phi + \overline{\Phi}$ stand for a ${\bf 16 + \overline{16}}$  or 
a ${\bf 126 + \overline{126}}$ pair of Higgs fields, and the field ${\cal S}$ 
is a scalar field singlet under {\rm SO}(10). The constants $\alpha$ and $\mu$ 
are assumed to be both positive and must satisfy ${\mu \over \sqrt{\alpha}} 
\sim (10^{15} - 10^{16})$ GeV. 

It is easy to see that the superpotential given in Eq. (\ref{eq:inflation}), 
used to break the rank of the group by one unit, is the same 
superpotential used by Dvali {\em et al.}\cite{ShafiDvaSch,Dvali1} to
implement a false vacuum hybrid inflationary scenario, identifying the
scalar field ${\cal S}$ with the inflaton field. Hence, as shown
below,  in supersymmetric {\rm SO}(10) models, the superpotential
used to break ${\rm U}(1)_{B-L}$ can also lead to a  
period of inflation. Inflation is then just a consequence of the theory. 
In order to understand  
the symmetry breaking and the inflationary dynamics, we can study the 
scalar potential. From Eq. (\ref{eq:scalarpot1}), the latter is given by (keeping the same notation 
for the bosonic component of the superfields as for the superfields):
\begin{equation}
V  = |F_S|^2 + |F_{\Phi}|^2 + |F_{\overline{\Phi}}|^2  + D-{\rm terms}
\end{equation}
where the $F$ terms are such that $F_{\Psi_i} = |{ \partial W \over
\partial \Psi_i} |$, for $\Psi_i = {\cal S}, \Phi$ and
$\overline{\Phi}$. The $D$-terms vanish if $|\Phi | =  |\overline{\Phi}|$. 
Therefore the Higgs potential
 \begin{equation}
V = \alpha^2 |{\cal{S}} \overline{\Phi} |^2 + \alpha^2 |{\cal{S}} \Phi |^2 +
|\alpha \overline{\Phi} \Phi - \mu^2|^2 \: .
\end{equation}
It is minimised for $\arg( \Phi ) + \arg(\overline{\Phi}) = 0$,
$(\alpha > 0)$, and it is independent of ${\rm\arg}({\cal{S}}) + \arg({\Phi})$
and ${\rm\arg}({{\cal{S}}} ) + \arg({\overline{\Phi}} )$. Thus 
we can rewrite the scalar potential with the new fields which minimise 
the potential, keeping the same notation for the old and new fields,
\begin{equation}
V = 4 \alpha^2 |{\cal{S}}|^2 |\Phi|^2 + (\alpha |\Phi|^2 - \mu^2)^2
\: \label{eq:V1} . 
\end{equation}
The potential has a unique supersymmetric minimum corresponding to  
$< \! |\Phi | \!> = < \! | \overline{\Phi} | \! > = {\mu \over \sqrt{\alpha }}$ and  
${\cal S} = 0$. The potential 
has also a local minimum corresponding   ${\cal S} > {\mu \over
\sqrt{\alpha}}$ and $< \! |\Phi |\! > = \\ 
< \! | \overline{\Phi} | \! > = 0$. We identify the scalar field ${\cal S}$
with the inflaton field and we assume chaotic initial conditions. All
the fields are  
thus supposed to take initial values the order of the Planck scale, and
 hence the initial value of the inflaton field $ {\cal S} \gg {\mu \over
\sqrt{\alpha}}$. Since the potential is flat in the ${\cal S}$
direction,  
we can minimise it at a fixed value of ${\cal S}$. The $\Phi$ and
${\overline{\Phi}}$ fields roll down their local minimum corresponding
to $<\! |\Phi |\! > = <\! | \overline{\Phi} |\! > = 0$. The 
vacuum energy density is then dominated by a non vanishing $F_{\cal S}$ term, 
$|F_{\cal S}| = \mu^2$. $F_{\cal S} \neq 0$ implies that supersymmetry 
is broken. The inflationary epoch takes place as the inflaton field 
slowly rolls down the potential. Quantum corrections to the effective potential will help the 
fields to slowly roll down their global minimum \cite{ShafiDvaSch}. At the end 
of inflation, the phase transition mediated by the $\Phi$ and
${\overline{\Phi}}$ fields takes place.

Now, in order to break {\rm SO}(10) down to the standard model gauge group, 
we need more than a  ${\bf 16 + \overline{16}}$ or a ${\bf 126 +
\overline{126}}$  
pair of Higgs fields. We need Higgs in other representations, like the
45, 54 or 210-dimensional  
representations if the $Z_2$ parity is to be kept unbroken down to low
energy, as required from proton lifetime measurements. Thus the full
superpotential needed to break {\rm SO}(10) down to  
the standard model must, apart of Eq. (\ref{eq:inflation}), contains 
terms involving the other Higgs needed to implement the symmetry 
breaking. Due to the nonrenormalisation theorem in 
supersymmetric theories, we can write down the full superpotential 
which can implement the desired symmetry breaking pattern, just adding to 
Eq. (\ref{eq:inflation}) terms mixing the other Higgs needed to 
implement the symmetry breaking pattern. There can be no mixing between 
the latter Higgs and the pair of Higgs used to break ${\rm U}(1)_{B-L}$ (see 
Sec. \ref{sec-building} for example) and the superpotential can be 
written as follows :
\begin{equation}
W_{\rm tot} = W({\cal S}, \Phi, \overline{\Phi} ) + W_1(H_1, H_2, ..)
\end{equation}
where ${\cal S}$ is a scalar field singlet under {\rm SO}(10) identified with 
the inflaton field, the $\Phi$ and $\overline{\Phi}$ fields are the 
Higgs fields used to break ${\rm U}(1)_{B-L}$ and the $H_i$ fields, $i = 
1, .. ,m$,  are the m other Higgs fields needed to implement the full 
symmetry breaking pattern  from {\rm SO}(10) down to the standard model 
gauge group. $W$ is given by Eq. (\ref{eq:inflation}) and 
$W + W_1$ has a global supersymmetric  minimum such that the 
{\rm SO}(10) symmetry group is broken down to the standard model gauge 
group. The scalar potential is then given by
\begin{equation}
V_{\rm tot} = V({\cal S},\Phi, \overline{\Phi}) + V_1(H_i) \: .
\end{equation}
  $V$ is given by Eq. (\ref{eq:V1}) and $V + V_1$ has a 
global minimum such that the {\rm SO}(10) symmetry is broken down to the 
standard model gauge group.
 The evolution of the fields 
is then as follows. The fields take random initial values, just
subject to the constraint that the energy density is at the Planck
scale. The inflaton field is distinguished from the other fields
from the fact that the {\rm GUT} potential is
flat in its direction; the potential can be minimised for fixed 
${\cal S}$.  Chaotic initial conditions imply that the initial value 
of the inflaton field is greater than $\mu \over \sqrt{\alpha}$. 
Therefore, the non-inflaton fields will roll very
quickly down to their global (or local) minimum, at approximately a
fixed value for the inflaton, $< \! |H_i|\! > \neq 0$, for $i = 1,...,n$,
$< \! |H_j|\! > =  
0$, for $j = n+1,...,m$,  and $<\! |\Phi |\! > = <\! |\overline{\Phi} |\! > = 0$; a 
first symmetry breaking,  implemented by the $n$ Higgs fields $H$ acquiring 
 VEV, takes place, {\rm SO}(10) breaks down to an intermediate
symmetry group G.  
Then inflation occurs as the inflaton rolls slowly down the potential. 
The symmetry breaking implemented with the $\Phi + \overline{\Phi}$ 
fields occurs at the end of inflation, and the the intermediate 
symmetry group G breaks down to the standard model gauge group. 

In the scenario described above, the rank of the intermediate symmetry 
group G is equal to the rank of {\rm SO}(10), which is five, and hence involves 
an unbroken ${\rm U}(1)_{B-L}$ symmetry. If the rank of the intermediate 
symmetry group were that of the standard model gauge group, that is 
if ${\rm U}(1)_{B-L}$ were broken at the first stage of the symmetry breaking, 
the inflationary scenario would unable to solve the monopole problem, 
since the later would form at the end of inflation or once inflation 
completed. Finally, in models where supersymmetric  {\rm SO}(10) is broken 
directly down to the standard model gauge group, such hybrid
inflationary scenarios cannot cure the monopole problem. 
    
We conclude that if inflation has to occur during the evolution of 
the universe 
described by a spontaneous symmetric breaking pattern from the 
supersymmetric grand unified gauge group {\rm SO}(10) down to the 
minimal supersymmetric standard model, it can thus just be a 
consequence of the theory. No external field and no external 
symmetry has to be imposed. One can use the superpotential 
given in Eq. (\ref{eq:inflation}) to lower the rank of the 
group by one unit and then identify the scalar field ${\cal S}$, 
singlet under {\rm SO}(10), with the inflaton field. A false vacuum 
hybrid inflationary scenario will be implemented. It emerges from the theory.

\section{The supersymmetric {\rm SO}(10) model and the standard cosmology}
\label{sec-modelcosmo}

We now construct a supersymmetric {\rm SO}(10) model which aims to
agree with observations. {\rm SO}(10) is broken  
down to the standard model gauge group with unbroken matter parity 
$3_c 2_L 1_Y Z_2$, via the intermediate symmetry group $3_c 2_L 1_R 1_{B-L}$.
We study the symmetry breaking pattern of the model and deduce general
impacts of the model on observations. We look for topological defects
formation.  

The model initially assumes that the symmetries between particles, forces 
and particles, are described by a supersymmetric {\rm SO}(10) theory. The 
{\rm SO}(10) symmetry is then broken down to the standard model gauge group 
via $3_c 2_L 1_R 1_{B-L}$,
\begin{equation}
{\rm SO}(10) \stackrel{M_{{\rm GUT}}}{\rightarrow} 3_c 2_L 1_R 1_{B-L}
 \stackrel{M_{{\rm G}}}{\rightarrow} 3_c 2_L 1_Y Z_2\stackrel{M_{{\rm Z}}}{\rightarrow} 
3_c 1_Q Z_2 ,
\label{eq:sym}
\end{equation}
$M_{{\rm GUT}} \sim 10^{16}$ GeV , $M_{{\rm G}} \sim M_{{\rm GUT}}$
with $M_{{\rm G}} \leq M_{{\rm GUT}}$ and $M_{{\rm Z}} \simeq 100$ GeV, and supersymmetry
is broken at $M_{{\rm s}} \sim 10^3$ GeV. Recall that the  
$Z_2$ symmetry, which appears at the second stage of the symmetry 
breaking in (\ref{eq:sym}),  is the discrete$\{1, -1\}$ symmetry, 
subgroup of both the $Z_4$ centre of ${\rm SO}(10)$ and of 
${\rm U}(1)_{B-L}$ subgroup of {\rm SO}(10). The $Z_2$ symmetry 
acts as R-parity. It preserves large values for the proton 
lifetime and stabilises the LSP; it is 
thus necessary that this $Z_2$ symmetry be kept unbroken down to 
low energies. 

In Sec. \ref{sec-three}, we discussed the formation of topological defect
in the symmetry breaking pattern given in Eq. (\ref{eq:sym}). We now summarise our results. Monopoles
form according to the Kibble mechanism during the first phase transition at $M_{{\rm GUT}}$ when SO(10) breaks down to $3_c 2_L 1_R 1_{B-L}$. Half of these monopoles are topologically stable down to low energies. 
 During the second phase transition, 
when the $3_c 2_L 1_R 1_{B-L}$ symmetry group breaks down to $3_c 2_L 
1_Y Z_2$ at $M_{\rm G}$, cosmic strings form. The strings connect half of the monopole-antimonopole pairs 
formed earlier.
Some closed strings can also form. The strings can break with 
monopole-antimonopole pair nucleation. The monopoles get attracted 
to each other and the whole system of strings disappears. Nevertheless,  
the other half of the monopoles, which do not get connected by strings, remain topologically stable, 
and are thus in conflict with the standard cosmology.

Now the rank of $3_c 2_L 1_R 1_{B-L}$ is equal to five, as the rank of
{\rm SO}(10),  
and is therefore greater than the rank of 
$3_c 2_L 1_Y Z_2$ from one unit. Thus we can couple the inflaton field
with the Higgs  
field mediating the breaking of $3_c 2_L 1_R 1_{B-L}$ down to 
$3_c 2_L 1_Y Z_2$, see Sec. \ref{sec-inflation2}, and the monopole 
problem can be cured. If the monopoles are pushed away before the
 phase transition leading to the strings formation takes place, 
then the evolution of the string network is quite different than 
previously said. It is that of topologically stable cosmic strings.

\section{Model building} 
\label{sec-building}
\subsection{Ingredients}

In this section, we explain how to implement the symmetry breaking pattern 
given in Eq. ({\ref{eq:sym}). The model solves the doublet-triplet
splitting and includes an  
inflationary scenario as described in Sec. \ref{sec-inflation2}.

In order to implement the symmetry breaking pattern given in
Eq. (\ref{eq:sym}) 
and in order to preserve the $Z_2$ symmetry unbroken down to low
energy, see Eq. (\ref{eq:sym}),  we must only use Higgs fields in
`safe' representations \cite{Martin}, such as the 
adjoint 45, the 54, the {\bf 126} or the 210-dimensional representations. 

In order to implement the first stage of the symmetry breaking, we
could use only one 
 Higgs in the 210-dimensional representation; unfortunately the model
would then not solve the doublet-triplet splitting problem. The latter
can be easily solved using the Dimopoulos-Wilczek mechanism \cite{DW},
using two Higgs, one in the adjoint 45-dimensional representation and 
one in the 54-dimensional one. The VEV of the adjoint 45, 
which we call $A_{45}$, which implements the Dimopoulos-Wilczek mechanism 
is in the $B-L$ direction, and breaks {\rm SO}(10) down to $3_c 2_L
2_R 1_{B-L}$.  
The Higgs in the 54 dimensional representation, which we call $S_{54}$, breaks 
{\rm SO}(10) down to $4_c 2_L 2_R$. Altogether the {\rm SO}(10) symmetry is 
broken down to $3_c 2_L 2_R 1_{B-L}$. 

We want to break {\rm SO}(10) directly 
down to $3_c 2_L 1_R 1_{B-L}$, we therefore need more
Higgs. We use another 54, which we call $S'_{54}$, and another 45, which
we call $A'_{45}$, in the $T_{3R}$ direction. The latter breaks {\rm
SO}(10) down to 
$4_c 2_L 1_R$.  $S'_{54}$ and $A'_{45}$ 
break together {\rm SO}(10) down to $4_c 2_L 1_R$.

The role of $S_{54}$ and $S'_{54}$ is to
force $A_{45}$ and $A'_{45}$ into $B-L$ and  $T_{3R}$ directions. {\rm SO}(10) 
breaks down to
$3_c 2_L 1_R 1_{B-L}$ with $A_{45}$, $S_{54}$, $A'_{45}$ and $S'_{54}$ 
acquiring VEVs, and as mentioned in Sec. \ref{sec-modelcosmo}, topologically 
stable monopoles form.

As discussed in Sec. \ref{sec-modelcosmo}, since the rank of $3_c 2_L 1_R 1_{B-L}$ is equal to the rank of ${\rm SO}(10)$ 
which is five whereas the rank of $3_c 2_L 1_{Y} Z_2$ is four, we can 
therefore implement a false vacuum hybrid inflationary scenario as 
described in Sec. \ref{sec-inflation2}, if we couple 
the inflaton field to the Higgs field used to break the intermediate 
symmetry gauge group $3_c 2_L 1_R 1_{B-L}$. The monopole problem can 
be solved and cosmic  strings can form at the end of inflation when 
the $3_c 2_L 1_R 1_{B-L}$ symmetry group breaks down to the standard 
model gauge group with unbroken matter parity, $3_c 2_L 1_Y Z_2$.

To break $3_c 2_L 1_R 1_{B-L}$, we use a ${\bf 126 + \overline{126}}$ pair 
of Higgs fields, which we call $\Phi_{126}$ and $\overline{\Phi}_{126}$. 
The latter are safe representations \cite{Martin} and therefore
keeps the $Z_2$ symmetry unbroken. A ${\bf 16 + \overline{16}}$ pair of 
Higgs fields usually used for the same purpose would break the $Z_2$ 
symmetry.  The VEV of the {\bf 126} and ${\bf \overline{126}}$ are in the $X$ direction, 
the {\rm U}(1) symmetry of {\rm SO}(10) which commutes with  {\rm
SU}(5). They break  
{\rm SO}(10) down to ${\rm SU}(5) \times
Z_2$. All together, i.e., with $A_{45}$, $S_{54}$, $A'_{45}$,
$S'_{54}$,  $\Phi_{126}$, and $\overline{\Phi}_{126}$ acquiring VEVs,
the {\rm SO}(10)  
symmetry group is broken down to $3_c 2_L 1_Y Z_2$. 

The symmetry breaking of the standard model is then achieved using 
two Higgs in the $10$-dimensional representation of {\rm SO}(10),
$H_{10}$ and $H'_{10}$.  

To summarise, the symmetry breaking is implemented as follows:
\begin{equation}
{\rm SO}(10) \stackrel{<A_{45}> <S_{54}><A'_{45}> <S'_{54}>}
{\rightarrow}3_c 2_L 1_R 1_{B-L} 
\stackrel{<\Phi_{126}> <\overline{\phi_{126}}>} {\rightarrow} 3_c 2_L
1_Y Z_2  \stackrel{<H_{10}> <H'_{10}>} {\rightarrow} 3_c 1_Q Z_2 \:
. 
\end{equation} 

\subsection{The superpotential}
\label{sec-super}

We now write down the superpotential involving the above mentioned 
fields. A consequence of the superpotential is the symmetry breaking
pattern given in  
Eq. (\ref{eq:sym}), which involves an inflationary sector.

As discussed above, our model involves four sectors. 
The first sector implements the doublet-triplet splitting and involves 
$A_{45}$,  with VEV in the ${\rm U}(1)_{B-L}$ direction. It also involves 
$S_{54}$ and two Higgs 10-plets, $H$ and $H'$. 
The superpotential in the first sector is given by $W_1 +W_2$, with, 
dropping the subscripts, 
\begin{equation}
W_1 = m_A A^2 + m_S S^2 + \lambda_S S^3 + \lambda_A A^2 S 
\end{equation}
and 
\begin{equation}
W_2 = H A H' + m_{H'} {H'}^2 \: .
\end{equation}
The Higgs potential $V_1$ associated with the superpotential $W_1$ has a global minimum such that the ${\rm SO}(10)$ symmetry group is 
broken down to $3_c 2_L 2_R 1_{B-L}$, with $A_{45}$ and $S_{54}$ 
acquiring VEVs. $W_2$ implements the doublet-triplet splitting; 
$H$ and $H'$ break 
${\rm SU}(2)_L \times {\rm U}(1)_Y$ down to ${\rm U}(1)_Q$.

The second sector involves $A'_{45}$, with VEV in the $T_{3R}$ direction, 
and $S'_{54}$. 
The superpotential in the second sector is given by 
\begin{equation} 
W_3 = m_{A'} {A'}^2 + m_{S'} {S'}^2 + \lambda_{S'} {S'}^3 +
\lambda_{A'} {A'}^2 {S'} \; .
\end{equation}
The associated Higgs potential $V_3$ has a global minimum such that the {\rm SO}(10) symmetry group
is broken down to 
 $3_c 2_L 2_R 1_{B-L}$, with $A'_{45}$ and $S'_{54}$ acquiring VEVs. 

The Higgs potential $V_1 + V_2 + V_3$ has a global minimum such that
the {\rm SO}(10) symmetry is broken down to $3_c 2_L 1_R 1_{B-L}$,
with $A_{45}$, $S_{54}$, $A'_{45}$ and $S'_{54}$ acquiring VEVs. 

The third sector 
involves $\Phi_{126}$ and $\overline{\Phi}_{126}$, and breaks ${\rm
SO}(10)$ down  
to ${\rm SU}(5) \times Z_2$. In order to force the $\Phi_{126}$ and
$\overline{\Phi}_{126}$ fields to get their VEVs the order of the {\rm
GUT} scale, we use a scalar field ${\cal S}$ singlet under {\rm
SO}(10). The superpotential is of the form, dropping the subscripts, 
\begin{equation}
W_4 = \alpha {\cal{S}} \overline{\Phi} \Phi - \mu^2 {\cal{S}} \; .
\end{equation}
$\alpha$ and $\mu$ are both positive and we must have 
${\mu \over \sqrt{\alpha}} = M_{{\rm G}}$, with $M_{{\rm G}} \simeq 10^{15} - 10^{16}$ GeV
for the unification of the gauge coupling constants. Identifying the
scalar field ${\cal S}$ with the inflaton field, $W_4$ leads to a
false vacuum hybrid inflationary scenario, as described in
Sec. \ref{sec-inflation2}.

The Higgs potential $V_1 + V_2 + V_3 + V_4$ has a global minimum such
that the $3_c 2_L 1_R 1_{B-L}$ symmetry group is broken  
down to the standard model gauge group with unbroken matter parity, 
$3_c 2_L 1_Y Z_2$, with $A_{45}$, $S_{54}$, $A'_{45}$, $S'_{54}$, 
$\Phi_{126}$ and $\overline{\Phi}_{126}$ acquiring VEVs. 

The full superpotential $W_{\rm tot} = W_1 + W_2 +W_3 + W_4$ does not involve terms
mixing $A'_{45}$ and $S_{54}$, $S'_{54}$ and $A_{45}$ etc... . In other 
words the three sectors are
independent. Thanks to the nonrenormalisable theorem, we are not obliged
to write down these terms, and it is not compatible with any extra
discrete symmetry \cite{babuandbarr}, therefore we do not have to fear
any domain wall formation when the symmetry breaks. Nevertheless, in
order to avoid any undesirable massless Goldstone Bosons, the three
sectors have to be related. The latter can be done introducing a third
adjoint $A"_{45}$, and adding a term of the form $A A' A''$ to the
superpotential \cite{babuandbarr}. The latter would neither affect the
symmetry breaking pattern, nor the inflationary scenario discussed
below. The full superpotential of the model is,
\begin{eqnarray}
W_{\rm tot } &=& m_A A^2 + m_S S^2 + \lambda_S S^3 + \lambda_A A^2 S + H A H' +
m_{H'} {H'}^2 \nonumber \\ 
&& + m_{A'} {A'}^2 + m_{S'} {S'}^2 + \lambda_{S'} {S'}^3 +
\lambda_{A'} {A'}^2 {S'} \nonumber \\ 
&& + \alpha {\cal{S}} \overline{\Phi} \Phi - \mu^2 {\cal{S}} 
\label{eq:super} \: . 
\end{eqnarray}
In Eq. (\ref{eq:super}), $A^2$ really means ${\rm Tr}(A^2)$, $A^2 S$
really means ${\rm Tr}(A^2 S)$, etc. 
The superpotential given in Eq. (\ref{eq:super}) leads to the desired
pattern of symmetry breaking and the VEVs of $A_{45}$, $S_{54}$,
$A'_{45}$, $S'_{54}$, $\Phi_{126}$ and $\overline{\Phi}_{126}$ are
given as follows  
(see Appendix \ref{chap-chapappC}). The adjoint $<\! A_{45}\! >$ is in the $B-L$ direction, 
\begin{equation}
<\! A_{45}\! > = J \otimes {\rm diag}(a,a,a,0,0) \label{eq:aaa}
\end{equation}
where $
J = \left (
\begin{array} {cc}
0 & 1 \\
-1 & 0
\end{array}
\right )
$ and $a \sim M_{{\rm GUT}}$. $<\! S_{54}\! >$ is a traceless symmetric
tensor given by,  
\begin{equation}
<\! S_{54}\! > = I \otimes {\rm diag}(x,x,x,-{3\over 2} x,-{3\over 2} x)
\label{eq:sss} 
\end{equation}
where $I$ is the unitary $2\times 2$ matrix and  $x = - {m_A
\over 2 \lambda_A}$.
$<\! A'_{45}\! >$ is in the $T_{3R}$ direction,
\begin{equation}
<\! A'_{45}\! > = J \otimes {\rm diag}(0,0,0,a',a') . \label{eq:ap}
\end{equation}
where $a' \sim M_{{\rm GUT}}$. $S'_{54}$ is a traceless antisymmetric tensor,
\begin{equation}
<\! S'_{54}\! > = I \otimes diag(x,'x',x',-{3\over 2} x',-{3\over 2} x')
\label{eq:spp} 
\end{equation}
where $x' = {2 m_{A'} \over 3 \lambda_{A'}}$. The only component of the {\bf 126} which acquires VEV is in the direction of the right-handed neutrino (it is the component which transforms as a singlet under SU(5))
\begin{equation}
<\! |\Phi_{126}|\! >_{\nu^c \nu^c} =
<\! |\overline{\Phi}_{126}|\! >_{\overline{\nu^c} \overline{\nu^c}} = d \:
. \label{eq:ppp} 
\end{equation}
Finally, we give VEV to the components of the 10 dimensional Higgs fields which correspond to the usual Higgs doublets. We do not use these, since we are interested in higher energies, where inflation and the GUT phase transitions take place.
 
With the VEVs above, if $<\! {\cal S} \! >= 0$ and $d = {\mu \over \sqrt{\alpha}}$, 
the Higgs potential has a global minimum such that the 
${\rm SO}(10)$ symmetry is broken down to the standard model gauge group with 
unbroken matter parity ${\rm SU}(3)_c \times {\rm SU}(2)_L \times {\rm U}(1)_Y 
\times Z_2$, and supersymmetry is unbroken (see App. \ref{chap-chapappC}).

\section{The inflationary epoch}
\label{sec-infl}
 
In this section we evaluate the details of the symmetry breaking pattern 
and of the inflationary scenario. We write down the scalar potential and 
find values of the scalar coupling constant and the mass scales $M_{{\rm G}}$
and $M_{{\rm GUT}}$ 
for which the inflationary scenario is successful.

We are interested in the dynamics of the symmetry breaking pattern and how 
the inflationary scenario fits in the symmetry breaking pattern.  We 
therefore need to study the scalar potential. In order fully to understand 
the dynamics of the model, one would need to use finite temperature
field theory.  
Nevertheless, the study of the scalar potential derived from the
superpotential given  
in Eq. (\ref{eq:super}) leads to a good understanding of the field evolution. We are mainly interested in what is happening above the electroweak scale, and hence we do not take into account the 10-dimensional Higgs multiplets $H$ and $H'$ which only break the standard model gauge group.
The scalar potential is then given by
\begin{eqnarray}
V &=& (2 m_A A + 2 \lambda_A A S)^2 + (2 m_S S + 3 \lambda_S S^2 +
\lambda_A A^2)^2 \nonumber \\ 
&&+ ( 2 m_{A'} A' + 2 \lambda_{A'} {A'} {S'})^2 + ( 2 m_{S'} {S'} + 3
\lambda_{S'} {S'}^2 + \lambda_{A'} {A'}^2 )^2 \nonumber \\ 
&& +  \alpha^2 |{\cal{S}} \overline{\Phi} |^2 + \alpha^2 | {\cal{S}} \Phi |^2 +
|\alpha \overline{\Phi} \Phi - \mu^2|^2. 
\end{eqnarray}
We remind the reader that $A$ and $A'$ are two Higgs in the 45-dimensional 
representation of ${\rm SO}(10)$ with VEV in the $B-L$ and $T_{3R}$
directions, respectively.  
$S$ and $S'$ are two Higgs in the 54-dimensional representation 
of {\rm SO}(10). $\Phi$ and $\overline{\Phi}$ are two Higgs in the
{\bf 126} and ${\bf \overline{126}}$ representations, with VEVs in the
right-handed neutrino direction. The scalar field ${\cal S}$  
is a singlet under {\rm SO}(10) and is identified with the inflaton field.  
$\alpha$ and $\mu$ are both positive constants which must satisfy the relation 
${\mu \over \sqrt{\alpha}} = M_{{\rm G}}$. The potential is minimised for
${\rm arg}( \Phi ) + {\rm arg}(\overline{\Phi}) = 0$, 
$(\alpha > 0)$, and it is independent of ${\rm\arg}({\cal{S}}) + {\rm
arg}({\Phi})$ 
and ${\rm\arg}({{\cal{S}}} ) + {\rm arg}({\overline{\Phi}} )$. We rewrite the
potential with the new fields which minimise the potential, keeping
the same notation for the old and new fields.  The Higgs potential becomes
\begin{eqnarray}
V &=& (2 m_A A + 2 \lambda_A A S)^2 + (2 m_S S + 3 \lambda_S S^2 +
\lambda_A A^2)^2 \nonumber \\ 
&&+ ( 2 m_{A'} A' + 2 \lambda_{A'} {A'} {S'})^2 + (2 m_{S'} {S'} + 3
\lambda_{S'} {S'}^2 + \lambda_{A'} {A'}^2)^2 \nonumber \\ 
&& + 4 \alpha^2 |{\cal{S}}|^2 |\Phi|^2 + (\alpha |\Phi|^2 - \mu^2)^2 +
{1 \over 2} m^2 |{\cal{S}}|^2 \label{eq:scalarpot} ,
\end{eqnarray}
where we have also introduced a soft
supersymmetry breaking term for ${\cal S}$, and $m \sim 10^3$ GeV.

The scalar potential is  flat in the
${\cal{S}}$ direction; we thus identify the scalar field ${\cal{S}}$
with the inflaton field. We suppose chaotic initial conditions; that
is 
we suppose that all the fields have initial values of the order of the
Planck scale. We then minimise the superpotential for fixed
${\cal{S}}$. We easily find that for $|{\cal{S}}| > {\mu \over
\sqrt{\alpha}} = s_c$, (recall that $\mu ,\alpha > 0$), there is a local 
minimum corresponding to  $|\Phi| = |\overline{\Phi}| = 0$, and $A$, $A'$, 
$S$ and $S'$ taking values as
given above in equations (\ref{eq:aaa}), (\ref{eq:sss}), (\ref{eq:ap}), and
(\ref{eq:spp}). 
Since all the fields are 
assumed to take initial values of the order of the Planck scale, the inflaton
field has an initial value greater than ${\mu \over \sqrt{\alpha}}$. 
Then, because the potential is flat in the inflaton direction, the
fields $\Phi$, 
$\overline{\Phi}$, $A$, $A'$, $S$ and $S'$ settle quickly to the local
minimum corresponding to $<S>$, $<A>$, $<S'>$ and $<A'>$ as in equations
(\ref{eq:aaa}), (\ref{eq:sss}), (\ref{eq:ap}) and
(\ref{eq:spp}) respectively, and $<\! |\Phi|\! > = <\! |\overline{\Phi}|\! > =
0$. The first phase transition takes place and the {\rm SO}(10)
symmetry group breaks 
down to the $3_c 2_L 1_R 1_{B-L}$ symmetry group. As 
shown in Sec. \ref{sec-modelcosmo}, topologically stable monopoles form 
according to the Kibble mechanism \cite{Kibble} during this first
phase transition.  

Once the fields $A$, $S$, $A'$ and $S'$ have settled down to their minimum, 
since the first derivatives $\partial V \over \partial A$, $\partial V
\over \partial S$, 
$\partial V \over \partial A'$ and $\partial V \over \partial S'$ are
independent of 
$\Phi$ and ${\cal{S}}$, the fields $A$, $A'$, $S$, and $S'$ will stay in 
their minimum independently of what the fields
$\Phi$ and ${\cal{S}}$ do.
When the VEV of the inflaton field is greater than ${\mu \over 
\sqrt{\alpha}} = s_c$, $|\Phi| = |\overline{\Phi}| = 0$, $F_{{\cal{S}}}$ 
term has a non-vanishing VEV, which means that
supersymmetry is broken in the ${\cal{S}}$ direction, by an amount
measured by the VEV of the ${\cal{S}}$ field. There is a 
non-vanishing vacuum energy density, $V = \mu^4$. An
inflationary epoch (an exponentially extending universe) can start.  

As has been pointed out
recently \cite{ShafiDvaSch}, the fact that supersymmetry is broken for 
$|{\cal S}|  > s_c$ implies that the one loop 
corrections to
the effective potential are non-vanishing. They are given by
\cite{ShafiDvaSch}  
\begin{equation}
\Delta V ({\cal{S}}) = \Sigma_i {(-1)^F \over 64 \pi^2} \,
M_i({\cal{S}})^4 \, \ln{({M_i({\cal{S}}) \over \Lambda^2})} \label{eq:effect}
\end{equation}
where the summation is over all helicity states for both fermions and
bosons. $\Lambda$ denotes a 
renormalisation mass and $(-1)^F$ indicates that the bosons and
fermions make opposite sign contributions to the sum; $(-1)$ stand 
for the fermions. Therefore the
one loop effective potential obtained from equations (\ref{eq:scalarpot})
 and (\ref{eq:effect}) is given by \cite{ShafiDvaSch}, 
\begin{eqnarray}
V_{eff} &=& \mu^4 [ 1 + {\alpha^2 \over 32 \pi^2} [2 \ln{({\alpha^2
s^2 \over \Lambda^2})} + ({\alpha s^2 \over \mu^2} - 1)^2 \ln{( 1 - {\mu^2
\over \alpha s^2})} \nonumber \\ 
&& + ({\alpha s^2 \over \mu^2 } + 1)^2 \ln{( 1 + {\mu^2 \over \alpha s^2})} ] 
+ {m^2 \over 2 \mu^4} s^2 ] \label{eq:veff}
\end{eqnarray}
where $s = |{\cal S}|$. Now $m \sim 10^3$ GeV and ${\mu \over 
\sqrt{\alpha}} \sim 10^{15 - 16}$ GeV, hence unless $\alpha \ll 1$, 
the soft supersymmetry breaking term can be neglected. Its contribution 
to the scalar potential is negligible. For $s > s_c$, the quantum corrections 
to the effective potential help ${\cal{S}}$ to roll down its minimum. 
Below $s_c$, the ${\cal{S}}$ field 
is driven to zero by the positive
term $\alpha^2 |{\cal{S}}|^2 |\Phi|^2$ which becomes larger
with increasing $|\Phi|$. Rapidly the  $\Phi$, $\overline{\Phi}$ and 
${\cal S}$ fields settle down the global minimum of the potential, 
corresponding to $<\Phi>_{\nu^c \nu^c} = 
<\overline{\Phi}>_{\overline{\nu^c} \overline{\nu^c}} = {\mu
\over \sqrt{\alpha}}$ and $s = 0$. This does not affect the VEVs of
the $S$, $A$, $S'$, and $A'$ fields which remain unchanged. The $3_c
2_L 1_R 1_{B-L}$  
symmetry group breaks down to $3_c 2_L 1_Y Z_2$. As shown in
Sec. \ref{sec-modelcosmo}, topological cosmic 
strings form during this phase transition. If inflation ends after the
phase transition, the strings may be inflated away. 

Inflation ends when the `slow roll' condition is violated. The slow
roll condition is characterised by \cite{Ed1} 
\begin{equation}
\epsilon \ll 1 \; , \;  \eta \ll 1 ,
\end{equation}
where
\begin{equation}
\epsilon = {M_{{\rm pl}}^2 \over 16 \pi} {({V' \over V})^2} \: ,\qquad  \: \eta
=  {M_{{\rm pl}}^2 
\over 8 \pi} {({V'' \over V})} \label{eq:epseta}
\end{equation}
and the prime refers to derivatives with respect to s. As pointed out by
Copeland {\it et al.} \cite{Ed1}, the slow-roll condition is a poor
approximation. But as shown in \cite{Ed1}, the number of {\it e}-foldings
which occur between the time when $\eta$ and $\epsilon$ reach unity
and the actual end of inflation is a tiny fraction of unity. It is
therefore sensible to identify the end of inflation with $\epsilon$ and
$\eta$ becoming of order unity. 

From the effective potential (\ref{eq:veff}) and the slow-roll parameters
(\ref{eq:epseta}) we have \cite{ShafiDvaSch}
\begin{eqnarray}
\epsilon = \left ({\alpha^2 M_{{\rm pl}} \over 8 \pi^2 M_{{\rm G}}}\right )^2 {x^2 \over 16 
\pi}
((x^2-1) \, \ln{(1-{1\over x^2})} + (x^2 +1) \ln{(1 + {1\over x})} )^2 \\
\eta = \left ({\alpha M_{{\rm pl}} \over 2 \pi M_{{\rm G}}}\right )^2 {1 \over 16 \pi} 
((3 x^2-1) \,
\ln{(1-{1\over x^2})} + (3 x^2 +1) \ln{(1 + {1\over x})} )^2 
\end{eqnarray}
where $x$ is such that ${\cal S} = x {\cal S}_c$. The phase transition down to the
standard model occurs when $x =1$. The results are as follows. We find 
the values of the scalar coupling $\alpha$, the scale $M_{{\rm GUT}}$ 
and the scale $M_{{\rm G}}$ which lead to successful inflation. For $\alpha 
\geq 35 - 43$, $M_{{\rm G}} \sim 10^{15} - 10^{16}$ GeV, $\epsilon$ is always 
greater than unity, and the slow roll condition is never satisfied. 
The scale $M_{{\rm GUT}}$ at which the monopoles form satisfies $M_{{\rm pl}} 
\geq M_{{\rm GUT}} \geq 10^{16} - 10^{17}$ GeV. For $\alpha \leq 0.02 - 
0.002$ and $M_{{\rm G}} \sim 10^{15} - 10^{16}$ GeV, neither $\eta$ nor 
$\epsilon$ ever reaches unity. ${\cal S}$ reaches ${\cal S}_c$ 
during inflation. Inflation must end by the instability of 
the $\Phi$ and $\overline{\Phi}$ fields. In that case, inflation 
ends in less than a Hubble time \cite{Ed1} once ${\cal S}$ reaches 
${\cal S}_c$. Cosmic strings, which form when $x=1$,
are not inflated away. The scale $M_{{\rm GUT}}$ at which the monopoles 
form must satisfy $M_{{\rm pl}} \geq M_{{\rm GUT}} \geq 10^{16} - 10^{17}$ GeV.
For the intermediates values of $\alpha$, inflation occurs, and ends 
when either $\epsilon$ or $\eta$ reaches unity; the string forming 
phase transition takes places once inflation completed.

\section{Formation of cosmic strings}
\label{sec-strings}

In this section we give general properties of the strings which form
at the end of inflation when the $3_c 2_L 1_R 1_{B-L}$ symmetry group
breaks down to $3_c 2_L 1_Y Z_2$. We find their  
width and their mass, give a general approach for their interactions 
with fermions and study their superconductivity.

\subsection{General properties}

Recall that, since the  first homotopy group 
$\pi_1({3_c 2_L 1_R 1_{B-L} \over 3_c 2_L 1_Y Z_2})$ is nontrivial,
cosmic strings form during the second phase transition [see Eq. (\ref{eq:sym})] when the $3_c 2_L 1_R 1_{B - L}$ symmetry breaks down to 
$3_c 2_L 1_Y Z_2$ .  
We note that the subspace 
spanned by $R$ and $B - L$ is also spanned by $X$ and $Y$. The generator 
of the string corresponds to the {\rm U}(1) of {\rm SO}(10) which
commutes with  
{\rm SU}(5), and the gauge field forming the string is the corresponding 
gauge field, which we call $X$. The strings are Abelian and physically 
viable. The model does not give rise to
Alice strings, like most of the non-Abelian {\rm GUT} phase transitions
where Abelian and non-Abelian strings form at the same time. This is a
good point of the model, since Alice strings give rise to quantum
number non-conservation, and are therefore in conflict with the
standard cosmology. The strings arising in our model can be related to the
Abelian strings arising in the symmetry breaking pattern of {\rm SO}(10)
down to to the standard model with ${\rm SU}(5) \times Z_2$ as intermediate
scale, since they have the same generator; the latter have been widely
studied in the nonsupersymmetric case \cite{Aryal87,Ma} (see also
Chap. \ref{chap-chap2}). Nevertheless, 
in our model, inside the core of the string, we do not have an {\rm SO}(10)
symmetry restoration, but an $3_c 2_L 1_R
1_{B - L}$ symmetry restoration. We therefore don't have any gauge fields
mediating baryon number violation inside the core of the strings, but one 
of the fields violates $B-L$. We also expect the supersymmetric strings to 
have different properties and different interaction with matter due 
to the supersymmetry restoration inside the core of the string. 
These special properties will be studied elsewhere.

The two main characteristics of the strings, their width and their mass, 
are determined through the Compton wavelength of the Higgs and gauge 
bosons forming the strings. The Compton wavelength of the Higgs and
gauge bosons are  
respectively 
\begin{equation}
\delta_{\Phi_{126}} \sim m_{\Phi_{126}}^{-1} = ( 2 \: \alpha \:  M_{{\rm G}}
)^{-1} \label{eq:deltaphi} 
\end{equation}
and 
\begin{equation}
\delta_X \sim m_X^{-1} = (\sqrt{2} \: e \: M_{{\rm G}})^{-1} \label{eq:deltaX} ,
\end{equation}
where $e$ is the gauge coupling constant in supersymmetric ${\rm SO}(10)$ and 
it is given by ${e^2 \over 4 \pi} = {1 \over 25}$ and $M_{{\rm G}}$ is the scale at 
which the strings form. 

As mentioned above, the strings formed in our model can be related
to those formed during the symmetry breaking pattern ${\rm SO}(10)
\rightarrow {\rm SU}(5) \times {\rm U}(1) \rightarrow {\rm SU}(5)
\times Z_2$. These 
strings have been studied by Aryal and Everett \cite{Aryal87} in
the nonsupersymmetric case. Using their results, with appropriate
changes in the gauge coupling constant and in parameters of the Higgs 
potential, we find that the string mass per unit length of the string 
is given by 
\begin{equation}
\mu \simeq (2.5-3) \times (M_{{\rm G}})^2 \label{eq:mu} ,
\end{equation}
for the scalar coupling $\alpha$  ranging from $5 \times 10^{-2}$ to 
$ 2 \times 10^{-1}$. Recall that the mass per unit length characterises 
the entire properties of a network of cosmic strings. 

\subsection{No superconducting strings}

One of the most interesting feature of {\rm GUT} strings is their
superconductivity. Indeed, if they become superconducting at the
{\rm GUT} scale, then vortons can form and dominate the energy density of the
universe; the model loses all its interest. The strings arising in
our model are not superconducting in Witten's sense
\cite{Witten}. They nevertheless can become current carrying with
spontaneous current generation at the electroweak scale through
Peter's mechanism \cite{Patrick}. But it is believed that this does 
not have any
disastrous impact on the standard cosmology. It has been shown in the
nonsupersymmetric case that the Abelian strings arising when {\rm SO}(10)
breaks down to ${\rm SU}(5) \times Z_2$ have right-handed neutrino zero
modes \cite{Das}. Since the Higgs field forming the string is a Higgs
boson in the {\bf 126} representation which gives mass to the right-handed
neutrino and winds around the string, we expect the same zero modes on
our strings. Since supersymmetry is restored in the core of the
string, we also expect bosonic zero modes of the superpartner of the
right-handed neutrino. Now, the question of whether or not the string
will be current carrying will depend on the presence of a primordial
magnetic field, and the quantum charges of the right-handed neutrino
with respect to this magnetic field. If there is a primordial magnetic
field under which the right-handed neutrino has a non-vanishing
charge, then the current will be able to charge up. On the other hand,
if such magnetic field does not exist, or if the right-handed neutrino is
neutral, then there will be nothing to generate
the current of the string. 
Although it is possible to produce a primordial magnetic field in
a phase transition \cite{Tanmay91}, we do not expect the fields produced
 through the mechanism of Ref. \cite{Tanmay91} to be able to charge up
 the current  on the string, since the latter are correlated on too large 
scales. Nevertheless, the aim of this section is to show that the
strings will not 
be superconducting at the {\rm GUT} scale in any case. We can therefore
assume a worse situation,
that is, suppose that the magnetic fields are correlated on smaller 
scales, 
due to any mechanism for primordial magnetic field production any time 
after the Planck scale. In our model, cosmic strings form when $3_c 
2_L 1_R 1_{B - L}$ breaks down to $3_c 2_L 1_Y Z_2$. Therefore the symmetric
phase  $3_c 2_L 1_Y Z_2$ will be associated with 
 colour, weak and hypercharge magnetic fields.  The colour and weak
magnetic fields formed when {\rm SO}(10) broke down to $3_c 2_L 1_R
1_{B - L}$,  
and and the hypercharge
magnetic field formed at the following phase transition, formed from
the $R$ and $B-L$ magnetic fields.  
Since the charges of the right-handed neutrino with respect to the colour, 
weak and hypercharge magnetic fields are all vanishing, no current will 
be generated.

We conclude that the strings will not be superconducting at the {\rm
GUT} scale.  
They might become superconducting at the electroweak scale, but this does 
not seem to affect the standard big-bang cosmology in any essential way. 

If the strings formed at the end of inflation are still present today,
they would affect temperature fluctuations in the CBR and have
affected large scale structure formation.

\section{Observational consequences}
\label{sec-observations}

We show here that the strings formed at the end of inflation may be present
today. 
We find the scale $M_{{\rm G}}$ at which cosmic strings form and 
the scalar coupling of the inflaton field which are consistent with
the temperature fluctuations observed by COBE. We then examine the 
dark-matter content of the model and make a qualitative discussion 
regarding large scale structure formation.

\subsection{Temperature fluctuations in the CBR}
\label{sec-temp}
If both inflation and cosmic strings are part of the scenario,
temperature fluctuations in the CBR are the result of the quadratic 
sum of the temperature fluctuations from inflationary perturbations
and cosmic strings.

The scalar density perturbations produced by the inflationary epoch 
induce temperature fluctuations in the CBR which are given by
\cite{ShafiDvaSch} 
\begin{eqnarray}
\left ({\delta T \over T} \right )_{inf} & \simeq & { \sqrt{ 32 \pi \over 
45}} {
V^{3 \over 2}  
\over V' M_{{\rm pl}}^3 } |_{x_q} \\
& \approx & (8 \pi N_q )^{1 \over 2} \left ({M_{{\rm G}} \over M_{{\rm pl}}} \right
)^2 , \label{eq:DTinfl} 
\end{eqnarray}
where the subscript indicates the value of ${\cal S}$ as the scale 
(which evolved to the present horizon size) crossed outside the 
Hubble horizon during inflation, and $N_q$ ($\sim 50 - 60$) denotes
the appropriate number of e-foldings. The contribution to the CBR
anisotropy due 
to gravitational waves produced by inflation in this
model is negligible.

The cosmic strings density perturbations also induce CBR anisotropies given 
by \cite{Allen}
\begin{equation}
\left ({\delta T \over T} \right )_{c.s.} \approx 9 \: G \mu , \label{eq:DTcs} 
\end{equation}
where $\mu$ is the strings mass per unit length, which is given by
Eq. (\ref{eq:mu}). It depends on the scalar coupling $\alpha$. Since
the later is undetermined, we can use the order of magnitude  
\begin{equation}
\mu \sim \eta^2 , \label{eq:app} 
\end{equation}
which holds for a wide range of the parameter $\alpha$; see Eq. (\ref{eq:mu}) and in Ref. \cite{Aryal87}. In Eq. (\ref{eq:app}),
$\eta$ is the symmetry breaking scale associated with the strings 
formation, here $\eta = M_{{\rm G}}$.

Hence, from equations (\ref{eq:DTinfl}) and (\ref{eq:DTcs}) the
temperature fluctuations in the CBR are given by 
\begin{eqnarray}
\left ({\delta T \over T} \right )_{tot } &\approx &  \sqrt{\left
({\delta T \over 
T}\right )_{inf}^2 + \left ({\delta T \over T}\right )_{c.s.}^2 }
\label{eq:DT1}\\  
 &\approx & \sqrt{ 8 \pi N_q + 81} \:
\left ({M_{{\rm G}} \over M_{{\rm pl}}} \right )^2 . \label{eq:DT}
\end{eqnarray}
The temperature fluctuations from both inflation and cosmic strings
add quadratically. Since they are both proportional to ${M_{{\rm G}} \over
M_{{\rm pl}} }$ their computation is quite easy. 

An estimate of the coupling
$\alpha$ is obtained from the relation \cite{ShafiDvaSch} 
\begin{equation}
{\alpha \over x_q} \sim { 8 \pi^{3\over 2} \over \sqrt{N_q}} {M_{{\rm G}} \over 
M_{{\rm pl}}} . \label{eq:alpha}
\end{equation}

With $x_q \sim 10$, using Eqs. (\ref{eq:DT}) and (\ref{eq:alpha}) and 
using the temperature fluctuations measured by COBE $\simeq 1.3 \times
10^{(-5)}$ \cite{COBE2} we get
\begin{eqnarray}
\alpha & \simeq & 0.03 , \\
M_{{\rm G}} & \simeq & 6.7 \times 10^{15} \: {\rm GeV} \: .
\end{eqnarray}
With these values, we find that $\eta$ reaches unity when $x \simeq 1.4$ 
and the scale $M_{{\rm GUT}}$ at which the monopoles form must satisfy, 
\begin{equation}
M_{{\rm pl}} \geq M_{{\rm GUT}} \geq 6.7 \times 10^{16} \: {\rm GeV}
\end{equation}
where $M_{{\rm pl}}$ is the Planck mass $\simeq 1.22 \times 10^{19} \: {\rm GeV}$. 
 
From the above results, we can be confident that the strings forming
at the end of inflation should still be around today.
 
Now that we have got values for the scalar coupling $\alpha$ and the 
scale $M_{{\rm G}}$ at which the strings form, the Compton wavelength of the 
Higgs and gauge bosons forming the strings given by
Eqs. (\ref{eq:deltaphi}) and (\ref{eq:deltaX}) can be computed. We
find  
\begin{equation}
\delta_{\Phi_{126}} \sim m_{\Phi_{126}}^{-1} \sim 0.42 \times 10^{-28} \: {\rm cm} 
\end{equation}
for the Compton wavelength of the Higgs field forming the string and 
\begin{equation}
\delta_X \sim m_X^{-1} \sim 0.29  \times 10^{-29} \: {\rm cm}
\end{equation}
for the Compton wavelength of the strings gauge boson. 
$\delta_{\Phi_{126}} > \delta_X$ thus the strings possess an inner 
core of false vacuum of radius $\delta_{\Phi_{126}}$ and a magnetic 
flux tube with a smaller radius $\delta_X$. The string energy per unit
length is given by Eq. (\ref{eq:mu}), 
thus, using above results, we have 
\begin{equation}
G \mu \sim 7.7 \times 10^{-7} ,
\end{equation}
where $G$ is Newton's constant. The results are slightly affected by the 
number of e-folding and by the order of magnitude (\ref{eq:app}) used to 
compute the temperature fluctuations in the CBR due to cosmic strings 
in Eqs. (\ref{eq:DT1}) and (\ref{eq:DT}). Once we have found the value 
for the scalar coupling $\alpha$ for successful inflation, we can redo 
the calculations with a better initial value for the string mass per 
unit length; see Eq. (\ref{eq:mu}); the scalar coupling $\alpha$ 
is unchanged. The results are summarised in Table 4.1.

\vspace{.5cm}

\begin{tabular} {||c||c|c|c|c||}
\hline\hline
Nq & 50 & 50 & 60 & 60  \\ \hline\hline
$\mu_{init}$ & $\eta^2$ & $2.5 \: \eta^2$ & $\eta^2$ & $2.5 \: \eta^2$
\\ \hline\hline 
$M_{{\rm G}}$ & $6.7 \times 10^{15}$ & $6.3 \times 10^{15}$ & $6.5 \times
10^{15}$ & $6.1 \times 10^{15}$  \\ \hline  
 $\alpha$ & 0.03 &  0.03 & 0.03 & 0.29 \\ \hline
$\delta_\Phi$ & $0.42 \times 10^{-28}$ & $0.44 \times 10^{-28}$ &$0.43
\times 10^{-28}$ &$0.46 \times 10^{-28}$ \\ \hline 
$\delta_X$ & $0.29 \times 10^{-29}$ & $0.31 \times 10^{-29}$ &$0.30
\times 10^{-29}$ &$0.32 \times 10^{-29}$ \\ \hline 
$G \mu$  & $7.7 \times 10^{-7}$ &$6.7 \times 10^{-7}$ & $7.1 \times
10^{-7}$ & $6.3 \times 10^{-7}$  \\ \hline\hline 
\end{tabular}

\vspace{.5cm}

\underline{Table 4.1} : The table shows the values obtained for the
scale $M_{{\rm G}}$  
at which the strings form, the scalar coupling $\alpha$, the Higgs and 
gauge boson Compton wavelengths $\delta_\Phi$ and $\delta_X$ of the 
strings, and $G \mu$, where $\mu$ is the strings mass per unit 
length and $G$ is the Newton's constant,
 for different values of the number of {\it e}-foldings $N_q$ and for different
initial values used for their computation for the string 
mass-per-unit length.

\subsection{Dark matter}
\label{sec-dark}

We specify here the nature of dark matter generated by the model. 

If we go back to 
the symmetry breaking pattern of the model given by Eq. (\ref{eq:sym}), we see that a discrete $Z_2$ symmetry remains 
unbroken down to low energy. This $Z_2$ symmetry is a subgroup of both 
the $Z_4$ centre of {\rm SO}(10) and of ${\rm U}(1)_{B-L}$ subgroup of
{\rm SO}(10).  
This $Z_2$ symmetry acts as matter parity. It preserves large values 
for the proton lifetime and stabilises the lightest superparticle. 
The LSP is a good cold dark matter candidate.

The second stage 
of symmetry breaking in Eq. (\ref{eq:sym}) is implemented with the 
use of a ${\bf 126 + \overline{126}}$ pair of Higgs multiplets, with
VEVs in the  
direction of the ${\rm U}(1)_X$ of ${\rm SO}(10)$ which commutes with
${\rm SU}(5)$. The  
${\bf {\overline{126}}}$ multiplet can couple with fermions and give 
superheavy Majorana mass to the right-handed neutrino, solving the
solar neutrino problem via the MSW mechanism \cite{MSW} and providing
a good hot dark matter candidate. This can be done if all fermions are
assigned to 
the 16-dimensional spinorial representation of {\rm SO}(10). In that
case, couplings of the form $f \overline{\Psi} \Psi {\bf
\overline{126}}$, 
where $\Psi$ denotes a 16-dimensional {\rm spin}or to which all
fermions belonging to a single family are assigned, provide
right-handed 
neutrinos masses of order $m_{R} \simeq  10^{12}$ GeV, 
if $f \sim 10^{-4}$ GeV. Neutrinos also get Dirac masses which are
typically of the order of the mass of the up-type quark of the
corresponding family; for instance $m^{\nu^e}_D \simeq m_u$. After
diagonalising the neutrino mass matrix, one finds that the
right-handed neutrino mass $m_{\nu_R} \simeq m_R$ and the left-handed
neutrino mass $m_{\nu_L} \simeq {m_D^2 \over m_R}$. With the above
values we get 
\begin{eqnarray}
m_{\nu_e} &\sim &  10^{-7} \: {\rm eV}\\
m_{\nu_\mu} &\sim & 10^{-3} \: {\rm eV}\\
m_{\nu_\tau} &\sim & 10 \: {\rm eV}
\end{eqnarray}
The tau neutrino is a good hot dark matter candidate.

Our model thus provides both CDM and HDM and is consistent with mixed
cold and hot DM scenarios. 

It is interesting to note that CDM and HDM are, in this model, related
to each other. Indeed, the $Z_2$ symmetry in Eq. (\ref{eq:sym}), which
stabilises the LSP, is kept unbroken because a ${\bf 126 +
\overline{126}}$ and not  ${\bf 16 + \overline{16}}$ pair of Higgs
fields are used to break ${\rm U}(1)_{B-L}$. If a 16 dimensional Higgs
representation were used,  
the right-handed neutrino could not get a superheavy Majorana mass 
and thus no HDM could be provided, also the $Z_2$ symmetry would 
have been broken, and thus the LSP destabilised. The ${\bf 126 + 
\overline{126}}$ pair of Higgs fields provide superheavy Majorana 
to the right-handed neutrino and keeps the $Z_2$-parity unbroken. 
It leads to both HDM and CDM. We conclude that, in this model, 
CDM and HDM are
intimately related. Either the model provides both cold and hot 
dark matter, or it does not provide any. Our model provides 
both CDM and HDM.

\subsection{Large scale structure}
\label{sec-structure}

We give here only a qualitative discussion of the consistency of the model 
with large scale structure. We do not make any calculations which would 
require a full study on their own. We can nevertheless use various results 
on large scale structure with inflation or cosmic strings. Since 
we determined the nature of dark matter provided by the model, we may make 
sensible estimations about the the consistency of the model with large 
scale structure.

Presently there are two candidates for large scale structure formation, the 
inflationary scenario and the topological defects scenario with cosmic 
strings. Both scenarios are always considered separately. Indeed, 
due to the difference in the nature of the density perturbations in 
each of the models, density perturbation calculations due to a mixed 
strings and inflation scenario are not straightforward. Indeed in the 
inflation-based models density perturbations are Gaussian adiabatic 
whereas in models based on topological defects inhomogeneities are 
created in an initially homogeneous background \cite{Andrew}.

In the attempt to explain large scale structure, inflation-seeded 
cold dark matter models or strings models with HDM are the most 
capable \cite{Andrew}. In adiabatic perturbations with hot dark 
matter small scale perturbations are erased by free streaming 
whereas seeds like cosmic strings survive free streaming. Small scale fluctuations in models with cosmic strings and HDM 
are not erased, but their growth is only delayed by free streaming \cite{leandros}. 

Our model involves both hot and cold dark matter, and both inflation 
and cosmic strings. It is therefore sensible to suggest that our model 
will be consistent with large scale structure formation, with the large 
scale structures resulting from the inflationary scenario and 
small scale structures (galaxies and clusters) being due to cosmic strings.

\section{Conclusions}
\label{sec-conclusion}

We have successfully implemented a false vacuum hybrid inflationary 
scenario in a supersymmetric {\rm SO}(10) model. We first argued that this 
type of inflationary scenario is a natural way for inflation to occur in 
global supersymmetric {\rm SO}(10) models. It is natural, in 
the sense that the inflaton field emerges naturally from the theory, no 
external field and no external symmetry has to be added. The scenario
does not require any fine  
tuning. In our specific model, the {\rm SO}(10) symmetry is broken via
the intermediate $3_c 2_L 1_R  
1_{B-L}$ symmetry down to the standard model with unbroken matter
parity $3_c 2_L 1_Y Z_2$. The model 
gives a solution for the doublet-triplet splitting via the
Dimopoulos-Wilczek mechanism. It also suppresses rapid proton decay .

The inflaton, 
a scalar field singlet under {\rm SO}(10), couples to the Higgs mediating the 
 phase transition associated with the breaking of $3_c 2_l 1_R
1_{B-L}$ down to the standard model. The scenario starts with chaotic
initial conditions.
The {\rm SO}(10) symmetry breaks at $M_{{\rm GUT}}$ down to $3_c 2_L 
1_R 1_{B-L}$ and topologically stable monopoles form. 
There is a non-vanishing vacuum energy density, supersymmetry is broken,
 and an exponentially extending epoch starts. Supersymmetry is broken,
 and therefore quantum corrections to the scalar potential can not be 
neglected. The latter help the inflaton field to roll down its minimum. 
At the end of inflation the $3_c 2_L 1_R 
1_{B-L}$ breaks down to $3_c 2_L 1_Y Z_2$, at a scale $M_{{\rm G}}$, and
cosmic strings form. They are not superconducting. 

Comparing the CBR temperature anisotropies measured by COBE with that 
predicted by the mixed inflation-cosmic strings scenario, we find  values 
for the scalar coupling $\alpha$ and for the scale $M_{{\rm G}}$ at which the 
strings form. $M_{{\rm GUT}}$ is calculated such that we get enough
{\it e}-foldings  
to push the monopoles beyond the horizon. The results are summarised
in Table 4.1. 
 The evolution of the strings is that of topologically stable cosmic 
strings.
The model is consistent with a mixed HCDM scenario. Left-handed neutrinos 
get very small masses and the tau neutrino may serve as a good HDM
candidate. They could also explain the solar neutrino problem via the
MSW mechanism.  
The unbroken matter parity stabilises the LSP, thus providing a good 
CDM candidate. A qualitative discussion leads to the conclusion that 
the model is consistent with large scale structures, very large scale 
structures being explained by inflation and cosmic strings explaining 
structures on smaller scales. An algebraic investigation for this
purpose would  
be useful, but will require further research.

\chapter{New Mechanism for Leptogenesis}
\label{chap-chap5}

\section{Introduction}

Big-bang nucleosynthesis predicts the present abundances of the
light-elements ${\rm He}^3$, D, ${\rm He}^4$ and $Li^7$ as a function
of an adjustable parameter, the baryon-to-photon ratio $\eta = {n_B
\over n_\gamma}$. Recent analysis \cite{Walker} shows that $\eta$ must
lie in the range 
\begin{equation}
\eta \simeq (2 - 7) \times 10^{-10} \label{eq:nucl}
\end{equation} 
to predict the light elements abundances which agree with
observations. $\eta$ is related to the baryon number of the universe
$B = {n_B \over s}$ via ${n_B \over
n_\gamma } = 1.80 g_* B$, where $s$ is the entropy of the universe and
$g_*$ counts the number of massless degrees of freedom. According to
the big-bang cosmology, the universe started on a baryon-symmetric
state. Hence there must have been out-of-equilibrium processes in the
very early universe which violated baryon number
plus C and CP \cite{Sakharov} to explain the matter/antimatter
asymmetry of the universe today.

The standard scenario for baryogenesis is provided by grand unified
theories (GUTs), via the out-of-equilibrium decays of heavy gauge and
Higgs 
bosons which violate baryon number ($B$) and/or lepton number ($L$). But it
has been realized 
a decade ago that, due to an anomaly in the baryonic current, and due to the
non trivial structure of the vacuum in non Abelian gauge theories, any
baryon asymmetry generated at the grand unified scale would be erased
by the electroweak anomaly \cite{KRS}.
Baryon and lepton number non-conservation in the standard electroweak 
theory had been known for a
while \cite{'t Hooft}. But in 1976, 't Hooft \cite{'t Hooft}
pointed out that the quantum tunnelling transition rate between two
topologically different vacua by instantons is exponentially
suppressed by the WKB factor $\exp({-4 \pi \over \alpha_w})$ at zero
temperature. It is only in 1985, after the discovery of sphalerons
\cite{Nick}, static but unstable solutions of the classical field
equations in the electroweak theory, that Kuzmin, Rubakov and
Shaposhnikov \cite{KRS} realized that at high temperature, the
transition rate between two neighbouring inequivalent vacua is
unsuppressed.  Since vacuum to vacuum transitions in the
electroweak theory conserve $B-L$ 
but violate $B+L$ and since $B$ and $L$ are related to $B+L$ and $B-L$ by
\begin{eqnarray}
B &=& {1\over 2} (B+L) + {1\over 2 } (B-L) \\
L &=& {1\over 2} (B+L) - {1\over 2 } (B-L)  \: ,
\end{eqnarray}
and therefore $B$ and $L$ are proportional to $B-L$:
\begin{eqnarray}
<B>_T & =  & \alpha \: <(B-L)>_T \label{eq:B} \\
<L>_T & = &\gamma \: <(B-L)>_T , \label{eq:L}
\end{eqnarray}
with $\alpha$ and $- \gamma$ $ \simeq 0.5$. Weak interactions
only involve left-handed fermion fields and hence the factors $\alpha$
and $- \gamma$ are not exactly ${1\over 2}$ \cite{HT}.
We see from Eqs. (\ref{eq:B}) and (\ref{eq:L}) that, unless the 
universe started with a non-vanishing $B-L$ asymmetry, any $B$ or $L$
asymmetry generated at the grand unified scale will be erased by the
electroweak anomaly. GUTs such as {\rm SU}(5) conserve $B-L$, and hence fail
in explaining the baryon number of the universe today. 

An initial $B-L$ asymmetry can be obtained in theories containing an 
extra gauge
${\rm U}(1)_{B-L}$ symmetry, such as the simple ${\rm
U}(1)$ extension 
of the standard model ${\rm SU}(3)_c \times {\rm SU}(2)_L \times {\rm
U}(1)_Y \times 
{\rm U}(1)_{Y'}$, where the charge $Y'$ is a linear combination of $Y$ and
$B-L$, left-right models or SO(10) and 
E(6) GUT's. These theories predict one or more extra fermions in
addition to the 
usual quarks and leptons. One of them, singlet under the
standard model gauge 
group, can be interpreted as a right-handed neutrino. 
Right-handed neutrinos acquire a heavy Majorana mass at
the scale of $B-L$ breaking. Out-of-equilibrium  
decays of heavy Majorana
right-handed neutrinos may provide the necessary primordial $B-L$
asymmetry, and hence explain the observed baryon asymmetry today
\cite{Y86,Luty,Y93}. This mechanism however requires either
very heavy neutrinos or extreme fine tuning of the parameters in the
neutrino mass matrix \cite{Y93}. Also, the masses of the new gauge
bosons must be bigger than the smallest heavy neutrino mass
\cite{Enqvist}. Hence there is a wide range of parameters for which 
the mechanism does not produce enough baryon asymmetry.

In this chapter, we show that in unified models involving an extra gauge
${\rm U}(1)_{B-L}$ symmetry, a primordial $B-L$ asymmetry can be
generated by the out-of-equilibrium decays 
of right-handed neutrinos released by collapsing cosmic string loops. 
As a consequence of ${\rm U}(1)_{B-L}$ breaking, cosmic strings may
form at the $B-L$ breaking scale according to the Kibble mechanism
\cite{Kibble}.  
We call them $B-L$ cosmic strings. The
Higgs field mediating the breaking of $B-L$ is the Higgs field forming
the strings and it is the same Higgs field  
that gives a heavy Majorana mass to the right-handed neutrinos. Hence,
due to the winding of the Higgs field around the 
string, we expect right-handed neutrino zero modes \cite{Jackiw}
trapped in the core of the strings. 
These zero modes are 
predicted by an index theorem \cite{Weinberg}. There are also
modes of higher energy bounded to the strings. We shall
consider only the zero modes, which are the most
favourable to be trapped. $B-L$ cosmic string loops lose their energy
by emitting 
gravitational radiation and rapidly shrink to a point, releasing
these right-handed neutrinos. This is an out-of-equilibrium process. Right-handed neutrinos
acquire a heavy Majorana mass and decay into massless leptons and 
electroweak Higgs particles to produce a lepton asymmetry. This lepton
asymmetry is converted into a baryon asymmetry via sphaleron
transitions.

\section{Unified theories with $B-L$ cosmic strings}

Theories beyond the standard model gauge group with rank greater or equal to five and containing an extra ${\rm U}(1)_{B-L}$ gauge symmetry predict
fermions in addition to the usual quarks and leptons of the standard
model. One of them, singlet under the standard model gauge group,
may be interpreted as a right-handed neutrino. Right-handed neutrinos 
may acquire a heavy Majorana mass at the $B-L$ breaking scale.
The simplest such extension of the standard model is the ${\rm SU}(3)_c \times
{\rm SU}(2)_L \times {\rm U}(1)_Y \times 
{\rm U}(1)_{Y'}$ symmetry group where the charge $Y'$ is a linear combination
of $Y$ and $B-L$. Left-right 
models and grand unified theories with rank greater or equal to five
also contain 
a ${\rm U}(1)_{B-L}$ gauge symmetry and predict right-handed
neutrinos. In such theories, as a consequence of ${\rm U}(1)_{B-L}$
breaking, cosmic strings may form. We call them $B-L$ cosmic
strings.  

Topological $B-L$ cosmic strings form when a gauge group 
${\rm G} \supset {\rm U}(1)_{B-L}$
breaks down to a subgroup ${\rm H} \not\supset {\rm U}(1)_{B-L}$ of G, if
the vacuum manifold ${\rm G}\over {\rm H}$ is 
simply connected, that is if the first homotopy group $\pi_1({{\rm G}\over {\rm H}})$
is non-trivial. If $\pi_1({{\rm G}\over {\rm H}}) =I$ but string solutions still
exist, then embedded strings \cite{Tanmay,embed} form when G breaks down to
H. Embedded strings are stable for a wide range of parameters. In
left-right models, embedded $B-L$ strings usually form. In the simple
U(1) extension of the standard model ${\rm SU}(3)_c \times  
{\rm SU}(2)_L \times {\rm U}(1)_Y \times {\rm U}(1)_{Y'}$ where $Y'$
is a linear combination of $Y$ and $B-L$ topological strings form. In
grand  
unified theories with rank greater than five, such as SO(10) or E(6),
$B-L$ cosmic strings may form, 
depending on the symmetry breaking pattern and on the set of Higgs
fields used to do the breaking down to the standard model gauge
group \cite{Kibble82,Witten}. There is a wide
range  
of theories which contain both ${\rm U}(1)_{B-L}$ and $B-L$ cosmic strings.

\section{Right-handed neutrino zero modes in $B-L$ cosmic strings}

We noticed that in unified theories with rank greater or equal to
five which contain an extra ${\rm U}(1)_{B-L}$ gauge symmetry $B-L$ cosmic
strings often form. Consider then a unified model with such a ${\rm U}(1)_{B-L}$ symmetry and stable $B-L$ cosmic strings. Such a theory predicts right-handed neutrinos.

The gauge and Higgs
fields forming the strings will be the $B-L$ associated gauge boson $A'$ and
the Higgs field $\phi_{B-L}$ used to break ${\rm
U}(1)_{B-L}$. Right-handed neutrinos acquire heavy Majorana mass 
via Yukawa couplings to $\phi_{B-L}$. So $\phi_{B-L}$ both winds around the strings and gives heavy Majorana mass to the right-handed neutrinos. Thus, following Jackiw and Rossi\cite{Jackiw}, we expect right-handed neutrino zero modes in the core of $B-L$ cosmic strings. The ${\rm U}(1)_{B-L}$ part
of the theory is described by the Lagrangian    
\begin{eqnarray}
L &=&  {1\over 4} f_{\mu \nu} f^{\mu \nu} + (D_\mu \phi_{B-L})^\dagger
(D^\mu \phi_{B-L}) - V(\phi_{B-L})  \nonumber\\ 
&& + i \overline{\nu}_L \gamma^\mu D_\mu \nu_L + i \overline{\nu}_R
\gamma^\mu D_\mu \nu_R \nonumber\\ 
&&+ i \lambda \phi_{B-L} \overline{\nu}_R \nu_L^c - i \lambda
\phi_{B-L}^* \overline{\nu}_L^c \nu_R  + L_f \: . \label{eq:Laglept}
\end{eqnarray}
The covariant derivative $D_\mu =
\partial_\mu -i e a^{Y'}_\mu$ where $e$ is the gauge coupling constant
and $Y'$ is a linear combination of $B-L$ and $Y$. 
$\lambda$ is a Yukawa coupling constant, and $V(\phi_{B-L})$ is the
Higgs potential. The spinor $N = \nu_R + 
\nu_L^c$ is a Majorana spinor satisfying the Majorana condition $N^c
\equiv C \gamma_0^T N^* = N$, where $C$ is the charge conjugation
matrix. Hence $N$ has only two independent components, two degrees of
freedom. $L_f$ 
is the fermionic $B-L$ part of the Lagrangian which does not contain
neutrino fields.

For a straight infinite
cosmic string lying along the $z$-axis, the Higgs field $\phi_{B-L}$ and
the $Y'$ gauge field $a$ in 
polar coordinates $(r,\theta )$ have the form 
\begin{eqnarray}
&&\phi_{B-L} = f(r) e^{i n \theta} \\
&&a_\theta = - n \tau {g(r) \over e r } \hspace{1cm}  a_z = a_r =0 ,
\end{eqnarray} 
where $n$ is the winding number; it must be an integer. Most strings have
winding number $n=1$; strings with winding number $|n| >1$ are
unstable. $\tau$ is the string's generator; it is the normalised ${\rm
U}(1)_{Y'}$ generator. It has different eigenvalues 
for different fermion fields.  
The functions $f(r)$ and $g(r)$ must satisfy the following boundary
conditions
\begin{eqnarray}
f(0) = 0 \hspace{.75cm} & {\rm and} & \hspace{1cm}  f \rightarrow
\eta_{B-L} \hspace{.5cm} {\rm as } \hspace{.5cm} r \rightarrow \infty ,
\\ 
g(0) = 0 \hspace{.75cm} & {\rm and}& \hspace{1cm} g \rightarrow 1
\hspace{.5cm} {\rm as }\hspace{.5cm}  r \rightarrow \infty 
\: ,
\end{eqnarray}
where $\eta_{B-L}$ is the scale of $B-L$ breaking.  The exact
forms of the functions $f(r)$ and $g(r)$ depend on the Higgs potential
$V(\phi_{B-L})$.

From the Lagrangian (\ref{eq:Laglept}) we derive the equation 
for the right-handed neutrino field: 
\begin{equation}
i \gamma^\mu D_\mu \nu_L^c - i \lambda \phi_{B-L}^* \overline{\nu}_L^c
= 0 \: \label{eq:mot1}  
\end{equation} 
where $\nu_L^c = C \gamma_0^T \nu_R^*$ .
Solving (\ref{eq:mot1}), we find that Majorana neutrinos trapped as
tranverse zero modes in the core of $B-L$ cosmic
strings have only one independent component. For an $n=1$ vortex it takes
the form :  
\begin{equation}
N_1 = \beta(r,\theta) \: \alpha(z+t) \label{eq:nuR}
\end{equation}
where $\beta(r,\theta)$ is a function peaked at $r=0$ which
exponentially vanishes outside the core of the string,
so that the fermions effectively live on the strings. The $z$ and $t$
dependence of $\alpha$ shows that the neutrinos travel at
the speed of light in the $-z$ direction, so that they are effectively
massless. In a $n = -1$ vortex, the function $ \alpha = \alpha(t-z)$, so that the fermions travel at the
speed of light in the $+z$ direction. These fermions can be described by
an effective theory in $1+1$ dimensions. The usual energy to 
momentum relation 
\begin{equation}
E = P \label{eq:E}
\end{equation}
holds. We have no boundary conditions in the 1 spatial dimension, and
the spectrum of states is continuous. In the ground state the Fermi 
momentum of the zero modes is $p_F = 0$.

The field solution (\ref{eq:nuR}) and the energy to momentum 
relation (\ref{eq:E}) have been
derived for fermions on a straight infinite string. However physical cosmic
strings are very wiggly and are not 
straight. Hence relations (\ref{eq:nuR}) and (\ref{eq:E}) do not hold in 
the physical case. Neither do they hold for 
cosmic string loops, even if the latter are assumed to be smooth. 
The behaviour of Dirac fermions on a cosmic string of finite radius of 
curvature has been analysed by Barr and Matheson\cite{BarrMathe}. On
such strings, fermions are characterised by  
their angular momentum $L$.  The energy 
relation becomes\cite{BarrMathe}   
\begin{equation}
E = {(L+{1\over 2})\over R} = P + {1\over 2R} \label{eq:E2}
\end{equation}
and hence the energy spectrum is 
\begin{equation}
E ={ \pm ({\rm n}  + {1\over 2})  \over R} \: \label{eq:E1} 
\end{equation}
where $n \in {\mathbb N}$. We see from
Eqs. (\ref{eq:E}), (\ref{eq:E2}), 
and (\ref{eq:E1}) that, when 
$R$ is very large, the string looks locally like a straight string. We
have an almost continuous spectrum of states. The fact that the string gets 
a finite curvature acts as a perturbation on the string bound states. The 
energy levels get quantised and the Fermi energy gets a non-vanishing 
value $E_F = {1\over 2 R}$. As the string loop shrinks, its radius $R$ 
decreases and we see from Eq. (\ref{eq:E1}) that the Fermi energy increases 
and that the separation between energy levels gets wider.

\section{Leptogenesis via decaying cosmic string loops}

Cosmic string loops form via the intercommuting of long strings. Some
loops are formed when the network initially forms. Cosmic strings
loops lose their energy via gravitational radiation and rapidly
decay, releasing right-handed neutrinos trapped as transverse zero
modes in their core.  

Assuming that a loop decays when its radius
$R$ becomes comparable to its width $w \sim 
\eta_{B-L}^{-1}$,  we deduce that the Fermi energy level
when the loop decays is $E_F \sim {1\over 2} \eta_{B-L} $, where the
$B-L$ breaking scale $\eta_{B-L}$ is proportional to the  
right-handed neutrino mass. $E_F$ is lower than the energy needed by
a neutrino to escape the string \cite{BarrMathe}. Hence, when a cosmic string loop decays, it 
releases at least $n_\nu = 1$ heavy Majorana neutrinos. Quantum
fluctuations and finite temperature corrections may increase
$n_\nu$. Part of the final  
burst of energy released 
by the decaying cosmic string loop is converted into mass energy for
the gauge and Higgs particles released by the string, and into mass energy for
the neutrinos. A decaying $B-L$ cosmic string loop releases heavy
$B-L$ Higgs particles which can decay into right-handed neutrino
pairs, and hence increase $n_\nu$. This is an out-of-equilibrium
process. Due to angular momentum conservation, the massive Majorana
neutrinos released by a decaying cosmic string loop 
which were trapped as transverse zero modes are spinning particles.

Heavy Majorana right-handed neutrinos interact with the standard model
leptons via the Yukawa couplings 
\begin{equation}
L_Y = h_{ij} \overline{l_i} H_{ew} \nu_{Rj} + h.c.
\end{equation} 
where $l$ is the usual lepton doublet; for the first family $l = (e,
\nu)_L$. $H_{ew}$ is the standard model doublet of Higgs
fields. Majorana right-handed neutrinos can decay via the diagrams shown in
Figs.1.a. and 1.b. 
\vspace{.5cm}

\begin{figure}[h]
\centering
\includegraphics[width=14cm]{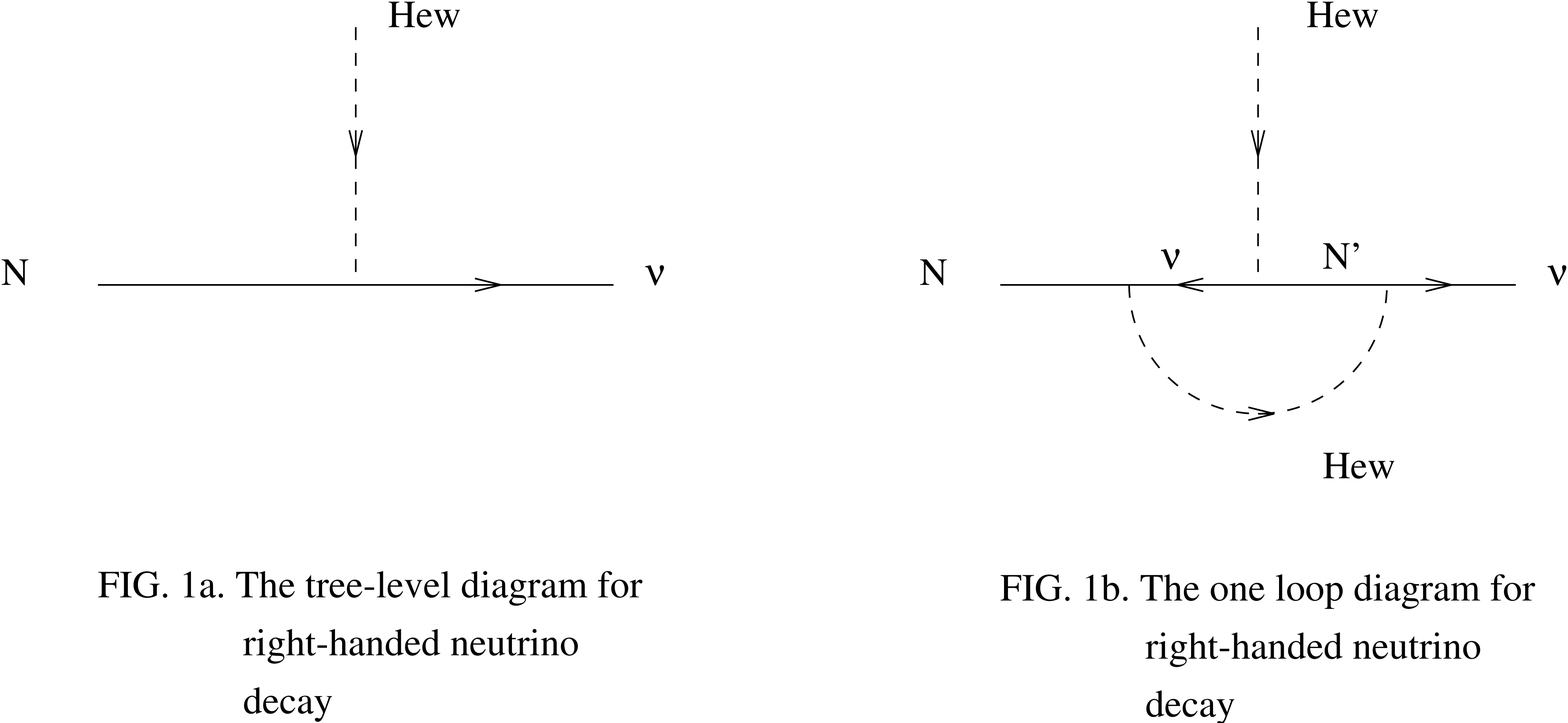}
\end{figure}

\vspace{.5cm}
CP is violated through the one loop radiative
correction involving a Higgs 
particle as shown in figure 1.b., that is, there is a difference between the
branching ratio of $N$ to the $\nu_L H^{0*}_{ew}$ final 
state and the branching ratio of $N$ to the $\overline{\nu_L}
H^0_{ew}$ final state. The right-handed neutrinos are
out-of-equilibrium, and hence a lepton asymmetry can be generated. The
lepton asymmetry 
is characterised by the CP violation parameter $\epsilon$ which,
assuming that the 
neutrino Dirac masses fall into a hierarchical pattern qualitatively
similar to that of the leptons and quarks, is estimated to
be \cite{Luty}  
\begin{equation}
\epsilon \simeq {m_{D_3}^2 \over \pi v^2} {M_{N1} \over M_{N2}} \sin{\delta
  } \label{eq:CP} 
\end{equation}
where $m_{D_3}$ is the Dirac mass of the third lepton generation, $v$
is the vacuum expectation value of the electroweak Higgs field $v = 
<H_{ew}> = 174$ GeV, $M_{N1}$ and $M_{N2}$ are the right-handed
neutrino Majorana masses of the first and second generation respectively and
$\delta $ is the CP violating phase.

The corresponding $B-L$ asymmetry (we use $B-L$ instead of $L$
since the ($B+L$)-violating electroweak anomaly conserves
$B-L$) must be calculated solving
Boltzmann equations which take into account all $B$, $L$ and $B+L$
violating interactions and their inverse decay rates. We can however
calculate the $B-L$ asymmetry produced taking into account only the
out-of-equilibrium decays of 
right-handed neutrinos released by decaying cosmic string loops and
assuming that 
the rates of inverse decays are 
negligible. Hence an upper limit on the baryon number  
 per commoving volume at temperature $T$ is then given by \cite{KolbTurner}
\begin{equation}
B(T) =  {1\over 2} {N_{\nu}(t) \epsilon \over s} , \label{eq:B-L}
\end{equation}
where s is the entropy at time $t$ and 
$N_\nu(t)$ is the number density of
right-handed neutrinos which have been released by 
decaying cosmic string loops at time $t$. Recall that the temperature
$T$ is related to the cosmic time $t$ via the relation  
\begin{equation}
t = 0.3 \: g_*^{-{1\over 2}} {M_{{\rm pl}} \over T^2}, \label{eq:tT}
\end{equation}
where $g_*$ counts the number of massless degrees of freedom in the
corresponding phase and $M_{{\rm pl}}$ is the Planck mass. $s$, the entropy
at time $t$, is given by 
\begin{equation}
s = {2\over 45} \pi^2 g_* T^3 .
\end{equation} 
$N_\nu(t)$
is approximately $n_\nu$ times the number density of cosmic string  
loops which have shrinked to a point at temperature $T$. Assuming that
sphaleron transitions are not in thermal equilibrium below $T_{ew}$,
and neglecting any baryon number violating processes which might have
occured below $T_{ew}$, the baryon number of the universe at
temperature $T \leq T_{ew}$ is then given by 
\begin{equation}
B = B (T_{ew}), \label{eq:today}
\end{equation}
which is also the baryon number of the universe today.
If sphaleron transitions are also rapid below $T_{ew}$, we should
include the neutrinos released below $T_{ew}$. However, below $T_{ew}$
the number density of decaying cosmic string loops is negligible, and
hence this 
would not affect the result in any sense.

The number density of decaying cosmic string loops can be estimated from
the three scales  model of Ref. \cite{ACK}. The model is based 
on the assumption that
the cosmic string network evolution is characterised by three length
scales $\xi (t)$, $\overline{\xi }(t)$ and $\chi (t)$ related to the
long string density, the persistence length along the long strings
(which is related to the fact that the typical loop size is much
smaller than $\xi$), and the small scale structure along the strings
respectively. 

Cosmic string loops lose their energy by emitting
gravitational radiation at a rate \cite{ACK}
$\dot{E} = - \Gamma_{loops} G \mu^2$
where $\mu \sim T_c^2$ is the string mass-per-unit-length and $T_c=
\eta_{B-L}$ is the critical temperature of the phase transition
leading to the string network formation. $G$ is Newton's 
constant. The numerical factor $\Gamma_{loops}\sim 50 - 100$ depends
on the loop's 
shape and trajectory, but is independent of its length. The mean size
of a loop born at $t_b$ is assumed to be $(k-1) \Gamma_{loops} G \mu
t_b$. At a later time $t$, it is then $\Gamma_{loops} G \mu (k t_b -
t)$.  The loop finally disappears at a time $t = k t_b$. Numerically,
k is found to lie between 2 and 10. 

The rate at which the string loops form in a volume $V$ is given by \cite{ACK}
\begin{equation}
\dot{N} ( t_b) = {\nu V \over (k-1) \Gamma_{loops} G \mu t_b^4}
\end{equation}
where the parameter $\nu$ can be expressed in terms of the various
length scales of 
the model which vary with time. We start with
$\xi  \sim \overline{\xi } \sim \chi $. Then $\xi $ and later
$\overline{\xi}$ will start to grow and will evolve to the scaling
regime characterised by $\xi (t)$ and $\overline{\xi }(t) \sim t
$. The length scale $\chi $ grows much less rapidly. Therefore $\nu $
varies with time. In the scaling regime, $\nu$ is estimated to lie in
the range $\nu = 0.1 - 10$ \cite{ACK}. 

Note that it has recently been shown that
cosmic string networks reach the scaling solution at a time $t_*$ much smaller 
than previously estimated \cite{Martins}. The authors of
ref.\cite{Martins} find that in the radiation dominated era   
\begin{equation}
t_* = \beta^2  f^3 {M_{{\rm pl}}^{-1} \over (G \mu)^2} 
\end{equation}
where $\beta $ is a numerical factor related
to the number of particle species interacting with the strings
(expected to be of order unity for minimal GUT strings) and $f = 0.3 \,
g_{*}^{-{1\over 2}}$ where $g_{*}$ counts the number of massless
degrees of freedom after the $H$ phase (which will usually be the
${\rm SU}(3)_c \times {\rm SU}(2)_L \times {\rm U}(1)_Y$ phase). For minimal
GUT strings $t_{*} = 8 \times 10^2 \, t_c$ \cite{Martins}, $t_c$ is
the time at 
which the strings form, and the associated temperature is $T_{*}
\simeq 10^{14.5}$ GeV. Hence our approximation for the 
rate of cosmic string loop formation is suitable; it leads to a lower
bound on the number of 
cosmic string loops. Only numerical simulations could lead us to  
a better estimate, but it is beyond the scope of this analysis.

Since it has been shown that most of the loops formed have relatively
small size, we shall assume that the number of loops
rejoining the network is negligible and thus that the number of
decaying loops is equal to the number of forming loops. Hence the
number density of right-handed neutrinos which have been released by
decaying cosmic string loops at time $t$ is given by  
\begin{equation}
N_\nu (t) = n_\nu \int_{k t_c}^t {\dot{N}(t_b) \over V} \Bigl({r(t_b)
\over r(t)}\Bigr)^3 \, dt_b 
\end{equation}
where where $r(t)$ is the cosmic scale factor and $n_\nu$ is the mean
number of right-handed neutrinos released 
by a single decaying loop. In the radiation dominated era $r(t) \sim
t^{1\over 2}$. After integration we obtain 
\begin{equation}
N_\nu (t) = {2\over 3} n_\nu \: {\nu \over (k-1) \Gamma_{loops} G
\mu} \: { 1 \over (0.3)^3 g_*^{-{3\over 2}}} \: \Bigl[{1\over k^{3\over 2}}
({T_c 
  \over M_{{\rm pl}}})^3 - ({T \over M_{{\rm pl}}})^3 \Bigr] T^3 
\end{equation}
where we have used the fact that the cosmic time $t$ is related to the
universe temperature via Eq. (\ref{eq:tT}). 
 Hence the baryon number per commoving volume today produced by the
decays of heavy Majorana right-handed neutrinos released by decaying
cosmic string loops given in Eq. (\ref{eq:B-L}) becomes   
\begin{eqnarray}
B \simeq {7.5  g_*^{1\over 2}\over ( 0.3 \: \pi)^3} \: {\nu \over (k-1)
k^{3\over 2 } \Gamma_{loops}} {T_c 
  \over M_{{\rm pl}}} \: {m^2_{D_3} \over v^2}\: {M_{N1} \over M_{N2}} \:
\sin{\delta } ,
\end{eqnarray}
where we have used Eq. (\ref{eq:CP}) for the CP violation parameter
$\epsilon$. The produced $B$ asymmetry depends on the cosmic string 
scenario parameters, on the neutrino mass matrix and on the strength
of CP violation. 

We now calculate the lower and
upper bounds on $B$ which correspond to different values of
the parameters in the cosmic string scenario. We fix the
neutrino mass matrix parameters and assume maximum CP violation,
i.e. $\sin { \delta } = 1$. We assume that the Dirac mass of
the third generation fermions lies in the range $m_{D_3} =1 - 100$
GeV, and that the ratio 
of the right-handed neutrino masses of the first and second generation
${M_{N1} \over M_{N2}} = 0.1$. With the above assumptions, we find $B$ to
lie in the range  
\begin{equation}
B \simeq (1 \times 10^{-10} - 5 \times 10^{-2}) \: g_*^{1\over 2} \:
({T_c \over M_{{\rm pl}}}) , 
\end{equation}
and we see that $B$ strongly depends on the cosmic string scenario
parameters.  Recall that the baryon number-to-photon ratio
${n_B \over n_\gamma}$ is related to the baryon number of the universe
$B$ by ${n_B \over n_\gamma} = 1.80 \, g_* \, B$. Hence, if $g_* 
\simeq 3.4$, our mechanism alone can explain the baryon-number-to-photon
ratio predicted by nucleosynthesis, ${n_B \over n_\gamma} = (2-7)
\times 10^{-10}$,  
with the $B-L$ breaking scale in the range 
\begin{equation}
\eta_{B-L} =  (1 \times 10^{6} - 2 \times 10^{15})  \; {\rm GeV} . 
\end{equation}
We point out that the result could be better
calculated solving Boltzmann equations, which take into account all $B$,
$L$, and $B+L$ violating interactions and do not neglect the inverse decay
rates. Furthermore, the rate of decaying cosmic string loops can be
calculated via numerical simulations, which would have to take into
account the different regimes of the network evolution which occur
during and after the friction dominated era. This may change the
allowed value for the $B-L$ breaking scale by a few orders of 
magnitude. Note also that if CP is not maximally violated, the $B-L$
breaking scale will be shifted towards higher values. Finally, we
recall that when a cosmic string loop decays, it also releases massive
Higgs bosons $\phi_{B-L}$ and massive gauge bosons $a$ which can decay
into right-handed neutrinos. This process is not taken into account
here because the masses of the Higgs and gauge bosons and of the
neutrinos are very close to each other and the Higgs and gauge fields
can also decay into other particles.

\section{Conclusions}

Unified theories with rank greater or equal to five containing an extra gauge ${\rm
U}(1)_{B-L}$ symmetry which predict heavy Majorana right-handed
neutrinos and $B-L$ cosmic strings are good candidates for baryogenesis.
$B-L$ cosmic strings are cosmic strings which form at the $B-L$ breaking scale so that the Higgs field used to break $B-L$ is also the Higgs field forming the strings. It is the same Higgs field which gives a heavy Majorana mass to the right-handed neutrinos. There are therefore right-handed neutrinos trapped as transverse zero modes in the core of $B-L$ cosmic strings. The
out-of-equilibrium decays of Majorana right-handed neutrinos released by
decaying cosmic string loops produce a lepton asymmetry which
is then converted into a baryon asymmetry via sphaleron transitions. 
 
The above scenario works for a wide range of parameters. But we have only made a quantitative analysis and therefore a more detailed study needs to be done. Numerical simulations for 
cosmic string network evolution should be used and full Boltzmann's 
equations should be solved.

\chapter{Conclusions and perspectives}
\label{chap-chap6}

The most appealing feature of particle physics-cosmology is that it
may give a reasonable description of the evolution of the universe
from very early times until today, and it is a testable
theory. The best recent success of the big-bang cosmology is perhaps
the perfect black-body spectrum of the cosmic background radiation
measured by COBE \cite{COBE1}. The particle physics standard model has
been widely tested at high energy colliders, and only one predicted
particle, namely the Higgs boson, is missing
\cite{partphysbook}. There is not yet direct experimental evidence for
physics beyond the standard model, but experiments are being
undertaken in search of exotic physics such as supersymmetry,
neutrino masses, and proton decay. The evidence for new physics may
come up soon. The best 
support for the existence of physics beyond the standard model are 
baryogenesis, the solar and atmospheric neutrinos problems, which
require non-zero neutrinos masses, the high energy gauge coupling
constants which require supersymmetry in order to merge 
in a single point, the necessity of non-baryonic dark-matter, as
required by nucleosynthesis, and the requirements of both hot and cold
dark-matter to explain structure formation. This thesis is concerned
with a small part of this broad subject.

In Chap. \ref{chap-chap2}, the scattering of fermions off cosmic
strings arising in a nonsupersymmetric SO(10) GUT 
model and baryon number violation processes due to the couplings of
fermions to GUT gauge bosons in the string core  were studied. The
elastic cross-sections 
were found to be Aharonov-Bohm type with a marked asymmetry for left
and right-handed fields. The catalysis cross-sections were 
found to be quite small, unlike previous toy model calculations
\cite{Perkins91} which suggested that they could be of the order of a 
strong interaction cross-section. The Callan-Rubakov effect is
suppressed, so these strings cannot catalyse proton decay. It 
is unlikely that grand unified cosmic strings could
erase a primordial baryon asymmetry as suggested in Ref. \cite{Branden89}.
 Also, if cosmic strings and the Aharonov-Bohm effect for cosmic
strings were observed, it could help tie down the underlying gauge 
group.

 In Chap. \ref{chap-chap3},
it was argued that 
supersymmetric 
SO(10) is a very attractive GUT, from both particle physics and
cosmological point of views. SO(10) is the minimal GUT which unifies all
kinds of matter and does not include any exotic particle, except a
right-handed neutrino. Also, supersymmetric SO(10)
models can solve many of the outstanding problems, like the gauge
hierarchy problem, the question of fermion masses and the solar
neutrino problem. It was shown that the conditions for topological defect formation in supersymmetric theories follow from those well-known in 
nonsupersymmetric ones. They are not affected by the presence of supersymmetry. By studying the formation of topological defects in all possible spontaneous
symmetry breaking (SSB) patterns from supersymmetric SO(10) down to
the standard  
model gauge group, and also using an inflationary scenario as described in Chap. \ref{chap-chap4}, and by requiring that the model be consistent with proton lifetime 
measurements,  we were able to show that only three of the SSB patterns were consistent with observations. SO(10) can either be broken via ${\rm SU}(3)_c\times 
{\rm SU}(2)_L\times {\rm SU}(2)_R\times {\rm U}(1)_{B-L}$, here SO(10) 
must be broken with a combination of a 45-dimensional Higgs 
representation and a 54-dimensional one, or  
via ${\rm SU}(3)_c\times 
{\rm SU}(2)_L\times {\rm U}(1)_R\times {\rm U}(1)_{B-L}$. In both cases,  
the intermediate symmetry group must be broken down to the standard 
model gauge group with unbroken matter parity, ${\rm SU}(3)_c\times 
{\rm SU}(2)_L\times {\rm U}(1)_Y\times Z_2$. In 
supergravity models, the breaking of SO(10) via flipped {\rm SU}(5) is also possible.

Using the above results, a specific supersymmetric SO(10) model  
was built. It was described in
Chap. \ref{chap-chap4}. 
The SO(10) symmetry group is broken down to the standard model 
gauge group, via an intermediate ${\rm SU}(3)_c\times {\rm SU}(2)_L\times 
{\rm U}(1)_R\times 
{\rm U}(1)_{B-L}$ symmetry. The model gives rise to a false vacuum hybrid 
inflationary scenario which solves the monopole problem. It is argued that false vacuum hybrid 
inflation is generic in supersymmetric SO(10) models. It 
arises naturally from the theory. Neither an external field nor an 
external symmetry has to be added. In our specific model, cosmic strings form. There are right-handed neutrinos
trapped as transverse zero modes in the core of the string. The
strings are not superconducting. The model produces a stable LSP and a very light left-handed neutrino which may serve as the cold and hot
dark-matter. A mixed cosmic strings and inflationary large scale structure scenario was briefly discussed.

In Chap. \ref{chap-chap5}, a new mechanism for leptogenesis was
described. The basic idea is that in unified theories with rank 
greater or equal to five which contain an extra ${\rm U}(1)_{B-L}$ symmetry
and  predict heavy Majorana right-handed neutrinos, 
cosmic strings may form as consequence of ${\rm U}(1)_{B-L}$
breaking. The Higgs 
field forming the strings is the Higgs field which breaks $B-L$ and it is
also the Higgs field which gives a Majorana mass to the right-handed
neutrinos. Due the winding of the Higgs around the strings,
there are right-handed neutrinos trapped as transverse zero modes in the
core of these strings.  When cosmic string loops 
decay they release these right-handed neutrinos. This is an
out-of-equilibrium process. The released neutrinos acquire heavy
Majorana mass and decay into massless leptons and   
electroweak Higgs particles to produce a lepton asymmetry. The lepton
asymmetry is converted into a baryon asymmetry via sphaleron
transitions. This mechanism could easily be implemented in the
supersymmetric SO(10) model 
described in Chap. \ref{chap-chap4}.

The results of Chapters \ref{chap-chap4} and \ref{chap-chap5} confirm
that supersymmetric SO(10) is a good GUT candidate. Of course, there
are other interesting GUTs, such as the trinification 
$({\rm {\rm SU}}(3))^3$, left-right models or E(6). It seems also that
only GUTs 
with rank greater or equal to five make 
sense phenomenologically.  We did not discuss gravity. To include
gravity, we should turn to supergravity or superstring
models. However, it is interesting to note that both global 
and local supersymmetric GUTs can be derived from string theories, and
hence, since we are working at energies below the string unification
scale, it makes sense not to  include gravity in our models.

 We have studied a mechanism for inflation in global
supersymmetric GUT models which is `natural'. There are many more to be
found which will probably lead to different low energy
phenomenology. In the model 
of Ref. \cite{Laz} for example, topological defects do not form at the
end of inflation, but baryogenesis via leptogenesis naturally
occurs. Inflation in supergravity models \cite{sugrainfl} is also the
next step to turn to. Inflation in supersymmetric models is usually
driven by the VEV of a 
$F$-term. Recently, $D$-term inflation has been proposed
\cite{Dvali}. $D$-term
inflation requires further study.

The baryon asymmetry produced by the leptogenesis scenario presented in
Chap. \ref{chap-chap5}, 
has been calculated 
assuming that the inverse decay rates were negligible and neglecting
all other $B$ and $L$ violating processes which may have occured. We
have also assumed that the cosmic string network scaling regime was a
good approximation to the study of the string network from very early
times. A more detailed study can be done, solving Boltzmann's equations and using numerical
simulations for cosmic string network evolution which would take into
account the different regimes. 

The scattering of fermions and the superconducting properties of
cosmic strings have been widely studied \cite{ShelVil}. It would be a
good idea to study cosmic strings arising from a supersymmetric
phase transition, before and after supersymmetry breaking. It
would be interesting to find signatures of low energy (below the
supersymmetry breaking scale) supersymmetric GUT strings, to
distinguish them from the usual ones.

Finally, particle physics-cosmology is an intriguing and fascinating
field of research and has a very promising future. New experimental
results in the search for 
physics beyond the  
standard model and new observations of the CBR anisotropy 
certainly will soon  have strong impact on particle physics and
cosmology.

\newpage
\thispagestyle{empty}
\cleardoublepage
\thispagestyle{empty}
\appendix

\chapter{Brief review of {\rm SO}(10)}
\label{sec-so10}

The fundamental representation of {\rm SO}(10) consists of
10 generalised gamma matrices. They can be written in
an explicit notation, in terms of cross products, 
\begin{eqnarray}
\Gamma_1 &=& \sigma_1 \times \sigma_3 \times \sigma_3 \times \sigma_3 \times \sigma_3 \nonumber\\
\Gamma_2 &=& \sigma_2 \times \sigma_1 \times \sigma_3 \times \sigma_3 \times \sigma_3 \nonumber\\
\Gamma_3 &=& I \times \sigma_1 \times \sigma_3 \times \sigma_3 \times \sigma_3 \nonumber\\
\Gamma_4 &=& I \times \sigma_2 \times \sigma_3 \times \sigma_3 \times \sigma_3 \nonumber\\
\Gamma_5 &=& I \times I \times \sigma_1 \times \sigma_3 \times \sigma_3 \nonumber\\
\Gamma_6 &=& I \times I \times \sigma_2 \times \sigma_3 \times \sigma_3 \nonumber\\
\Gamma_7 &=& I \times I \times I \times \sigma_1 \times \sigma_3 \nonumber\\
\Gamma_8 &=& I \times I \times I \times \sigma_2 \times \sigma_3 \nonumber\\
\Gamma_9 &=& I \times I \times I \times I \times \sigma_1 \nonumber\\
\Gamma_{10} &=& I \times I \times I \times I \times \sigma_2
\end{eqnarray}
where the $\sigma_i$ are the Pauli matrices and I denotes
the two dimensional identity matrix. They generate a
Clifford algebra defined by the anti-commutation rules 
\begin{equation}
\{ \Gamma_i , \Gamma_j \} = 2 \, \delta_{ij} \, \, \, \: \: i = 1,...,10 \; .
\end{equation}
One can define the chirality operator $\chi$, which is the generalised $\gamma_5$ of the standard model by
\begin{equation}
\chi = (-i)^5 \prod_{i = 1}^{10} \Gamma_i \; .
\end{equation}
In terms of the cross-product notation, $\chi$ has the form,
\begin{equation}
\chi= \sigma_3 \times \sigma_3 \times \sigma_3 \times \sigma_3 \times \sigma_3 \; .
\end{equation}
The 45 generators of {\rm SO}(10) are also given in terms of
the generalised gamma matrices 
\begin{equation}
M_{ab} = {1 \over 2i} \, [\Gamma_i, \Gamma_j] \, \, \, i,j = 1... 10 \; .
\end{equation}
They are antisymmetric, purely imaginary $32 \times
32$ matrices. One can write the diagonal M, 
\begin{eqnarray}
M_{12}&=& {1\over 2}\, \sigma_3 \times I \times I \times I
\times I \nonumber\\ 
M_{34}&=& {1\over 2}\, I \times \sigma_3 \times I \times I
\times I \nonumber\\ 
M_{56}&=& {1\over 2} \,I \times I \times \sigma_3 \times I
\times I \nonumber\\ 
M_{78}&=& {1\over 2}\, I \times I \times I \times \sigma_3
\times I \nonumber\\ 
M_{910}&=& {1\over 2}\, I \times I \times I \times I
\times \sigma_3  \; . 
\end{eqnarray}
In {\rm SO}(N) gauge theories fermions are conventionally
assigned to the spinor representation. For N even, the
spinor representation is $2^{N\over 2}$ dimensional
and decomposes into two equivalent spinors of
dimension $2^{{N\over 2} - 1}$ by means of the
projection operator $P = {1 \over 2} \, (1 \pm \chi)$,
where 1 is the $2^{N\over 2} \times  2^{N\over 2}$
identity matrix. Thus {\rm SO}(10) has got two irreducible
representations, 
\begin{equation}
\sigma^\pm = {1 \pm \chi \over 2}
\end{equation}
of dimension 16. Therefore {\rm SO}(10) enables us to put
all the fermions of a given  family in the same
spinor. Indeed, since each family contains eight
fermions, we can put all left and right handed
particles of a given family in the same 16 dimensional
spinor. This is the smallest grand unified group which
can do so. However, gauge interactions conserve
chirality. Indeed, 
\begin{equation}
\bar{\psi} \gamma_\mu A^\mu \psi = \bar{\psi_L} \gamma_\mu
A^\mu \psi_L + \bar{\psi_R} \gamma_\mu A^\mu \psi_R  \;
. \nonumber 
\end{equation}
Therefore $\psi_L$ and $\psi_R$ cannot be put in the
same irreducible representation. Hence, instead of
choosing $\psi_L$ and $\psi_R$, we chose $\psi_L$ and
$\psi_L^c$. The fields $\psi_L$ and $\psi_L^c$
annihilate left-handed particles and antiparticles,
respectively, or create right-handed antiparticles and
particles. As mentioned in Sec. \ref{sec-Majorana}, the fields $\psi_L$ and $\psi_L^c$ are
related to the fields $\psi_R$ and $\bar{\psi_R}$ by
the following relations, 
\begin{eqnarray}
\psi_L^c &\equiv & P_L \psi^c = P_L C \bar{\psi}^T = C
(\bar{\psi} P_L )^T = C \bar{\psi_R}^T = C \gamma_0^T
\psi_R^* \\
\bar{\psi_L^c} &\equiv & \psi_L^{c \dagger} \gamma_0 =
\psi_R^{* \dagger} \gamma_0^{T \dagger} C^\dagger \gamma_0 = -
\psi_R^T C^{-1} = \psi_R^T C 
\end{eqnarray}
where the projection operators $P_{L,R} = {\frac{1} {2}} (1
\pm \gamma_5)$ and C is the usual charge conjugation
matrix. For the electron family we get, 
\begin{equation}
\Psi^{(e)}_L = (\nu_{(e)}^c\, , u^c_r\, , u^c_y\, ,
u^c_b\, , d_b\, , d_y\, , d_r\, , e^-\, , u_b\, ,
u_y\, , u_r\, , \nu_{(e)}\, , e^+\, , d^c_r\, ,
d^c_y\, , d^c_b)_L\\ \label{eq:psiL} 
\end{equation}
where the upper index c means conjugate, and the
sub-indices refer to quark colour. We find similar
spinor $\Psi^{(\mu)}$ and $\Psi^{(\tau)}$ associated
with the $\mu$ and the $\tau$ family respectively: 
\begin{eqnarray}
\Psi^{(\mu)} &=& (\nu_{(\mu)}^c\, , c^c_r\, , c^c_y\,
, c^c_b\, , s_b\, , s_y\, , s_r\, , \mu^-\, , c_b\, ,
c_y\, , c_r\, , \nu_{(\mu)}\,, \mu^+\, , s^c_r\, ,
s^c_y\, , s^c_b)_L \nonumber \\ 
\Psi^{(\tau)} &=& (\nu_{(\tau)}^c\, , t^c_r\, ,
t^c_y\, , t^c_b\, , b_b\, , b_y\, , b_r\, , \tau^-\, ,
t_b\, , t_y\, , t_r\, , \nu_{(\tau)}\,, \tau^+\, ,
b^c_r\, , b^c_y\, , b^c_b)_L \; .  
\end{eqnarray}

\chapter{Scattering amplitude calculations}

In this appendix, we give the technical details of the external and
internal solutions calculations given in Chap. \ref{chap-chap2}. We also give a discussion of the
matching conditions at the core radius. 

\section{The external solution}
\label{sec-extap}

We want to solve equations (\ref{eq:ext}). We set
$\partial_t = -i \omega$, where $\omega$ is the energy of the
electron and take the usual Dirac representation $e_L
= (0,\xi_e)$ , $e_R = (\chi_e,0)$ , $q^c_L =
(0,\xi_q)$ and $q^c_R = (\chi_q,0)$. We use the usual
mode decomposition for the spinors $\xi_q$, $\xi_e$,
$\chi_q$ and $\chi_e$ : 
\begin{eqnarray}
\chi_{(e, {q^c})}(r,\theta) &=& \sum_{n=-\infty}^{n=+\infty}
\left ( \begin{array}{rl}
	&\chi_{1\, (e, {q^c})}^n (r)\\
         i &\chi_{2\, (e, {q^c})}^n (r) \, e^{i\theta}
\end{array}
\right )
e^{in\theta} \nonumber\\
\xi_{(e, {q^c})}(r,\theta) &=& \sum_{n=-\infty}^{n=+\infty}
\left ( \begin{array}{rl}
	&\xi_{1\, (e, {q^c})}^n (r)\\
         i &\xi_{2 \, (e, {q^c})}^n (r) \, e^{i\theta}
\end{array}
\right )
e^{in\theta} \; . 
\end{eqnarray}
 Then, using the basis,
\begin{equation}
\gamma^j = \left ( \begin{array} {lr}
              0 & -i\sigma^j\\
	      i \sigma^j & 0
		\end{array}
	\right )
\end{equation}
the equations of motion (\ref{eq:ext}) become,
\begin{eqnarray}
\left. \begin{array}{cccccccccc}
\omega \chi_{1,(e, {q^c})}^n &-& ( {d\over dr} &+& { n+1
\over r} &-& {\tau_{str}^{R\, (e, {q^c})} \over r} )
&\chi_{2,(e, {q^c})}^n &=& 0 \\ [0.1cm] 
\omega \chi_{2,(e, {q^c})}^n &+& ( {d\over dr}  &-& { n
\over r} &+& { \tau_{str}^{R \, (e, {q^c})} \over r} )
& \chi_{1,(e, {q^c})}^n &=& 0  \\ [0.1cm] 
\omega \xi_{1,(e, {q^c})}^n &+& ( {d\over dr} &+& { n+1
\over r} &-& {\tau_{str}^{L \, (e, {q^c})} \over r} )
& \xi_{2,(e, {q^c})}^n &=& 0  \\ [0.1cm] 
\omega \xi_{2,(e, {q^c})}^n &-& ( {d\over dr} &-& { n
\over r} &+& {\tau_{str}^{L \, (e, {q^c})}\over r} ) &
\xi_{1,(e, {q^c})}^n &=& 0 \label{eq:mot} 
	\end{array}
	\right. \; .
\end{eqnarray}
It is easy to show that the fields $\xi_{1,(e,
{q^c})}^n$, $\xi_{1,(e, {q^c})}^n$, $\chi_{1,(e,
{q^c})}^n$ and $\chi_{2,(e, {q^c})}^n$ satisfy Bessel
equations of order $n -  \tau_{str}^{R\, (e, {q^c})}$,
$n + 1 -  \tau_{str}^{R\, (e, {q^c})}$, $n -
{\tau_{str}^{L \, (e, {q^c})}}$ and $n -
\tau_{str}^{R\, (e, {q^c})}$ respectively. Hence the
external solution is, 
\begin{eqnarray}
&&	\left ( \begin{array}{l}
\xi_{(e, {q^c})} (r,\theta)\\
\chi_{(e, {q^c})} (r, \theta)
	\end{array}
	\right )
 = \\
&& \sum_{n=-\infty}^{n=+\infty}
	\left ( \begin{array}{rlcrl}
&(v_n^{(e, {q^c})} Z^1_{n - \tau_{str}^{R\, (e,
{q^c})}} (\omega r) &+& v_n^{(e, {q^c})'} Z^2_{n -
\tau_{str}^{R\, (e, {q^c})}} (\omega r)) & e^{i n \theta}
\nonumber\\ 
i&(v_n^{(e, {q^c})} Z^1_{n + 1 - \tau_{str}^{R\, (e,
{q^c})}} (\omega r) & +& v_n^{(e, {q^c})'} Z^2_{n + 1 -
\tau_{str}^{R\, (e, {q^c})}} (\omega r)) & e^{i (n + 1)
\theta} \nonumber\\ 
&( w_n^{(e, {q^c})} Z^1_{n - \tau_{str}^{L \, (e,
{q^c})}} (\omega r) & +& w_n^{(e, {q^c})'} Z^2_{n -
\tau_{str}^{L \, (e, {q^c})}} (\omega r)) & e^{i n \theta}
\nonumber\\ 
i &(w_n^{(e, {q^c})} Z^1_{n + 1 - \tau_{str}^{L \, (e,
{q^c})}} (\omega r) &+& w_n^{(e, {q^c})'} Z^2_{n + 1 -
\tau_{str}^{L \, (e, {q^c})}} (\omega r)) & e^{i (n + 1)
\theta} 
       \end{array}
	\right ) \; \nonumber .
\end{eqnarray}
The order of the Bessel functions will always be
fractional. We therefore take $Z^1_\nu = J_\nu$ and
$Z^2_\nu = J_{-\nu}$.

\section{The internal solution}
\label{sec-intap}

We get solutions for fields which are linear
combinations of the quark and electron fields. Indeed,
we get solutions for the fields $\sigma^\pm = \xi_q \pm
\xi_e$ and $\rho^\pm = \chi_q \pm \chi_e$. Using the
mode decomposition (\ref{eq:mode}), the upper
components of the fields $\rho^\pm$ and  $\sigma^\pm$ are
respectively $\rho_{n1}^\pm =  \chi_{1\, {q^c}}^n \pm
\chi_{1\, e}^n$ and $\rho_{n2}^\pm =  \chi_{2\, {q^c}}^n
\pm \chi_{2\, e}^n$ whilst the lower components are
$\sigma_{n1}^\pm = \xi_{1\, {q^c}}^n \pm \xi_{1\, e}^n$
and $\sigma_{n2}^\pm =  \xi_{2\, {q^c}}^n \pm \xi_{2\,
e}^n$ respectively. The equations of motions
(\ref{eq:fint}) become 
\begin{eqnarray}
\left.   \begin{array} {cccccccccc}
\omega \rho_{n1}^\pm & - &({d\over dr} & + & {n+1 \over r}
&\mp & e A')&\rho_{n2}^\pm &=&0 \\ [0.2cm] 
\omega \rho_{n2}^\pm & + &({d\over dr} &-& {n \over r} &\pm
& e A' )&\rho_{n1}^\pm &=&0  \\ [0.2cm] 
\omega \sigma_{n1}^\pm &+& ({d\over dr} &+& {n+1 \over r} &\mp
& e A )&\sigma_{n2}^\pm &=&0 \\ [0.2cm] 
\omega \sigma_{n2}^\pm &-& ({d\over dr} &-& {n \over r} &\pm &
e A ) &\sigma_{n1}^\pm &=&0 
\end{array}  
	\right. \; . \label{eq:int}
\end{eqnarray}
Combining the two first equations of
 ({\ref{eq:int}), one can see that $\rho_{n1}^\pm$
satisfy an hyper-geometric equation giving, 
\begin{equation}
\rho_{n1}^\pm = (kr)^{|n|} e^{-ikr} \sum_{j =
0}^{n=+\infty} \alpha^\pm_j {(2ikr)^j \over j!} 
\end{equation}
where $k^2 = \omega^2 - (e A)^2$, $e = {g \over 2
\sqrt{2}}$. $\alpha^\pm_{j+1} = {(a^\pm + j) \over (b
+ p)} \alpha^\pm_j$ with $a^\pm = {1\over 2} + |n| \pm
{e A (2n+1) \over 2ik}$ and $b = 1 +
2|n|$. $\rho_{n2}^\pm$ can be obtained using the coupled
equation (\ref{eq:int}.2). We find 
\begin{equation}
\rho_{n2}^\pm = - {1\over \omega} (kr)^{|n|} e^{-ikr} \sum_{j
= 0}^{n=+\infty} \alpha^\pm_j {(2ikr)^j \over j!} \,
\,({|n| - n \over r} -ik + {j\over r} \pm e A) \; . 
\end{equation}
$\sigma_{n2}^\pm$ are also solutions of hyper-geometric
equations, and using the coupled equation
(\ref{eq:int}.4) we get, 
\begin{eqnarray}
\sigma_{n1}^\pm &=& (kr)^{|n|} e^{-ikr} \sum_{j =
0}^{n=+\infty} \beta^\pm_j {(2ikr)^j \over j!} \\ 
\sigma_{n2}^\pm &=& - {1\over \omega} (kr)^{|n|} e^{-ikr}
\sum_{j = 0}^{n=+\infty} \beta^\pm_j {(2ikr)^j \over
j!} \, \,({|n| - n \over r} -ik + {j\over r} \pm e A')
\end{eqnarray}
where $k^2 = \omega^2 - (e A')^2$, $\beta^\pm_{j+1} =
{(c^\pm + j) \over (b + p)} \beta^\pm_j$ with $c^\pm =
{1\over 2} + |n| \pm {e A' (2n+1) \over 2ik}$. 
And the internal solution is,
\begin{equation}
\left ( \begin{array} {rl}
	&\rho_{n1}^\pm \, e^{in\theta}\\
	i & \rho_{n2}^\pm \, e^{i(n+1)\theta}\\
	&\sigma_{n1}^\pm \, e^{in\theta}\\
	i & \sigma_{n2}^\pm \, e^{i(n+1)\theta}
	\end{array}
	\right ) \; .
\end{equation}
Therefore the internal solution is giving by a linear
combination of the quark and electron fields.

\section{The matching conditions}
\label{sec-matchap}

The continuity of the solutions at $r=R$ lead to,
\begin{eqnarray}
\lefteqn{(kR)^{|n|} \, e^{-ikR} \sum_{j =
0}^{n=+\infty} \alpha^\pm_j \, {(2ikR)^j \over j!}}
\label{eq:matchun} \nonumber \\ 
&& = \: (v_n^q \pm v_n^{e }) J_{n - \tau_R } (\omega R) \:
+ \: (v_n^{q'} \pm v_n^{e'}) J_{-(n - \tau_R)} (\omega R) 
\end{eqnarray}
\begin{eqnarray}
\lefteqn{- {1\over w} \,  (kR)^{|n|} \, e^{-ikR}
\sum_{j = 0}^{n=+\infty} \alpha^\pm_j \, {(2ikr)^j
\over j!} \, \,({|n| - n \over R} -ik + {j\over R} \pm
e A)} \nonumber \\ 
&& = \: (v_n^q \pm v_n^{e }) J_{n + 1 - \tau_R} (\omega R)
\:  + \: (v_n^{q'} \pm v_n^{e'}) J_{-(n + 1 - \tau_R)}
(\omega R) \; . \label{eq:matchde} 
\end{eqnarray}
Nevertheless, we will have discontinuity of the first
derivatives. Indeed, inside we have 
\begin{eqnarray}
\left.   \begin{array} {cccccccccc}
\omega \rho_{n1}^\pm & - &({d\over dr} & + & {n+1 \over r}
&\mp & e A')&\rho_{n2}^\pm &=&0 \\ 
\omega \rho_{n2}^\pm & + &({d\over dr} &-& {n \over r} &\pm
& e A' )&\rho_{n1}^\pm &=&0 
\end{array}  
	\right.
\end{eqnarray}
whereas outside we have
\begin{eqnarray}
\left. \begin{array}{cccccccccc}
\omega (\chi_{1,{q^c}}^n \pm \chi_{1,e}^n) &-& ( {d\over
dr} &+& { n+1 \over r} &-& {\tau_{str}^{R\, (e,
{q^c})} \over r} ) &( \chi_{2,{q^c}}^n \pm
\chi_{2,e}^n) &=& 0 \\ 
\omega (\chi_{2,q}^n \pm \chi_{2,e}^n) &+& ( {d\over dr}
&-& { n \over r} &+& { \tau_{str}^{R\, (e, {q^c})}
\over r} ) & (\chi_{1,{q^c}}^n \pm \chi_{1,e}^n) &=& 0

	\end{array}
	\right. \; .
\end{eqnarray}
Now,
\begin{eqnarray}
(\chi_{1,{q^c}}^n \pm \chi_{1,e}^n)^{out} &=&
\rho_{n1}^{\pm \: in} \\ 
(\chi_{2,{q^c}}^n \pm \chi_{2,e}^n)^{out} &=&
\rho_{n2}^{\pm \: in} 
\end{eqnarray}
giving us the relations for the first derivatives,
\begin{eqnarray}
({d\over dr} \mp eA ) \, \rho_{n2}^{\pm \: in} &=&
({d\over dr} - {\tau_{str}^{R\, (e, {q^c})} \over R})
\, (\chi_{2,{q^c}}^n \pm \chi_{2,e}^n)^{out}
\label{eq:matchdif}\\ 
({d\over dr} \pm eA ) \, \rho_{n1}^{\pm \: in} &=&
({d\over dr} + {\tau_{str}^{R\, (e, {q^c})} \over R})
\, (\chi_{1,{q^c}}^n \pm \chi_{1,e}^n)^{out} \; . 
\end{eqnarray}
Dividing equation (\ref{eq:matchun}) by equation
(\ref{eq:matchde}) or either replacing equation
(\ref{eq:matchun}) in equation (\ref{eq:matchdif}), we
get the following relations 
\begin{equation}
{v_n^{q'} \pm v_n^{e'} \over v_n^{q} \pm v_n^{e}} = {
w \, \lambda_n^\pm J_{n + 1 - \tau_R}(\omega R) + J_{n -
\tau_R}(\omega R) \over w \, \lambda_n^\pm J_{-(n + 1 -
\tau_R)}(\omega R) + J_{-(n - \tau_R)}(\omega R)}  
\end{equation}
where 
\begin{equation}
\lambda_n^\pm = {\sum_{j = 0}^{n=+\infty} \alpha^\pm_j
{(2ikr)^j \over j!} \over \sum_{j = 0}^{n=+\infty}
\alpha^\pm_j {(2ikr)^j \over j!} ({|n| - n \over r}
-ik + {j\over r} \pm e A)} \; . 
\end{equation}

\chapter{Minimising the superpotential}
\label{chap-chapappC}

In this Appendix, we find the true minimum of the superpotential of 
the model studied in Chap. \ref{chap-chap4}. We calculate the $F$-terms 
and find the VEVs of the Higgs 
fields which correspond to the global minimum.

The full superpotential of the model is given by Eq. (\ref{eq:super}),
\begin{eqnarray}
W &=& m_A A^2 + m_S S^2 + \lambda_S S^3 + \lambda_A A^2 S + H A H' +
m_{H'} {H'}^2 \nonumber \\ 
&& + m_{A'} {A'}^2 + m_{S'} {S'}^2 + \lambda_{S'} {S'}^3 +
\lambda_{A'} {A'}^2 {S'} \nonumber \\ 
&& + \alpha {\cal{S}} \overline{\Phi} \Phi - \mu^2 {\cal{S}}  \; .
\end{eqnarray}
where $S$ and $S'$ are 54-dimensional Higgs representations and are  
traceless second rank symmetric tensors. The $S$ and $S'$ Higgs field must implement the Dimopoulos-Wilczeck meachanism \cite{DW}. Therefore, in the 10-dimensional representation 
of {\rm SO}(10) they are of the form \cite{babuandbarr}, with appropriate subscripts,
\begin{equation}
<\! S_{54}\! > = I \otimes {\rm diag}(x,x,x,-{3\over 2} x,-{3\over 2} x) \label{eq:vevs}
\end{equation}
where $I$ is the $2 \times 2$ unitary matrix and $x \sim M_{\rm GUT}$ and
\begin{equation}
<\! S'_{54}\! > = I \otimes {\rm diag}(x,'x',x',-{3\over 2} x',-{3\over 2} x')
\label{eq:vevsp} 
\end{equation}
where $x' \sim M_{{\rm GUT}}$ and are determined by the vanishing condition of the $F$ terms. 
The Higgs $A_{45}$ and $A'_{45}$ are 45-dimensional representations and must be in 
the $B-L$ and $T_{3R}$ directions respectively (see Sec. 
\ref{sec-building}). Therefore in the 10-dimensional representation of 
{\rm SO}(10) $A_{45}$ and $A'_{45}$ are antisymmetric and are given by
\begin{equation}
<\! A_{45}\! > = J \otimes diag(a,a,a,0,0) \label{eq:veva}
\end{equation}
where $J = \left (
\begin{array} {cc}
0 & 1 \\
-1 & 0
\end{array}
\right )
$ and $a \sim M_{{\rm GUT}}$ and
\begin{equation}
<\! A'_{45}\! > = J \otimes diag(0,0,0,a',a') . \label{eq:vevap}
\end{equation}
where $a' \sim M_{{\rm GUT}}$. The $\Phi$ and $\overline{\Phi}$ fields
are Higgs fields in  
the ${\bf 126}$ and ${\bf \overline{126}}$-dimensional representations. The
$\Phi$ and $\overline{\Phi}$ fields must break the ${\rm U}(1)_X$
symmetry which commutes with {\rm SU}(5), and thus acquire VEVs in the
right-handed neutrino direction. From the vanishing condition for the
$D$ terms, $< \! \Phi \! > = <\! \overline\Phi \! >$ and thus, with appropriate
subscripts, 
\begin{equation}
<\! \Phi_{126}\! >_{\nu^c \nu^c} = <\! \overline{\Phi}_{126}\! >_{\overline{\nu^c}
\overline{\nu^c}} = d \label{eq:vevphi} 
\end{equation}
where $d \sim M_{\rm G}$. We do not take into account the 10-multiplets 
which are used to break the standard model gauge group, since we are 
interesting in what is happening at much higher energies. Their VEVs do not affect the VEVs of the other Higgs fields.

The true vacuum, with unbroken supersymmetry, corresponds to $F$ terms vanishing. Using the same notation for the scalar component than for
the superfield, the $F$ terms are given by 
\begin{eqnarray}
F_A &=& 2 m_A A + 2 \lambda_A A S ,\\
F_S &=& 2 m_S S + 3 \lambda_S S^2 + \lambda_A A^2 ,,\\
F_{A'} &=& 2 m_{A'} A' + 2 \lambda_{A'} {A'} {S'} \\
F_{S'} &=& 2 m_{S'} {S'} + 3 \lambda_{S'} {S'}^2 + \lambda_{A'} {A'}^2 ,\\
F_\Phi &=& \alpha {\cal{S}} \overline{\Phi} ,\\
F_{\overline{\Phi}} &=& \alpha {\cal{S}} \Phi, \\
F_{{\cal{S}}} &=& \alpha \overline{\Phi} \Phi - \mu^2 .
\end{eqnarray}
Using the VEVs of the Higgs fields given above and also the vanishing
conditions for the $F$ terms, we get the following relations, for 
each
term respectively:
\begin{eqnarray}
 m_A a + 2 \lambda_A a x &=& 0 \label{eq:une},\\
 - m_S x + {3 \over 4} \lambda_S x^2 + {1 \over 5} \lambda_A a^2 &=& 0,
\\ 
2 m_{A'} a' - 3 \lambda_{A'} a' x' &=& 0 ,\\
-m_{S'} x' + {3 \over 4} \lambda_{S'} {x'}^2 - \lambda_{A'} {a'}^2 &=&
0 ,\\ 
\alpha s d &=& 0 , \\
\alpha d^2 - \mu^2 &=& 0 . \label{eq:deux}
\end{eqnarray} 
where $s = |{\cal S}|$. We note that the roles 
of the 54 dimensional representations $S_{54}$ and $S'_{54}$ are to force the 
adjoint $A_{45}$ and $A'_{45}$ into $B-L$ and $T_{3R}$ directions. With the VEVs chosen above, see Eqs. (\ref{eq:vevs})-(\ref{eq:vevphi}), if $s=0$ and $d={\mu \over \sqrt{\alpha}}$ the potential 
has a global
minimum, such that the {\rm SO}(10) symmetry is broken down to ${\rm SU}(3)_c
\times {\rm SU}(2)_L \times {\rm U}(1)_R \times {\rm U}(1)_{B - L}$ and supersymmetry is
unbroken and we have $x = {2 m_{A} \over 3 \lambda_{A}}$ and $x' = 
{2 m_{A'} \over 3 \lambda_{A'}}$.  $a \sim M_{{\rm GUT}}$, $a' \sim M_{{\rm GUT}}$, 
and ${\mu \over \sqrt{\alpha}} \sim
M_{\rm G}$, where $M_{\rm G} \sim 10^{15 - 16}$ GeV and $M_{\rm G} \leq M_{{\rm GUT}} 
\leq M_{\rm pl}$
and $M_{\rm pl}$ is the Planck mass $\sim 10^{19}$ GeV.

\cleardoublepage


\begin{thebibliography}{99}

\bibitem{paper1} A-C. Davis and R. Jeannerot, {\em ``Scattering off 
an SO(10) cosmic string ''}, Phys. Rev. {\bf D 52}, 1944-1954 (1995).


\bibitem{paper2}R. Jeannerot and A-C. Davis, {\em ``Constraining 
supersymmetric SO(10) models through cosmology''}, Phys. Rev. {\bf D 52}, 7220-7231 (1995).

\bibitem{paper3} R. Jeannerot, {\em ``Supersymmetric SO(10) model 
with inflation and cosmic strings''}, Phys. Rev. {\bf D 53}, 5426-5436 (1996).

\bibitem{paper4} R. Jeannerot, {\em ``New mechanism for leptogenesis''}, 
Phys. Rev. Lett. {\bf 77}, 3292-3295 (1996).


\bibitem{KolbTurner} E.W. Kolb and M.S. Turner, {\em ``The early
universe''}, Redwood City, USA: Addison-Wesley, Frontiers in
physics, 1990.

\bibitem{QFT} C. Ittzykson and J-B. Zuber, {\em ``Quantum field
theory''}, McGraw-Hill International Editions, Physics Series,
International Edition, 1985; L. H. Ryder, {\em ``Quantum field
theory''}, Cambridge University Press, 1985.

\bibitem{ShelVil} A. Vilenkin and E.P.S. Shellard, {\em ``Cosmic strings 
and other topological defects''}, Cambridge monographs on mathematical 
physics, Cambridge University Press, England, 1994; M.B. Hindmarsh 
and T.W. Kibble, Rept. Prog. Phys. {\bf 58}, 477 (1995).

\bibitem{Langacker81} P. Langacker, Phys. Rep. {\bf 72}, No. 4, 185
(1981); G.G. Ross, {\em ``Grand Unified Theories''}, Reading, USA :
Bejamin/Cummings, Frontiers in Physics, 1984. 

\bibitem{SUSY} H.P. Nilles, Phys. Rep. {\bf 110}, No 1, (1984);
J. Wess and J. Bagger, {\em ``Supersymmetry and supergravity''}, Princeton
series in physics, Princeton University Press, 1992; D. Bailin and
A. Love, {\em ``Supersymmetric gauge field theory and string theory''},
Graduate student series in physics, IOP Pub. Ltd, 1994; J.L Lopez,
Rept. Prog. Phys. {\bf 59}, 819 (1996).

\bibitem{Omega} J.L. Tonry, in {\em Relativistic astrophysics and
particle cosmology}, Ed. C.W. Akerlof and M. Srednicki, New York
Academy of Sciencess, New York, 1993. 

\bibitem{Walker} C.J. Copi, D.N. Schramm and M.S. Turner, Science {\bf 267}, 192 (1995); {\em ibid.}, Phys. Rev. Lett. {\bf 75}, 3981 (1995).
 

\bibitem{neutrinoDM} P. Langacker, Invited talk presented at Beyond
the Standard Model IV, Lake Tahoe, CA, December 1994, Preprint No.
hep-ph/9503327; A.D. Dolgov,  Nucl. Phys. Proc. Suppl. {\bf 48}, 5 (1996).
J.W.F. Valle, Nucl. Phys. Proc. Suppl. {\bf 48}, 137 (1996).

\bibitem{LSPDM} For a review on supersymmetric dark-matter see
G. Jungman, M. Kamionkowski and K. Griest, Phys. Rep. {\bf 267}, No 5,
195 (1996). 

\bibitem{axion} M. S. Turner, Pys. Rep. {\bf 197}, 67 (1990);
G. Raffelt, Phys. Rep. {\bf 198}, 1 (1990). 

\bibitem{COBE1} J.C. Mather {\em et al.}, Astor. J. {\bf 354}, L37 (1990).

\bibitem{COBE2} C.L. Bennett {\em et al.}, Astro. J. {\bf 436}, 423 (1994).  

\bibitem{inflconstr1} A.H. Guth, Phys. Rev. {\bf D 23}, 347 (1981).

\bibitem{inflconstr2} A.D. Linde, Phys. Lett. {\bf 108B}, 389 (1982);
{\bf 114B}, 431 (1982); A. Albrecht aand P.J. Steinhardt,
Phys. Rev. Lett. {\bf 48}, 1220 (1982). 

\bibitem{meet} S. Dimopoulos and H. Georgi, Nucl.\ Phys.\ {\bf 
B 193}, 150 (1981); J. Ellis, S. Kelley and D.V. Nanopoulos, Phys.\
Lett.\ {\bf 249B}, 441 (1990); U. Amaldi, W. de Boer and H.
Furstenau, Phys.\ Lett.\ {\bf 260B}, 447 (1991); P. Langacker and
M.X. Luo, Phys.\ Rev.\ {\bf D 44}, 817 (1991); M. Carena, S. Pokorski,
M. Olechowski and C.E.M. Wagner,     
Nucl.\ Phys.\ {\bf B 406}, 59 (1993); M. Shifman, Talk given at the 4th international workshop on {\em Supersymmetry and unification of fundamental interactions}, SUSY96 (May 29-June 1), University of Maryland, USA, 1996, Preprint No. TPI-MINN-96/07-T, hep-ph/9606281.


\bibitem{Linde} A.D. Linde, Phys.\ Lett.\ {\bf 259B}, 38 (1991);
Phys.\ Rev.\ {\bf D 49}, 748 (1994).

\bibitem{Lindechaos} A.D. Linde, Phys. Lett. {\bf 129B}, 177 (1983).

\bibitem{partphysbook} R.M. Barnett {\em et al.}, Phys. Rev. {\bf D 54}, 1 (1996); for updated data see PDG WWW pages (URL: http://pdg.lbl.gov/).

\bibitem{Y86} M. Fukugita and T. Yanagida, Phys. Lett. {\bf 174B}, 45
  (1986). 

\bibitem{Luty} M. A. Luty, Phys. Rev. {\bf D 45}, 455 (1992).

\bibitem{Y93} W. Buchm\"{u}ller and Y. Yanagida, Phys. Lett. {\bf 302B},
  240 (1993); M. Pl\"{u}macher, Prepint No. DESY 96-052, hep-ph/9604229; L. Covi, 
  E. Roulet and F. Vissani, Phys. Lett. {bf 384B}, 169 (1996).

\bibitem{G74} H. Georgi, in {\em Particles and fields-1974}, 
Proceedings of the Williamsburg Meeting, edited by C. Carlson, 
AIP Conf. Proc. No. 23 (AIP, New York, 1975); H. Fritsch and
P. Minkowski, Ann. Phys. (N.Y.) {\bf 93}, 193 (1975).


\bibitem{Kibble} T.W.B. Kibble, J. Phys. {\bf A 9}, 387 (1976).

\bibitem {Brandon} B. Carter, Rencontres de Moriond: Euroconferences:
Dark-matter in cosmology, Clocks and testes of fundamental laws, Villars sur Ollon, Switzerland, 21-28 Jan
1995; R. Brandenberger, B. Carter, A.C. Davis, M. Trodden, 
Phys. Rev. {\bf D 54}, 6059 (1996).


\bibitem{ShafiDvaSch}  G. Dvali, Q. Shafi and R. Schaefer,
Phys. Rev. Lett. {\bf 73}, 1886 (1994). 


\bibitem{Ed1} E.J. Copeland, A.R. Liddle, D.H. Lyth, E.D. Stewart and
D. Wands, Phys. Rev. {\bf D 49}, 6410 (1994).

\bibitem{Laz} G. Lazarides and C. Panagiotakopoulos, Phys. Rev. {\bf D
52}, 559 (1995); {\em ibid.}, Phys. Rev. {\bf D 54}, 1369 (1996).

\bibitem{sugrainfl} G. Dvali, Phys. Lett. {\bf 355B}, 78 (1995);
G. Dvali, Preprint No. CERN-TH-96-129, hep-ph/9605445; D.H. Lyth and
E.D. Stewart, Phys. Rev. Lett. {\bf 75}, 201 (1995); D.H. Lyth,
E.D. Stewart, Phys. Rev. {\bf D 53}, 1784 (1996); G.G. Ross and
S. Sakar, Nucl. Phys. {\bf B 461}, 597 (1996); L. Randall, M. Soljacic
and A. Guth, Preprint No. MIT-CTP-1499, hep-ph/9601296. 

\bibitem{Dvali} G. Dvali, Preprint No. IFUP-TH 09/95, hep-ph/9503259; 
P. Binetruy and G. Dvali, Phys. Lett. {\bf 388B}, 241 (1996); 
E. Halyo, Phys. Lett. {\bf 387B}, 43 (1996).

\bibitem{infltop} A. Shafi and A. Vilenkin, Phys. Rev. {\bf D 29},
1870 (1984); G. Lazarides and Q. Shafi, Phys. Lett. {\bf B 148}, 35
(1984); T. Barreiro, E.J. Copeland, D.H. Lyth, T. Prokopec, Phys. Rev. 
{\bf D 54}, 1379 (1996);
G. Lazarides and Q. Shafi, Phys. Lett. {\bf 372B}, 20 (1996). 

\bibitem{Rubakov} V. Rubakov JETP Lett. {\bf 33}
(1981), Nucl. Phys. {\bf 203}, 311 (1982); C. Callan
Phys. Rev. {\bf D 25}, 2141 (1982).

\bibitem{Perkins91} W.B. Perkins, L. Perivolaropoulos,
A.C. Davis, R.H. Brandenberger and A. Matheson,
Nucl. Phys. {\bf B 353}, 237 (1991); M.G. Alford,
J. March-Russel and F. Wilczek, Nucl. Phys. {\bf B
328}, 140 (1989). 

\bibitem{Kayserbook} B. Kayser, {\em ``The physics of massive neutrinos''},
World Scientific Lecture in Physics, Vol. 25, World Scientific
Publishing Co. Pte. Ltd. 1989. 

\bibitem{Bilenky} S. M. Bilenky and S. T. Petcov, Rev. of Mod. Phys.
{\bf 59}, 671 (1987). 

\bibitem{HCDM} A.N. Taylor and M. Rowan-Robinson, Nature {\bf 359},
356 (1993); R.K. Schaefer and Q. Shafi, Phys. Rev. {\bf D 47}, 1333
(1993); J.A. Holtzman and J.R. Primack, Astrophys. J. {\bf 405} 428
(1993); A.R. Liddle and D.H. Lyth, Mon. Not. Roy. Ast. Soc. {\bf 265}, 379
(1993). 

\bibitem{MSW} S.P. Mikheyev and A.Y. Smirnov, Yad.\ Fiz.\, {\bf 42},
1441 (1985) [Sov. J. Nucl. Phys. {\bf 42}, 913 (1985)]; L. Wolfenstein, Phys.\ Rev.\ {\bf D 17}, 2369 (1978); 
  ibid.\ {\bf D 18}, 958 (1979);  ibid.\ {\bf D 20}, 2634 (1980).

\bibitem{Dar} A. Dar and G. Shaviv, Astro J. {\bf 468}, 933 (1996).

\bibitem{seesaw} M. Gell-Mann, P. Ramond, and R. Slansky,
in {\em Supergravity}, Proceedings of the Workshop, Stony Brook,
edited by P. van Nieuwenhuizen and D. Freedman, North-Holland, 
 Amsterdam, 1980; T. Yanagida, in  {\em  Proceedings of the workshop
on unified theories and baryon number in the universe}, Tsukuba,
Japan, 1979, edited by O. Sawada and A. Sagamoto, KEK Report
No. 79-18, Tsukuba, 1979; R. N. Mohapatra and G. Senjan\'{o}vic. Phys.\ Rev.\
 Lett.\ {\bf 44}, 912 (1980). 
 
 
\bibitem{Kayser84} B. Kayser, Phys. Rev. {\bf D 30}, 1023 (1984).

\bibitem{Martin}  S.P. Martin, Phys.\ Rev.\ {\bf D 46}, 2769 (1992).

\bibitem{colliders} S. Ambrosio, B. Mele, M. Carena and C.E.M. Wagner,
Phys. Lett. {\bf 373B}, 107 (1996); G.F. Giudice {\em et al.},
preprint No. HEPPH-96022207, hep-ph/9602207; F. Abe {\em et al.}
Phys. Rev. Lett. {\bf 76}, 2006 (1996); S. Aid. {\em et al.},
preprint No. DESY-96-082;{\em ibid.}, hep-ex/9605002, DESY-96-056,
hep-ex/9604006; D. Buskulic {\em et al.}, Phys. Lett. {\bf 373B}, 246
(1996); S. Abachi {\em et al.} Phys. Rev. Lett. {\bf 76}, 2222 (1996); H. Baer
{\em et al.}, Preprint No. FSU-HEP-950401, hep-ph/9503479; S. Kuhlman
{\em et al.}, Preprint No. SLAC-R-0485, hep-ex/9605011. 

\bibitem{beta} M. Hirsch, H.V. Klapdor-Kleingrothaus, S.G. Kovalenko,
Phys. Lett. {\bf 372B}, 181 (1996). {\em ibid.}, Phys. Rev. {\bf D 53},
1329 (1996). 

\bibitem{Suzuki} Y. Suzuki, Talk given at the 4th international workshop on {\em Supersymmetry and unification of fundamental interactions}, SUSY96 (May 29-June 1), University of Maryland, USA, 1996.

\bibitem{DMexpt} M. Mori {\em et al.}, Phys. Rev. {\bf D 48}, 5505
(1993); D.O. Calwell, Nucl. Phys. {\bf B}, Proc. Suppl. {\bf 38}, 394
(1995); V. A. Bednyakov, S.G. Kovalenko, H.V. Klapdor-Kleingrothaus,
Y. Ramachers, Preprint No. JINR-E2-96-198, hep-ph/9606261.


\bibitem{Warren} S. Dimopoulos, J. Preskill and
F. Wilczeck, Phys. Lett. {\bf 119B}, 320 (1982). 

\bibitem{Branden89} R.H. Brandenberger, A.C. Davis and
A. Matheson, Phys. Lett. {\bf 218B}, 304 (1989);
A.Matheson. L. Perivolaropoulos, W. Perkins,
A.C. Davis and R.H. Brandenberger, Phys. Lett. {\bf 248B}, 263 (1990).
 
\bibitem{Alf89} M.G. Alford and F. Wilczek,
Phys. Rev. Lett. {\bf 62}, 1071 (1989).  

\bibitem{Aharonov} Y. Aharonov and D. Bohm,
Phys. Rev. {\bf 115}, 485 (1959).


\bibitem{Adrian} T.W.B. Kibble Acta. Phys. Pol. {\bf B
13}, 723 (1982); R. Rohm, Ph.D. THESIS, Princeton
University (1985) unpublished.

\bibitem{Kibble82} T.W.B. Kibble, G. Lazarides and
Q.Shafi, Phys. Lett. {\bf 113B}, 237 (1982).

\bibitem{Aryal87} M. Aryal and A. Everett,
Phys. Rev. {\bf D 35}, 3105 (1987)

\bibitem{Ma} Chung-Pei Ma,
Phys. Rev. {\bf D 48}, 530 (1993). 

\bibitem{Branden88} R.H. Brandenberger, A.C. Davis and
A. Matheson Nucl. Phys. {\bf B 307}, 909 (1988). 

\bibitem{Bucher} M. Bucher and A. Goldhaber,
Phys. Rev. {\bf D 49}, 4167 (1994). 


\bibitem{Michael} M.A. Earnshaw and A.C. Davis,
Nucl. Phys. {\bf B 407}, 412 (1993).

\bibitem{Marie} M. Machacek, Nucl. Phys. {\bf B 159}, 37 (1979).

\bibitem{Hindmarsh} R. Brandenberger, A.C. Davis and
M. Hindmarsh, Phys. Lett. {\bf 263B}, 239 (1991).  


%chap3


\bibitem{tanB} M. Olechowski and  S. Pokorski, Phys.\ Lett.\  {\bf 
214B}, 393 (1988). 

\bibitem{DW}  S. Dimopoulos and F. Wilczek, in Proceedings Erice Summer
School, Ed. A. Zichichi, Subnuclear Series Vol. 19 (Plenum, New York, 1983); Dae-Gya Lee and 
R.N. Mohapatra, Phys.\ Lett.\ {\bf 324B}, 376 (1994).

\bibitem{masses} G. Anderson, S. Dimopoulos, L. Hall, S. Raby and
G. Starkman, Phys.\ Rev.\  {\bf D 49}, 3660 (1994); S. Raby, in {\em 
Physics from Planck scale to electroweak scale}, Proceedings of the Workshop,  Warsaw, Poland, 1994, edited by P.Nath {\em et al.} (World Scientific, Singapore, 1995); M. Carena, S. Dimopoulos, C.E.M. Wagner,
 S. Raby, Phys. Rev. {\bf D 52}, 4133 (1995); K.S. Babu and Q. Shafi, Phys. Lett. {\bf 357B}, 365 (1995).


\bibitem{shafiun} see last paper in Ref.\cite{masses}


\bibitem{baryon} H. Murayama, H. Suzuki and T. Yanagida, Phys. Rev. Lett. {\bf 70}, 1912 (1993); H. Murayama and T. Yanagida, Phys. Lett. {\bf 322B}, 349 (1994)

\bibitem{moha} R.N. Mohapatra and M.K. Parida, Phys.\ Rev.\ {\bf D
47}, 264 (1993); Dae-Gyu Lee, R.N. Mohapatra, M.K. Parida, and
M. Rani, {\em ibid.} {\bf 51}, 229 (1995).

\bibitem{Kib} T.W.B. Kibble, G. Lazarides and Q. Shafi, Phys.\ Rev.\
{\bf D 26}, 435, (1982); A. Vilenkin and A.E. Everett, Phys.\ Rev.\
Lett.\ {\bf 48}, 1867 (1982); A.E. Everett and A. Vilenkin, Nucl.\
Phys. {\bf B 207}, 43 (1982). 

\bibitem{Vil} A. Vilenkin, Nucl.\ Phys.\ {\bf B 196}, 240 (1982);
G. Lazarides, Q. Shafi and T. Walsh, Nucl.\ Phys.\ {\bf B 195}, 157
(1982).


\bibitem{Srivastava} P.P. Srivastava, Lett.\ Nuovo Cimento {\bf 12},
 161 (1976). 

\bibitem{Albert}  A.A. Albert, Trans.\ Ann.\ Math.\ Soc.\ {\bf 64}, 552 (1948).

\bibitem{Santilli} R.M. Santilli, Hadronic Journal {\bf 1}, 223
(1978). 


\bibitem{nano} J.P. Derendinger, J.E. Kim and D.V. Nanopoulos, Phys. Lett. 
{\bf 139B}, 170 (1984).

\bibitem{54} G. Lazarides, M. Magg and Q. Shafi, Phys. Lett. {\bf 97B}, 87
(1980); F. Bucella, L. Cocco and C. Wetterich, Nucl. phys. {\bf B 248}, 273 
(1984).

\bibitem{210} D. Chang and A. Kumar, Phys. Rev. {\bf D 33}, 2695 (1986);
J. Basaq, S. Meljanac and L. O'Raifeartaigh, Phys. Rev. {\bf D 39}, 3310
(1989); X.-G. He and S. Meljanac, Phys. Rev. {\bf D 40}, 2098 (1989).

\bibitem{45} O. Kaymakcalan, L. Michel, K.C. Wali, W.D. McGlinn and 
L. O'Raifeartaigh, Nucl. Phys. {\bf B 267}, 203 (1986); R. Thornburg 
and W.D. McGlinn, Phys. rev. {\bf D 33}, 2991 (1986); R. Kuchimanchi, 
Phys. Rev. {\bf D 47}, 685 (1993);

\bibitem{Dvali1} See first paper by G. Dvali in Ref.\cite{sugrainfl}.

\bibitem{Tanmay} T. Vachaspati, Phys.\ Rev.\ Lett.\ {\bf 68}, 1977 (1992).

\bibitem{Nathan} A.C. Davis and N. Lepora, Phys. Rev. {\bf D 52}, 7265 (1995).

\bibitem{babuandbarr} K.S. Babu and S.M. Barr, Phys. Rev. {\bf D 48},
5354 (1993). 

\bibitem{Allen} B. Allen, R.R. Caldwell, E.P.S. Shellard, A.Stebbins
and S. Veeraraghavan, Preprint No. FERMILAB-Conf-94/197-A.

\bibitem{Witten} E. Witten, Nucl. Phys. {\bf B 249}, 557 (1985).

\bibitem{Patrick} P. Peter, Phys. Rev. {\bf D  49}, 5052 (1994).

\bibitem{Das} R. Das, Nucl. Phys. {\bf B 227}, 462 (1983).

\bibitem{Tanmay91} T. Vachaspati, Phys. Lett. {\bf 265B},  258 (1991).

\bibitem{Andrew} A.R. Liddle and D.H. Lyth, Phys. Rep. {\bf 231}, 1, (1993).

\bibitem{leandros} L. Perivolaropoulos, published in the Proceedings of the 1994 Summer School in High Energy Physics and Cosmology, Trieste, Italy, Jun 13 - Jul 29.

\bibitem{Sakharov} A.D. Sakharov, JETP Let. {\bf 5}, 24 (1967).

\bibitem{KRS} V. A. Kuzmin, V. A. Rubakov and M. E. Shaposhnikov,
  Phys. Lett. {\bf 155B}, 36 (1985).

\bibitem{'t  Hooft} G. 't Hooft, Phys. Rev. Lett. {\bf 37}, 8 (1976);
  Phys. Rev. {\bf D 14}, 3432 (1976). 

\bibitem{Nick} N. Manton. Phys. Rev. {\bf D 28}, 2019 (1983);
  F. Klinkhamer and N. Manton, Phys. Rev. {\bf D 30}, 2212 (1984).

\bibitem{HT} J. A. Harvey and M. S. Turner, Phys. Rev. {\bf D 42},
  3344 (1990).

\bibitem{Enqvist} K. Enqvist and I. Vilja, Phys. Lett. {\bf 295B},
  281 (1993). 

\bibitem{Jackiw} R. Jackiw and P. Rossi, Nucl. Phys. {\bf B 190}, 681
  (1981). 

\bibitem{Weinberg} E. Weinberg, Phys. Rev. {\bf D 24}, 2669 (1981).

\bibitem{embed} M. Barriola, T. Vachaspati and M. Bucher,
Phys. Rev. {\bf D 50}, 2819 (1994); N.F. Lepora and A.C. Davis,
Preprint No. CERN-TH-95-185, hep-ph/9507457.  

\bibitem{BarrMathe} S.M. Barr and A.M. Matheson, Phys. Lett. {\bf
198B}, 146 (1987).

\bibitem{ACK} D. Austin, E.J. Copeland and T.W.B. Kibble,
  Phys. Rev. {\bf D 51}, 2499 (1993). 

\bibitem{Martins} C.J.A.P. Martins and E.P.S. Shellard,
  Phys. Rev. {\bf D 53}, 575 (1996). 


\end{thebibliography}
\end{document}